\documentclass[usenatbib,useAMS]{mnras}
\usepackage{graphicx}	
\usepackage{amsmath}	
\usepackage{amssymb}	
\usepackage{multicol}        
\usepackage{bm}		
\usepackage{pdflscape}	
\usepackage{adjustbox}
\usepackage{array}
\usepackage{footnote}
\usepackage{float}
\usepackage{xcolor}
\usepackage{comment}
\usepackage{verbatim}

\defcitealias{ShengAISTATS}{S22}

\newcolumntype{C}{>{\centering\arraybackslash}m{0.8cm}}
\newcolumntype{A}{>{\centering\arraybackslash}m{1.5cm}}

\newcommand{\OIII}{[O\tiny{ }\footnotesize{III}\normalsize{] }}

\newcommand{\Ha}{H$\alpha$ }

\newcommand{\kms}{\ifmmode\,{\rm km}\,{\rm s}^{-1}\else km$\,$s$^{-1}$\fi}

\usepackage{lipsum}  

\usepackage[T1]{fontenc}
\usepackage{ae,aecompl}
\usepackage{newtxtext,newtxmath}

\title[Machine Learning Discovery of Strong Lenses in DLS]{Optimizing machine learning methods to discover strong gravitational lenses in the Deep Lens Survey}

\author[Vasan G.C., Sheng, Jones, Choi,  Sharpnack]{
Keerthi Vasan G.C.,$^{1}$\thanks{E-mail: kvch@ucdavis.edu}
Stephen Sheng,$^{2}$
Tucker Jones,$^{1}$
Chi Po Choi,$^{3}$
and James Sharpnack$^{2,3}$
\\
$^{1}$Department of Physics and Astronomy, University of California, Davis, 1 Shields Avenue, Davis, CA 95616, USA\\
$^{2}$Amazon AWS AI\thanks{Work done prior to joining Amazon} \\
$^{3}$Department of Statistics, University of California, Davis, 1 Shields Avenue, Davis, CA 95616, USA
}

\date{Last updated \today; in original form \today}

\pubyear{2022}

\begin{document}
\label{firstpage}
\pagerange{\pageref{firstpage}--\pageref{lastpage}}
\maketitle

\begin{abstract}
 Machine learning models can greatly improve the search for strong gravitational lenses in imaging surveys by reducing the amount of human inspection required. In this work, we test the performance of supervised, semi-supervised, and unsupervised learning algorithms trained with the ResNetV2 neural network architecture on their ability to efficiently find strong gravitational lenses in the Deep Lens Survey (DLS). We use galaxy images from the survey, combined with simulated lensed sources, as labeled data in our training datasets. We find that models using semi-supervised learning along with data augmentations (transformations applied to an image during training, e.g., rotation) and Generative Adversarial Network (GAN) generated images yield the best performance. They offer 5--10 times better precision across all recall values compared to supervised algorithms. Applying the best performing models to the full 20 deg$^2$ DLS survey, we find 3 Grade-A lens candidates within the top 17 image predictions from the model. This increases to 9 Grade-A and 13 Grade-B candidates when $1$\% ($\sim2500$ images) of the model predictions are visually inspected. This is $\gtrsim10\times$ the sky density of lens candidates compared to current shallower wide-area surveys (such as the Dark Energy Survey), indicating a trove of lenses awaiting discovery in upcoming deeper all-sky surveys. These results suggest that pipelines tasked with finding strong lens systems can be highly efficient, minimizing human effort. We additionally report spectroscopic confirmation of the lensing nature of two Grade-A candidates identified by our model, further validating our methods. 

\end{abstract}



\section{ Introduction}  \label{sec:intro}
Under rare alignment configurations, the gravitational potential of a massive galaxy can cause light from a distant galaxy located behind it to take multiple paths around it. This results in the formation of several distinct images of the distant galaxy around the massive galaxy, a phenomenon known as strong gravitational lensing \citep[e.g.,][]{tomasso-ara-review}. These multiple images are magnified by factors that can reach $>$10 times, making them appear brighter and more spatially extended. Such magnification makes these systems ideal for studying the formation and evolution of galaxies across cosmic time \citep[e.g.,][]{Wuyts2014,Pettini2002,Swinbank2009,SLACS2006,Leethochawalit2016}, while analysis of the lensing mass distribution enables insight into the nature of dark matter \citep[e.g.,][]{Masashi2002,Bradac2002,Marco2007,Daniel2019,Shajib2022}. 

The main current challenge in working with strong lens systems is their scarcity on the sky. Therefore, methods which are able to efficiently identify lensed galaxies from wide-area sky surveys are extremely beneficial. Automated methods will be especially valuable for lens searches in upcoming wide-area sky surveys to be carried out by the Vera Rubin Observatory, Euclid, and Roman \citep[e.g.,][]{LSST2009,euclid2011,Roman2015},  whose improvements in sensitivity, angular resolution and sky coverage will enable detection of far more lens samples than are currently known. 

Early approaches to finding strong lens systems included various algorithms searching for multiple lensed images or arc shapes, manual searches around massive galaxies, and citizen science projects \citep[e.g.,][]{MoustakasAEGIS,pca-lensfinder,Arcfinder,ringfinder,threshold-elongationmap,manualsearch-GOODS,spacewarps-ii,Belokurov2009,Diehl2009,HubbleCitizenscience2022}. While successful, these methods are time-consuming and difficult to incorporate into an automated framework. 
Convolutional Neural Networks \citep[CNNs;][]{lecun1989, alex2012}, which have been successfully developed into a standard tool in the field of computer vision in the past decade, are a promising approach to solving image recognition problems. 
Depending on the problem, there are various neural network architectures that can be optimized for the desired objectives. CNNs and machine learning techniques in general have indeed been used with success in the past few years to uncover gravitationally lensed candidates in wide-area imaging surveys \citep[e.g.,][]{Colin-CFHTLS,colin-des-2019,hsc-2018,Milad-HST,Huang_DESI-decam_2020,kids-lens-search,panstaars-2020}.

Most machine learning searches for lenses have relied primarily on supervised learning methods (i.e., using a data set consisting of labeled lensed and non-lensed galaxies to train a model). 
However, while non-lensed galaxies are plentiful, current surveys have very few known lenses to be used as positive labels. 
Instead, machine learning models are trained on simulated lenses, which can be generated in abundance \citep[e.g.,][]{Colin-CFHTLS}.
However, this presents a new problem, that the training data distribution (i.e., the simulated lenses) differs from the test data distribution (i.e., the real lenses) -- a problem called distribution shift \citep[][]{quinonero2008dataset}.
To overcome distribution shift, machine learning researchers have repurposed semi-supervised learning methods, which use unlabeled data and data augmentation to adapt the trained model to the test data \citep[][]{berthelot2021adamatch}. 

An advantage to the semi-supervised learning approach is that it can learn from the abundance of unlabeled images from the survey, which allows models to generalize better to unseen images. This is particularly useful to improve performance given millions of galaxy images that are detected in sky surveys but not included in the training data. The model performance is further improved through augmentations applied to images during training (e.g., translation and rotation). In addition to conventional transformations, a rich source of data augmentation can be derived by making use of unsupervised learning algorithms \citep[e.g.,][]{GAN2014, kingma2014autoencoding, pmlr-v9-erhan10a}. Given the range of methodologies available, we now seek to address the question of which combination of machine learning methods (supervised and semi-supervised) and augmentations are best suited for finding strong gravitational lenses.

We seek efficient models which minimize human effort by reducing the number of images that must be visually inspected to recover a given sample of lenses. In this work we apply CNN models to the Deep Lens Survey (DLS; \citealt{DLS-Wittman-2002}), which has relatively good image quality and also remains relatively unexplored in terms of machine learning searches, thus serving as a good testbed for this study. Also, because of the small size of known lenses from the DLS survey, we reserve those for use only in our test dataset. Training and validation datasets will only contain simulated lenses. In our previous methodology paper \citep[][hereafter \citetalias{ShengAISTATS}]{ShengAISTATS}, we discussed the CNN models and lens detection techniques used in this work. 
Herein, we describe our training data in detail and focus on evaluating the performance of the different models on the DLS dataset.

This paper is organized as follows.
In Section~\ref{sec:DLS-details} we give an overview of the Deep Lens Survey and our source selection used for this work. We summarize our machine learning architecture and learning methods in Section~\ref{sec:methods}.
Section~\ref{sec:TrainingData} describes the method used to generate training, validation, and testing data from DLS images. Section~\ref{sec:pr-curve} discusses our metric to evaluate the performance of the different CNN models. We discuss the results from our experiments in Section~\ref{sec:results}, including the sample of new lens candidates from DLS and spectroscopic confirmation of two systems. Finally, we summarize the main conclusions in Section~\ref{sec:conclusion}. Throughout this paper we use the AB magnitude system and a $\Lambda$CDM cosmology with $\Omega_{M}=0.3$, $\Omega_{\Lambda}=0.7$ and $\mathrm{H}_0=70$~\kms~Mpc$^{-1}$.

\section{Deep Lens Survey Data}\label{sec:DLS-details}

Here we give a brief overview of imaging data from the Deep Lens Survey (DLS) which we use to test and optimize strong lens detection methods. 
The DLS consists of relatively deep imaging over 20 square degrees in five independent $2^{\circ} \times 2^{\circ}$ fields which are widely separated in the sky \citep{DLS-Wittman-2002}. Each field was imaged in {\it BVRz} photometric filters \citep{sam-DLS-BVRz-paper} using the 4-meter Mayall telescope at Kitt Peak National Observatory or Blanco telescope at Cerro Tololo Inter-American Observatory, depending on declination. The survey was carried out over $\sim$120 nights.
The survey was designed for weak gravitational lensing measurements, with stringent requirements on image quality and limiting magnitude, such that the data are naturally well suited for identifying strong lens systems. 
Typical $5\sigma$ point-source detection limits are 25.8, 26.3, and 26.9 AB magnitude in the $B$, $V$, and $R$ filters respectively \citep{sam-DLS-BVRz-paper}. 
The $R$ band limit is only $\sim$0.6 magnitudes shallower than the expected depth to be reached by Rubin observatory's 10-year survey \textbf{\citep{Ivezic2019}}. The seeing is by design best in the R band (FWHM$\lesssim$0\farcs9) and is typically $\gtrsim$0\farcs9 in the B, V, and z bands  \citep{DLS-Wittman-2002}. Images in the $z$ band are shallowest and typically subject to worse seeing conditions. In this paper, we use only the $BVR$ data.

\subsection{Source selection and regions of interest} \label{sec:making-png}
The DLS catalog includes $\sim$5 million detected galaxies across 20 square degrees. However, only those of moderate redshift and relatively high mass will act as detectable strong lenses (i.e., with Einstein radii $\Theta_E \gtrsim 1$ arcsecond). We applied a magnitude cut of $17.5<R<22$ (similar to that used by \citealt{Colin-CFHTLS}) in order to remove objects which are unlikely to produce a detectable lensing effect. Additionally, we use SExtractor \citep{SExtractor1996} flags to eliminate saturated low-redshift galaxies, and exclusion masks to remove galaxies around bright stars or at the edge of the field. This results in 281,425 objects (hereafter referred to as the SurveyCatalog). We find that SExtractor flags and exclusion masks remove $\sim5\%$ of the galaxies from the survey which  reduces the effective sky area probed by our SurveyCatalog to $\sim19$ square degrees. We set aside 2277 ($\sim$0.8\%) randomly sampled object images from this catalog to experiment and tune the HumVI scaling parameters (discussed in Section~\ref{sec:nonlenses-generation}). All model training analysis in this paper pertains to the remaining set of 279,149 objects (hereafter referred to as the TrainCatalog).

For our analysis we extract image cutouts spanning 25\farcs7 $\times$ 25\farcs7 (100 $\times$ 100 pixels) centered on each object. This size is sufficient for galaxy- and group-scale lenses ($\Theta_E \lesssim 12$''); we do not focus on the most massive cluster lenses which are already well cataloged \citep{WittmanDLSClusters} and simpler to identify. We create color-composite images from the source $BVR$ FITS files for all targets in the SurveyCatalog (Figure~\ref{fig:lens-nonlens-example}; discussed in detail in Section~\ref{sec:nonlenses-generation}). These color composite images have smaller file sizes compared to original data, enabling us to keep the rest of the analysis computationally efficient. These images are still able to capture the detected low-suface brightness features, while not saturating the brightest objects of interest for this work.

Additionally, they are better suited for the machine learning architecture and methods used in this work (discussed in Section~\ref{sec:methods}).


\section{Deep Learning Architecture and learning methods used}\label{sec:methods}

The task at hand is to establish a machine learning (ML) algorithm that efficiently classifies the 281,425 color-composite images from the survey into lensed and non-lensed galaxies. 
Furthermore, by ranking the images from highest predicted probability of being a lens to lowest, we can order the images for human inspection.
This requires the selection of an architecture (i.e., a function that takes images as input and gives prediction probabilities as output) and learning methods (i.e., a way for our function to learn from the data). 
The key components of our ML training pipeline are a supervised convolutional neural net (CNN), domain adaptation with semi-supervised learning, and augmenting training samples with generative adversarial nets (GAN).
A more detailed account of our ML method can be found in \citetalias{ShengAISTATS}.

\subsection{Convolutional neural network architecture}\label{sec:ResNetV2}

\begin{figure}
\centerline{
\includegraphics[width=\linewidth]{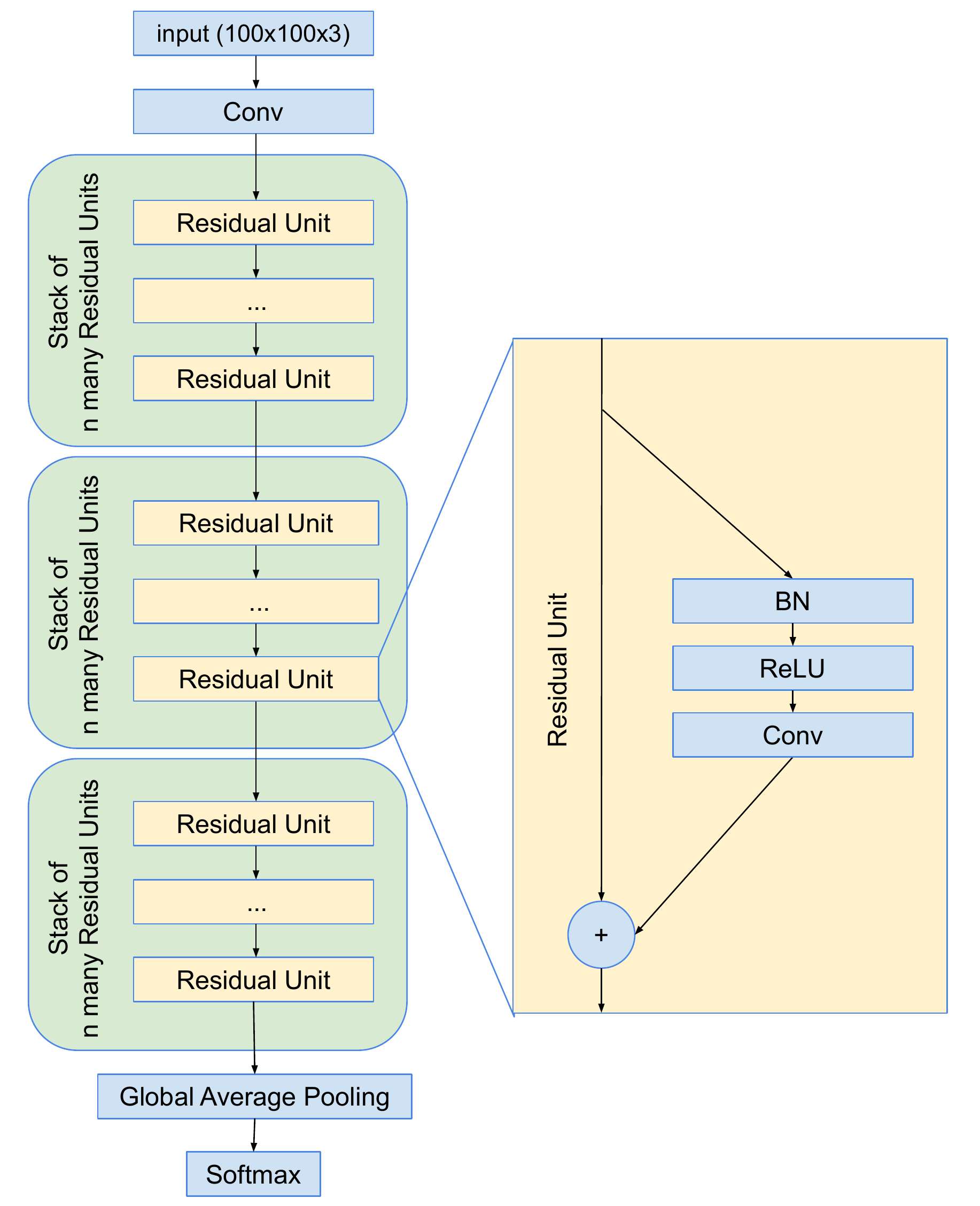}
}
\caption{Schematic depiction of the ResNetV2 deep learning architecture used in this work. The input to the network is an RGB color-composite image generated from the BVR fits files (Section.~\ref{sec:nonlenses-generation}), and the output is a value between 0 and 1 indicating the probability of the input image being a lens. The general network consists of three stacks, each containing $3n$ residual units. In this work, we use a stack size of $n=1$ resulting in a total of three residual units. Each residual unit consists of two sets of Batch Normalization (BN), Rectified Linear Unit activation function (ReLU), and Conv units, where Conv denotes a convolutional layer with kernel size $3\times 3$ and appropriate stride size. The network ends with global average pooling and a softmax layer. 
} \label{fig:resnet}
\end{figure}

CNNs have previously been used for classifying and identifying lens candidates \citep[e.g.,][]{Colin-CFHTLS}.
They are a specific form of neural network that learns translation invariant representations via trainable convolution kernels.
This is particularly well suited to astronomical images where patterns are repeated throughout the sky.
Deep CNNs are models where these learned non-linear representations of the image (called layers) are stacked on top of one another.
Deep CNNs are trained using variations of stochastic gradient descent, where an objective function is evaluated on small subsets of the data, called mini-batches, and the parameters are updated by subtracting some fraction of the objective's gradient. 

There are many choices of how precisely these layers are constructed and combined, such as selection of the convolutional kernel size, number of output channels for each convolution layer, the non-linear activation function, and the incorporation of other layers that improve performance such as Batch Normalization \citep[][]{ioffe2015batch}.
All of these details together are called the model architecture.

We make use of the ResNet version-2 architecture \citep[ResNetV2;][]{he2016identity,he2016deep} designed for the CIFAR10 dataset \citep{Krizhevsky09learningmultiple}, shown schematically in Figure~\ref{fig:resnet}. 
It is one of the widely used industry standard networks for image classification problems \citep[e.g.,][]{LITJENS201760, GU2018354, Madireddy2019}. The ResNetV2 used in this work consists of three stacks (see Figure~\ref{fig:resnet}; green blocks) and each stack consists of $n$ residual unit blocks, where $n$ is a parameter to be chosen that controls the depth of the neural network. A deeper neural network has more learning capacity but requires more computational power and training samples. Each residual unit block consists of three convolution layers of kernel size $3 \times 3$ and one skip connection. To match the feature map dimensions (width, height) and the number of channels between stacks, a few extra convolution layers are included at the input and the beginning block of each stack. Therefore, \smash{$9n + 4$} convolution layers are present in the network in total. For all the models used in this work, we adopt $n=1$. With strided convolutions, the feature map dimensions to each stack decrease by a factor of \smash{$1/2$}. The number of input and output channels to each stack are: \smash{$(16 \rightarrow 64),~(64 \rightarrow 128),~(128 \rightarrow 256)$}.

The network ends with global average pooling, a fully-connected layer and softmax. The global average pooling constrains the output to be rotationally invariant. The softmax transforms the output to be a value between 0 and 1 which can be interpreted as a probability. Throughout this work, a value of 1 is designated for lensed candidates (referred to herein as Lenses) and 0 for nonlensed candidates (referred to as NonLenses).

\subsection{Domain adaptation with semi-supervised learning}\label{sec:learning-methods}

In supervised learning, our algorithm is trained via mini-batches of images $X$ and corresponding labels $y$ ($1$ for Lenses and $0$ for NonLenses). The algorithm then tries to learn the neural network parameters, collectively referred to as $\Theta$.
The output of the neural network after the softmax activation produces a prediction $p_\Theta(X)$, which is our predicted probability of $X$ being a lens.
Our supervised learning objective function is the cross-entropy loss function, denoted $\ell_S$, which is a measure of the quality of our predictions, $p_\Theta(X)$, when compared to the true labels, $y$. Merely using supervised learning does not perform well in the face of distributional shift, and we turn to semi-supervised learning (SSL) methods which make use of the unlabeled test data to adapt to this domain.

\begin{table*}
\begin{tabular}{|c|c|}
\hline
RGB-shuffle & Randomly perturb the order 
            of the channels in the images \\
\hline
JPEG-quality & 50-100\% \\
\hline
Rot90 & Randomly rotate the 
      images by a multiple of 90 degrees    \\ 
\hline
Translations & Randomly translate the images by 
              at most 20 pixels in the up, down,
             left and right directions    \\
\hline
Horizontal flips & Randomly flips the images
                 across the x-axis \\
\hline
Color augmentation &  Randomly perturb the 
                    brightness(-0.1-0.1), saturation(0.9-1.3),
                    hue(0.96-1.00), and gamma(1.23-1.25) 
                    of the images \\
\hline
\end{tabular}
\caption{\label{tab:dataAugmentationUsed} Data augmentations used on images in the semi-supervised training pipeline.}
\end{table*}

There are many semi-supervised approaches to deep learning. The methods we explore are FixMatch\footnote{FixMatch was not part of the original lens search study since this technique had not been published at the time. We are including it in our results here to be thorough.} \citep{fixMatch2020}, MixMatch \citep{MixMatch}, Virtual Adversarial Training \citep{VAT}, Mean Teacher \citep{mean_teacher}, $\Pi$-Model \citep{pi_model}, and Pseudo-Labeling \citep{pseudo-label}. 

Most SSL algorithms follow the same template. We minimize an objective function consisting of a supervised component (i.e. $\ell_S$ losses), where the label is provided, plus an unsupervised component (i.e. $\ell_U$ losses). 
Both are optimized together over mini-batches, now consisting of labeled and unlabeled data, but without significant modification to the stochastic gradient descent algorithm.
The main feature that distinguishes our setting from typical SSL is that our training NonLenses and test set come from the same pool of data, while the simulated Lenses do not exist in the test data.
This is in contrast to \citet{Colin-CFHTLS} for example, in which they produce simulated NonLenses as well, but do not attempt domain adaptation.

In the Pseudo-Label algorithm \citep{pseudo-label}, we assign pseudo-labels to unlabeled data by taking the model's predicted class as the label. We can then use the same loss as in the supervised task (i.e., $\ell_S = \ell_U$). 
The motivation is that we are implicitly enforcing {\em entropy minimization} by forcing the model to be confident on unlabeled samples. 
An alternative approach to SSL is {\em consistency regularization}, where two independently augmented samples of the same test image are encouraged to produce similar predictions.
The $\Pi$-model algorithm \citep{pi_model} directly uses consistency regularization. 
The idea is to take two random augmentations of the same sample data point, $X$, and compute the squared difference of the model outputs for the augmented copies. 
We use $\text{aug}, \widetilde{\text{aug}}$ to denote two independent augmentations, which can be produced by selecting different randomization seeds. The unsupervised loss is then
\begin{equation}
\label{eq:consistency_regularization}
    \ell_U(X) = \left\| p_\Theta(\text{aug}(X)) - p_\Theta(\widetilde{\text{aug}}(X))\right\|^2.
\end{equation}
The choice of stochastic augmentation function is up to the modeler and will often be domain specific. 

The Mean Teacher algorithm \citep{mean_teacher} also uses consistency regularization, but replaces one of the augmentations in Equation~\ref{eq:consistency_regularization} with the output of the model using an exponential moving average (the teacher model) of model parameters, $\Theta$.
FixMatch \citep{fixMatch2020} and MixMatch \citep{MixMatch} employ both consistency regularization and entropy minimization.
MixMatch was originally proposed as a heuristic approach, and FixMatch was later derived as a more principled simplification of MixMatch and other related SSL methods. 
Virtual adversarial training \citep[VAT;][]{VAT} uses an adversarial, worst-case, augmentation.
This adversarial augmentation pushes the image in the direction which will cause the greatest increase in loss.
One downside to VAT is that the adversarial augmentations are not able to encode the domain specific prior information that random augmentations can provide (see Table \ref{tab:dataAugmentationUsed}).

\subsection{Data augmentation and GANs}
\label{sec:GANs}

Data augmentation serves as a crucial regularizer in semi-supervised learning (SSL) algorithms. Several SSL algorithms, including those mentioned in this paper such as pi-model \citep{pi_model}, MixMatch \citep{MixMatch}, and fixMatch \citep{fixMatch2020}, utilize data augmentation techniques. The data augmentation techniques we employed in our study are provided in Table \ref{tab:dataAugmentationUsed}, and are particularly well-suited for DLS images.

\begin{figure*}
\centerline{
\includegraphics[width=0.95\linewidth]{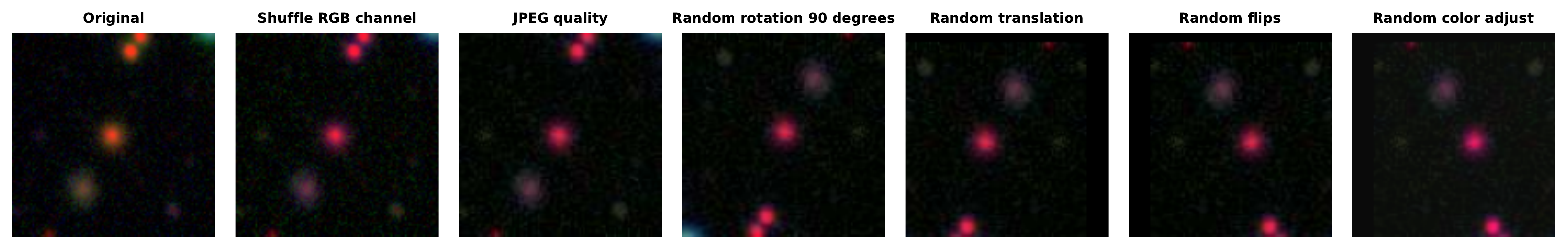}
}
\centerline{
\includegraphics[width=0.95\linewidth]{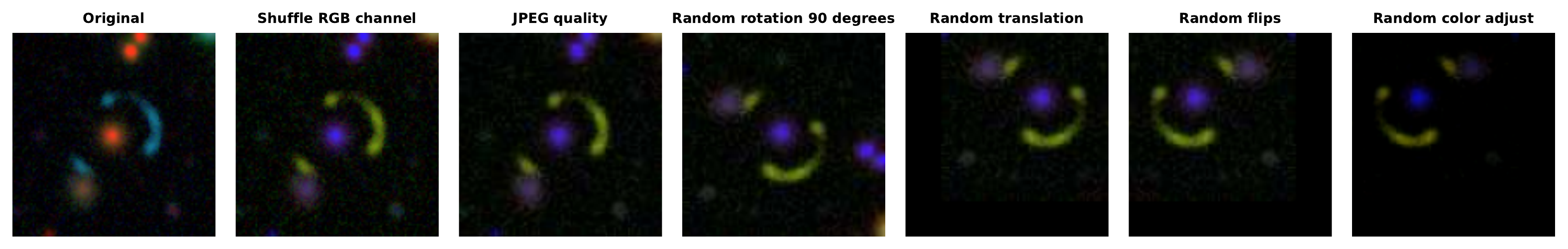}
}\caption{ Example of augmentations used during training. From left to right, the original RGB color composite image undergoes the series of augmentations described in Table~\ref{tab:dataAugmentationUsed}: RGB-shuffle, JPEG quality, Rot90, Translation, Flip, Color adjustment. The final image is then passed as input to the model.
} \label{fig:augmentation-example}
\end{figure*}

RGB-shuffle randomizes the order of channels and Color augmentation perturbs the colors in the images.
These have the effect of accounting for systematic bias in channel and color information introduced by the simulation pipeline.
JPEG-quality augmentation accounts for varying levels of noise and image quality, and applies to any color composite image 
irrespective of the format that the image is saved in (e.g., in this case we use png format instead of jpeg). Rot90, Translations, and Horizontal flips induce translational and rotational invariance in the predictions.
Examples of these augmentations are shown in Figure~\ref{fig:augmentation-example}. We note that even though some augmentations (e.g., RGB-shuffle) result in unrealistic images, our empirical tests described in Section~\ref{subsec:ablationstudy} indicate that these augmentations yield improved model performance. Domain adaptation problems employing semi-supervised algorithms (SSLs) have been shown to benefit greatly from data augmentations in general \citep[e.g.,][]{fixMatch2020}, suggesting that this effect is not specific to our lens search. 

A second tool that we use to augment our data is to generate new images that mimic the simulated lenses.
In deep learning, the state-of-the-art method to produce generative models is by using Generative Adversarial Networks \citep[GANs;][]{GAN2014,WGAN-2017}. 
GANs generate unseen samples that are distinct from the original images, but are distributionally quite similar.
These generative models are trained along with an adversarial discriminator that is attempting to distinguish between the fake and real images.

\begin{figure*}
\centerline{
\includegraphics[width=\linewidth]{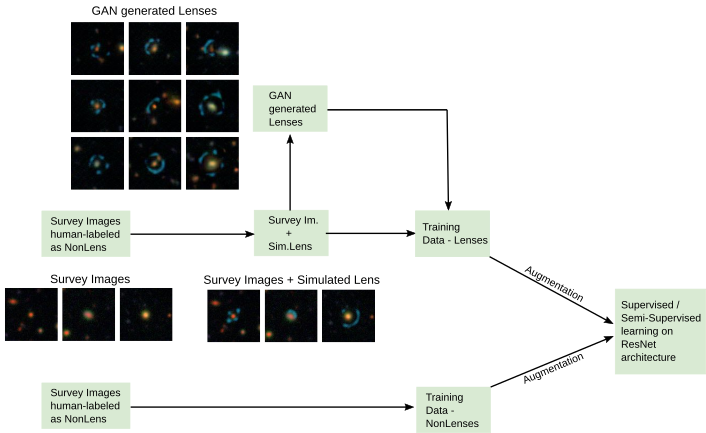}
}
\caption{Schematic of the pipeline used in this work to test the performance of different learning methods described in Sections~\ref{sec:learning-methods} and \ref{sec:GANs} (see text for details). The GAN generated lenses are only included in the training data for unsupervised learning methods (e.g., GAN+MixMatch).} \label{fig:pipeline}
\end{figure*}

We trained a WGAN-GP \citep[Wasserstein GAN + Gradient Penalty;][]{wgan_gp} on simulated lenses and add the generated images (see examples in Figures~\ref{fig:pipeline} and \ref{fig:lens-nonlens-example}) to our training set as another form of data augmentation. 
The motivation is that GANs can provide a rich source of more exotic data augmentations. 

Figure~\ref{fig:pipeline} gives a brief summary of the steps discussed thus far. The training, testing, and validation data along with the model checkpoints used in this paper are made available on our GitHub repository \footnote{\href{https://github.com/sxsheng/SHLDN}{https://github.com/sxsheng/SHLDN}}.

\section{Training and Validation data }\label{sec:TrainingData}

One of the challenges that we face in gravitational lens searches is trying to generate a training and testing dataset when having limited knowledge of the type of strong lenses that we might find in a survey. Prior to this work, \citet{kubo_deeplenssurvey_ML} used a semi-automated method to search for lensed candidates in one of the DLS fields (F2) and uncovered two lens candidates. But in order to train a machine learning model to recognize lenses, we require Lens and NonLens image samples on the order of a few thousand. This is not a problem for NonLens galaxies, as they are abundant. But this is challenging for Lenses, as the known samples are extremely small compared to training requirements. We note that although the DLS area overlaps with other surveys used for strong lens searches (e.g., SDSS), no lens candidates have been published from these other surveys within the DLS footprint. This is likely due to the shallower depth of other surveys (see Section~\ref{sec:FutureSurveys}). We must therefore generate an artificial lens training set. We describe our process of generating the training and testing datasets in this section.

\subsection{Generating the NonLenses dataset} \label{sec:nonlenses-generation}
\begin{figure}
    \centering
    \includegraphics[width=\linewidth]{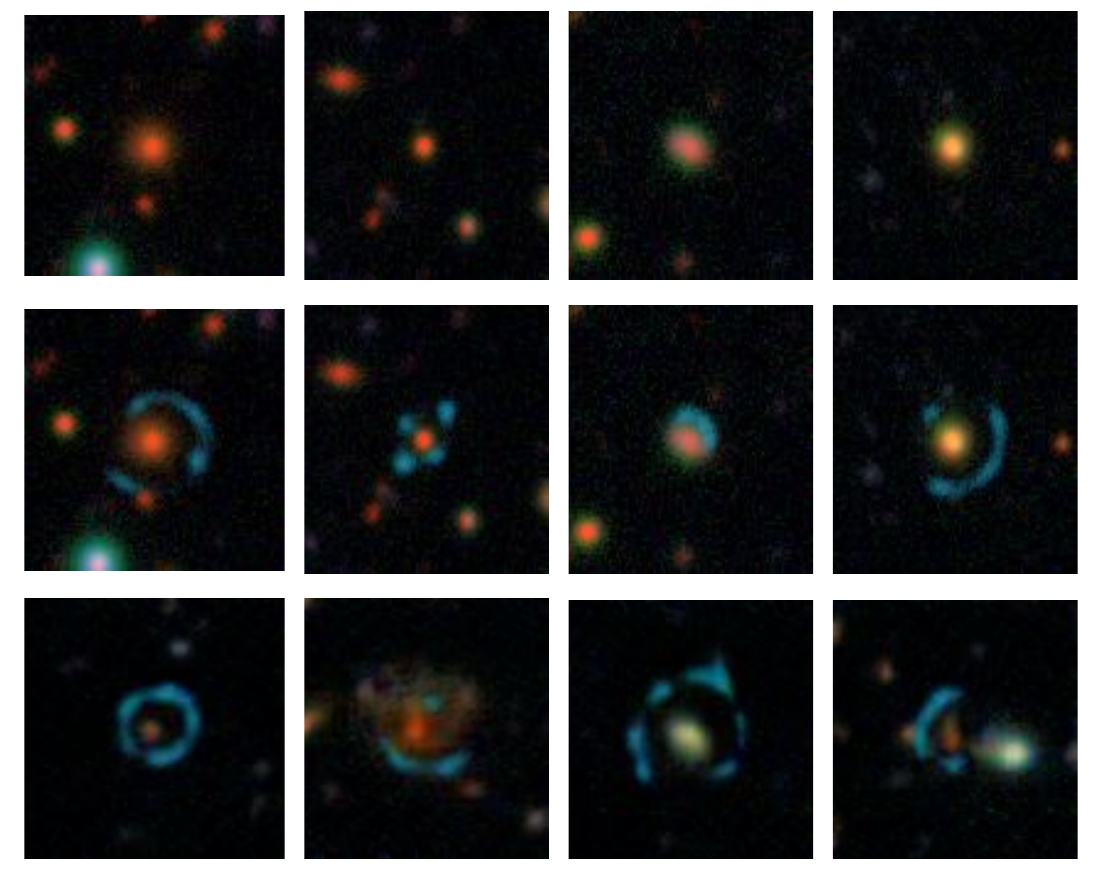}
 \caption{ \emph{Top row:} Four randomly selected color composite survey images generated by running HumVI on their respective BVR FITS files. These images are examples of NonLenses used for training the network. Each image spans 25\farcs7 $\times$ 25\farcs7 on the sky. Table~\ref{tab:glafic-params} lists the HumVI parameters used to generate these images.  \emph{Middle row:} The same set of survey images as in the top row, but superimposed with simulated lens configurations generated with glafic. Section~\ref{sec:simulating_arcs} discusses the steps involved in detail. These images are examples of Lenses used during training. \emph{Bottom row:} GAN generated simulated lenses. These are added to the training data as Lenses for our unsupervised models (e.g., GAN+MixMatch; Section.~\ref{sec:GANs}). } \label{fig:lens-nonlens-example}
\end{figure}

Color png images centered on each object in the SurveyCatalog are constructed from $BVR$ fits files using HumVI \citep{phil-humvi}. HumVI is based on the color composition algorithm described in \citet{Lupton2003} and offers several tunable parameters to control the output image (e.g., contrast). We randomly sample objects from the SurveyCatalog and visually inspect the effect of changing the HumVI parameters $s$ and $p$ which control the contrast and  color balance respectively. Although there is a degeneracy in the choice of these values, we pick ones that reasonably represent both the bright and dim features in the data (i.e., spanning the range of detectable surface brightness). Table~\ref{tab:glafic-params} lists our chosen HumVI parameters and Figure~\ref{fig:lens-nonlens-example} (top panel) shows 4 randomly selected color-composite survey images generated using these values. The chosen HumVI parameters are kept constant and applied to all images in the survey. It is beyond the scope of this work to explore the effect of choosing  different HumVI parameters on the performance of the models, but we note that color augmentations applied during training (Table~\ref{tab:dataAugmentationUsed}; Section~\ref{subsec:ablationstudy}) have the effect of making our models invariant to small perturbations in color.

\addtolength{\tabcolsep}{-5pt}    
\addtolength{\tabcolsep}{1pt}
 
\begin{table}
	\centering
	\begin{tabular}{|cc|c|} 
	    \hline 
	     & Parameter & Value  \\
	    \hline 
	    & glafic   & \\
	    \hline 
	    Position & $x_{\text{def}}, y_{\text{def}}, x_{\text{src}},y_{\text{src}}$ & U(-0.5,0.5)\\ 
	    (arcseconds) & &\\
	    PA & $\theta_{\text{def}},\theta_{\text{src}}$ & U(0,180)\\	
	    (degrees) & &\\
	    Ellipticity &$e_{\text{def}},e_{\text{src}}$ & U(0.3,0.7) \\ 
	    Dispersion &$\sigma_{\text{def}}$ & U(250,450) \\ 
	    (\kms) & & \\
	    &$r_{\text{core, def}}$ & U(0,0.5) \\
	    Brightness & & U(200,600) \\
	    (counts/$\text{pix}^2$) & &\\
	    Redshift & $z_{\text{def}}$ & U(0.3,0.7) \\
	    Redshift & $z_{\text{src}}$ & U($z_{\text{def}}$ + 0.5, $z_{\text{def}}$ + 2.5) \\
	    \hline
	     \hline
	    & HUMVI  & \\
	    \hline 
	     & -s & 0.2,0.7,1.3  \\
	     & -p & 2.5, 0.01  \\
	     & -m & 0.1\\
	    \hline 
	\end{tabular}
	\caption{Values for the glafic and HumVI parameters used to generate the simulated arcs and png color-composite images respectively. $U(x_{min},x_{max})$ indicates that the value was sampled from a uniform distribution with $x_{min}$ and $x_{max}$ being the minimum and maximum values. }
	\label{tab:glafic-params}
\end{table}

\subsection{Generating the simulated Lenses dataset} \label{sec:simulating_arcs}

As described above, the scarcity of known lensed galaxies requires us to generate simulated lens samples for training ML models. Our approach is to add simulated lensed galaxies onto survey images, as has been used successfully in prior work \citep[e.g.,][]{Colin-CFHTLS, colin-des-2019}. 
For this work, we adopt an agnostic procedure for simulating lensed arcs which does not rely on photometric measurements of the deflector galaxy. We consider all galaxies which satisfy the magnitude cut criteria described in Section~\ref{sec:making-png} (regardless of their color) for simulating the lensed arcs. We note that $\sim 50\%$ of the galaxies in our SurveyCatalog have a BPZ best fit photometric template from \cite{sam-DLS-BVRz-paper} indicating that they are massive early-type galaxies at intermediate redshifts, and are indeed likely to act as strong lenses.
We discuss the actual color distribution for lens candidates in Section~\ref{subsec:colorcolorspace}.

Given any object from the training dataset, we assume that the central galaxy (``deflector'') is at a redshift $z_{\text{def}}\in[0.3,0.7]$ and is characterized by a Singular Isothermal Ellipsoid (SIE) mass density \citep{kormann-SIE-paper}. The mass profile is dependent on the galaxy's position ($x_{\text{def}},y_{\text{def}}$), ellipticity ($e_{\text{def}}$), position angle ($\theta_{\text{def}}$), velocity dispersion ($\sigma_{\text{def}}$), and  choice of $r_{\text{core,def}}$. The values for these parameters are sampled from a uniform distribution spanning the ranges listed in Table~\ref{tab:glafic-params}. These values ensure that the resulting mass profile of the deflector is sufficient to produce a detectable lensing effect (i.e., $\Theta_E\gtrsim 1$ arcsecond). A background galaxy (``source'') is assumed to lie at a redshift $z_{\text{src}}$ with morphology given by a S\'ersic profile parameterized by its position $(x_{src},y_{src})$, central brightness (in units of counts/$\text{pix}^2$), ellipticity ($e_{\text{src}}$), position angle ($\theta_{\text{src}}$), and a S\'ersic index of 1. The value for $z_{\text{src}}$ is randomly chosen from a uniform distribution between $z_{\text{def}}+0.5$ and $z_{\text{def}}+2.5$. These values for the deflector and source redshifts are typical of spectroscopically measured values from previous strong lens surveys \citep[e.g.,][]{SL2S,SLACS,AGELpaper}.

The light from the background galaxy is traced using glafic \citep{glafic} to produce a simulated lensed arc in the image plane. The simulated lensed arcs are convolved with the point spread function (PSF) of the survey, scaled by a factor of (1,1.5,3) for the BVR filters, and then added to the $BVR$ fits images of the galaxy. We model the PSF of the survey in all the three filters as a 2D Gaussian kernel with a FWHM of $\sim$1 arcsecond corresponding to the approximate average seeing conditions. In addition to smoothing, we add Poisson noise in order to produce more realistic simulated arc images. The fits images are converted to a color png image using HumVI (as described in Section~\ref{sec:nonlenses-generation}). For this paper, we focus on generating moderately bright blue lensed arcs, and the parameter ranges that produce these configurations are listed in  Table~\ref{tab:glafic-params}.  Figure~\ref{fig:lens-nonlens-example} illustrates common configurations of the arcs produced using this method. 
However, we note that the RGB-shuffle augmentation which is applied during training produces arcs of different colors (e.g., Figure~\ref{fig:augmentation-example}). 
We find that such an approach, where the simulated arcs are not dependent on the photometric properties of the central deflector galaxy, likely serves as an additional form of augmentation. This approach prevents over-fitting of our deep learning models while allowing for rapid prototyping and testing.

\subsection{Generating the training datasets: TrainingV1 and TrainingV2} \label{sec:trainv1andv2}

Using the Lenses and NonLenses datasets, we construct two training sets: TrainingV1 and TrainingV2. The main difference between the two training sets is the number of labeled images used as Lenses and NonLenses. Prior work using CNNs \citep[e.g.,][]{colin-des-2019} have favored large training datasets (i.e., $\gtrsim$150,000 galaxies). Therefore, for TrainingV1 we use 266,301 images for non-lenses and 257,874 corresponding simulations as lenses (described in Section~\ref{sec:simulating_arcs}). Since semi-supervised training requires both labeled and unlabeled data, TrainingV1 cannot be used to test semi-supervised learning methods.

For TrainingV2, we choose the number of images for each class to be similar to those used in standard computer vision datasets such as Canadian Institute for Advanced Research-10  \citep[CIFAR-10;][]{CIFAR10-dataset} and Street View House Numbers \citep[SVHN;][]{SVHN-dataset} dataset. We use a set of 7,074 human-labeled objects as NonLenses and 6,929 corresponding simulations as Lenses. The human labeling was carried out on randomly chosen images from Field-1 (F1) of the DLS. We note that the choice of labeling the data only from F1 does not affect the results presented in the rest of the paper (see Appendix~\ref{sec:modelPerformanceByField}). The 259,248 NonLens images which are not part of TrainingV2 serve as unlabeled data for our semi-supervised learning methods (e.g., MixMatch; Section~\ref{sec:learning-methods}).

Counter-intuitively, we find that too much training data from simulated lenses and randomly selected NonLenses can hurt the performance of our algorithms. 
We refer readers to Section~\ref{subsec:largerNonLenses} and \citetalias{ShengAISTATS} for further discussion of sample size effects, which can also contribute to differences in performance between the training sets. We note that the TrainingV2 labeled datasets are comparable to the size where we find peak performance.

We performed a 90-10 split for both TrainingV1 and TrainingV2, where 90\% of the data was allocated for training the ResNetV2 model and 10\% was kept aside for validation. We chose the maximum number of epochs (passes through the training dataset) for each training combination as 100, since this was sufficient to observe a plateau in the validation metrics. For each of the training combination described in Section~\ref{sec:methods}, we conducted four independent trials and selected the checkpoint with the best validation metrics for testing it on the survey data.

\section{Metric to evaluate model performance}\label{sec:pr-curve}

We have described several models which are each tuned to optimize validation accuracy, which is measured on the validation dataset  (Section~\ref{sec:trainv1andv2}) consisting of simulated Lenses and survey NonLenses. In order to gauge the performance of the models on their ability to find real lenses from the survey, we require a testing dataset consisting of lenses from the survey, as well as a metric to evaluate them on.

\subsection{Generating the Testing dataset} \label{sec:TestingData}

Curating testing data in our case is a challenging task. As discussed earlier, only two strong lenses in the entire survey were known prior to this work, which is insufficient for meaningful evaluation. Therefore, we use an ensemble of 5 ResNet models trained on simulated lenses but using polar transformed images as input to the network. The exclusive task of this model is to find real lens candidates to add to our test dataset. We emphasize that this model is independent of the rest of the models discussed so far in this paper, and does not influence their performance in any way. Details of its implementation are discussed in \citetalias{ShengAISTATS}. It is beyond the scope of this paper to quantify the performance of ensemble models or the effect of polar transformation during training, but it is an interesting avenue for future work.

We find 52 likely lens candidates from this model, of which 27 are deemed to be good candidates upon visual inspection. Therefore, we create two testing datasets: TestV1 and TestV2. TestV1 contains all the 52 lens candidates found using our ensemble model approach, while TestV2 contains the 27 best visual candidates. NonLenses for both TestV1 and TestV2 were formed by randomly selecting 874 of our 8734 human-labeled non-lenses (Section~\ref{sec:TrainingData}).

\subsection{Precision and Recall}\label{subsec:PRCurve}

A standard metric widely used in machine learning to evaluate the performance of test data on a trained model is the Precision-Recall curve (PR curve), where precision and recall are defined as follows:
\begin{equation}
    \text{Precision} = \frac{\text{TP}}{\text{TP}+\text{FP}},
\quad    \text{Recall} = \frac{\text{TP}}{\text{TP}+\text{FN}}.
\end{equation}
Here TP, FP, and FN are the number of True Positive, False Positive, and False Negative images respectively. These values are computed by passing a labeled test dataset (TestV1 and TestV2 in this case) through a trained model (e.g., GAN+Mixmatch) and setting different prediction thresholds. 

Since the primary goal of this work is to find models which minimize the number of nonlensed images that an investigator encounters while maximizing the number of lensed images found (i.e., less FP and FN values), we seek models which have high precision at high recall. We present the results from our PR curve analysis in the next section.


\section{Results and Discussion} \label{sec:results}

\subsection{Semi-supervised algorithms with GANs and Augmentations have superior performance}\label{subsec:semi-supervised-performance}

We consider 17 variations on the learning approaches described in Section~\ref{sec:methods}: 4 supervised, 6 semi-supervised, and 7 semi-supervised with GANs. SupervisedV1 and SupervisedV2 are our baseline models. They were trained using a supervised learning approach with no data augmentation on TrainingV1 ($\sim$250,000 Lenses and NonLenses) and TrainingV2 ($\sim$7000 Lenses and NonLenses) respectively. On the other hand, SupervisedV1+DA and SupervisedV2+DA were trained using supervised learning with data augmentation (DA). The rest of the models were trained on TrainingV2 using semi-supervised learning methods with DA or with DA + GANs. In this subsection, we summarize the performance of these different models. We primarily use the PR curve (Section~\ref{subsec:PRCurve}) evaluated on our TestV1 and TestV2 sets to gauge which models perform best. We note that our methodology paper \citetalias{ShengAISTATS} includes an additional discussion of these results.

\begin{figure*}
    \centering
    \begin{tabular}{cc}
      \includegraphics[scale=0.36]{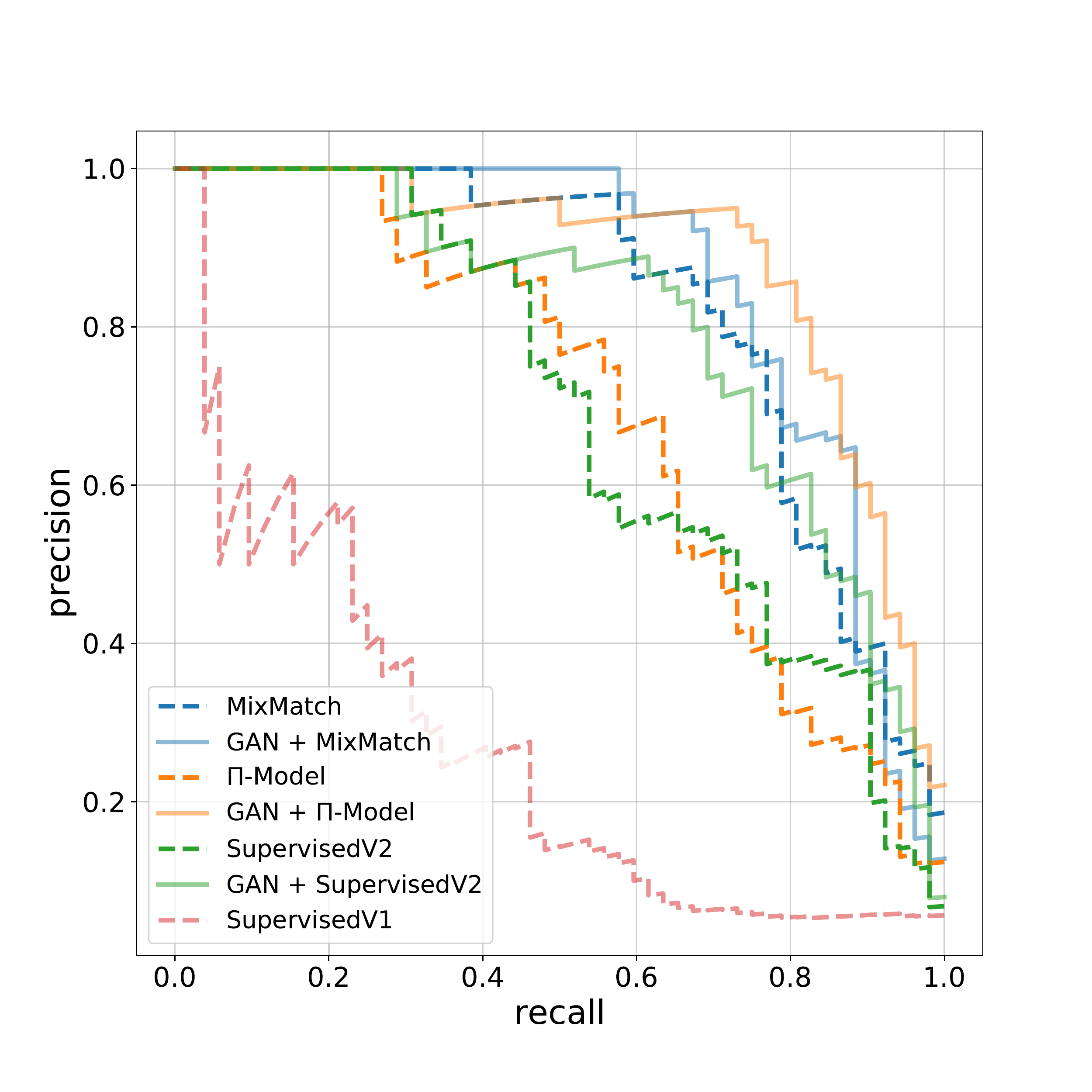}   &  \includegraphics[scale=0.36]{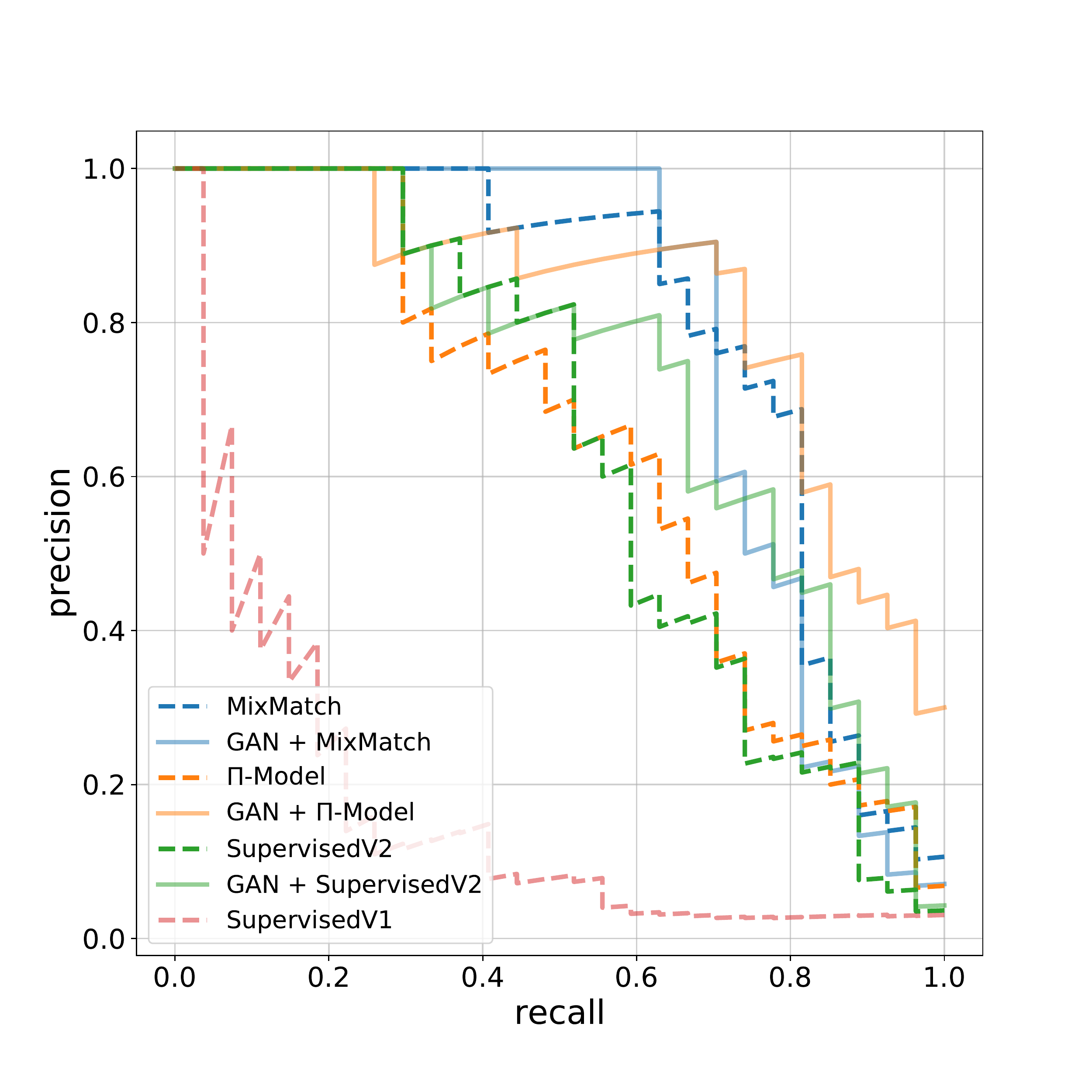} \\
    \end{tabular}
    \caption{Precision-Recall curves (PR curves) for a subset of the models described in Section~\ref{sec:methods} obtained using TestV1 (\emph{Left}) and TestV2 (\emph{Right}). TestV1 contains 52 lens candidates found using our ensemble model approach, while TestV2 contains the 27 best visual candidates (Section~\ref{sec:TestingData}). SupervisedV1 and SupervisedV2 are our baseline models. They were trained using a supervised learning approach with no data augmentation on  TrainingV1 ($\sim$250,000 Lenses and NonLenses) and TrainingV2 ($\sim$7000 Lenses and NonLenses) respectively. The rest of the models were trained on TrainingV2 with augmentations. MixMatch and $\Pi$-Model are semi-supervised learning approaches, whereas GAN+MixMatch and GAN+$\Pi$-Model use GAN generated images along with semi-supervised learning (see Figure~\ref{fig:pipeline} and Section~\ref{sec:methods} for details). GAN+SupervisedV2 uses supervised learning with GAN generated images. 
    Models which use semi-supervised learning along with GANs clearly outperform our baseline supervised learning models at all recall values, with GAN+$\Pi$-model having the highest precision at 100\% recall (see results in Table~\ref{tab:best_recall100}; we note that Table~\ref{tab:best_recall100} reports the average of our four runs while this figure shows the runs with the best precision). 
    }\label{fig:PRCurve}
\end{figure*}

We plot the PR curve obtained for our best-performing baseline models (SupervisedV1, SupervisedV2) along with a subset of semi-supervised and GAN+semi-supervised models in Figure~\ref{fig:PRCurve} (see Tables~3 and 4 of \citetalias{ShengAISTATS} for additional model results).  Table~\ref{tab:best_recall100} lists the precision value obtained for a subset of models at 100\% recall. We find that our models tend to generalize poorly when trained without any augmentations. Our baseline models, trained without any data augmentation, performed worst out of all models at every recall level. For example, at 100\% recall, the baseline SupervisedV1 and SupervisedV2 have a precision of $\sim3\%$ on our TestV2 set, whereas the GAN+$\Pi$-model has a precision of $\sim22\%$. The poor precision values of our supervised models may reflect challenges in simulating the characteristics of lenses from a survey given limited priors. Fortunately, we find that data augmentation methods are able to address this problem. We find a factor $\sim$5-10$\times$ improved precision across almost all recall levels when applying the full set of augmentations (Table~\ref{tab:dataAugmentationUsed}) to our supervised models. 

The improvement of semi-supervised over supervised algorithms suggests that valuable features can in fact be extracted from the mostly unlabeled NonLenses, providing benefits in the classification of real lenses. Adding GAN images to our training pipelines had a seemingly profound impact at all recall levels, especially at higher recalls where more difficult-to-classify images come into play. This suggests that GAN-generated images contain subtle variations which, while not necessarily significant to the naked eye, do in fact produce a strong regularizing effect when used in training.

\subsubsection{Ablation study on data augmentations}\label{subsec:ablationstudy}
We investigated the impact of each of the data augmentations we used by doing an ablation study using TrainingV2. The results from this study are tabulated in Table~\ref{tab:ablation_tests}. We find that removing GAN images from the training sets causes a noticeable decrease in model performance at all recall levels, which agrees with our earlier conclusion. It also appears that color augmentations and JPEG quality play a very significant role in model performance. Including these three augmentations in our training pipelines is apparently what allows our model to generalize so well, despite relying on simulated lenses for training. A curious result from this ablation study is that multiples of 90-degree rotations actually had a negative effect on model performance. The difference in performance is relatively small compared to that seen for other augmentations (e.g., GANs), but persists at all recall rates. A possible reason for this could be our small validation and test sets. Because the validation set is small, model selection may be biased towards certain orientations of the image. Likewise, an equally small test set may have preferred orientations that the model does not generalize to, resulting in degraded performance.

\begin{table*}
    \centering

    \begin{tabular}{|c|c|c|c|}
        \hline
        {Model} & Training data used & {TestV1 Precision(\%)} & {TestV2 Precision(\%)} \\ 
        \hline
        SupervisedV1 & TrainingV1 w/ no augmentation & 5.62$\pm$0.01 & 3.01$\pm$0.02\\
        SupervisedV2 & TrainingV2 w/ no augmentation & 5.65$\pm$0.02 & 3.06$\pm$0.04\\
        MixMatch & TrainingV2 w/ augmentation & 12.28$\pm$5.09 & 6.84$\pm$3.00\\
        $\Pi$-Model & TrainingV2 w/ augmentation  & 13.41$\pm$2.33 & 8.68$\pm$1.49\\
        GAN + Supervised & TrainingV2 w/ augmentation  & 8.25$\pm$2.85 & 6.05$\pm$2.69\\
        GAN + MixMatch & TrainingV2 w/ augmentation  & 14.13$\pm$6.53 & 7.97$\pm$3.93\\
        GAN + $\Pi$-Model & TrainingV2 w/ augmentation  & \textbf{15.2$\pm$6.21} & \textbf{22.27$\pm$7.71}\\
        \hline 
    \end{tabular}
     \caption{Average precision values were obtained for a subset of the models tested at $100\%$ recall. We note that a table with the performance of all the models at various recall values is presented in \citetalias{ShengAISTATS}. Here the average is computed from the performance of four independent runs on the test sets. The uncertainties are $1\sigma$ standard deviations from the mean.} \label{tab:best_recall100}
\end{table*}

\begin{figure*}
    \centering
    \includegraphics[width=\textwidth]{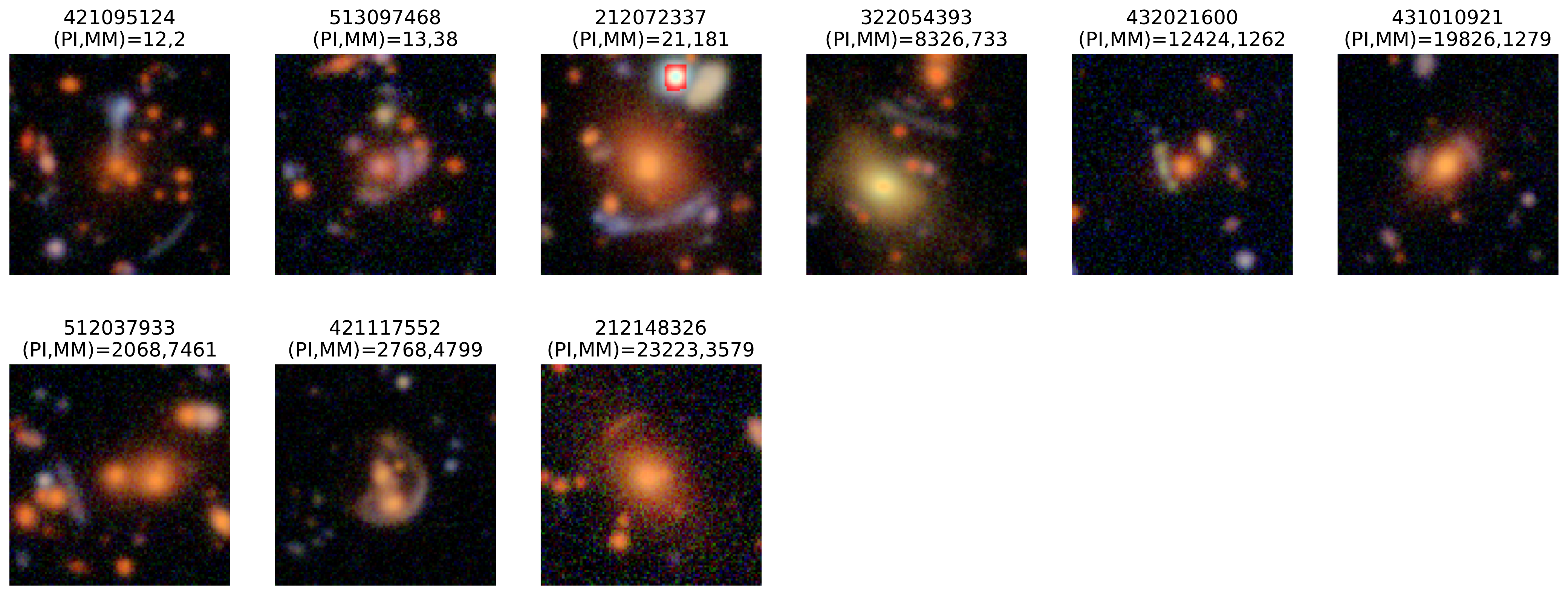}
    \caption{Grade-A lenses found in the DLS along with the rank (Section~\ref{sec:catalog}) assigned to them by GAN+MixMatch(MM) and GAN+$\Pi$-model(PI) models. All Grade-A lenses have a clear arc morphology and are located near a moderately massive galaxy or group, making them convincing lens candidates. Among these candidates, 212072337 and 432021600 have been spectroscopically confirmed to be true strong lens systems (Section~\ref{sec:specz}).
    }\label{fig:GradeA-lensesfonud}

\end{figure*}
\begin{figure*}
    \centering
    \includegraphics[width=\textwidth]{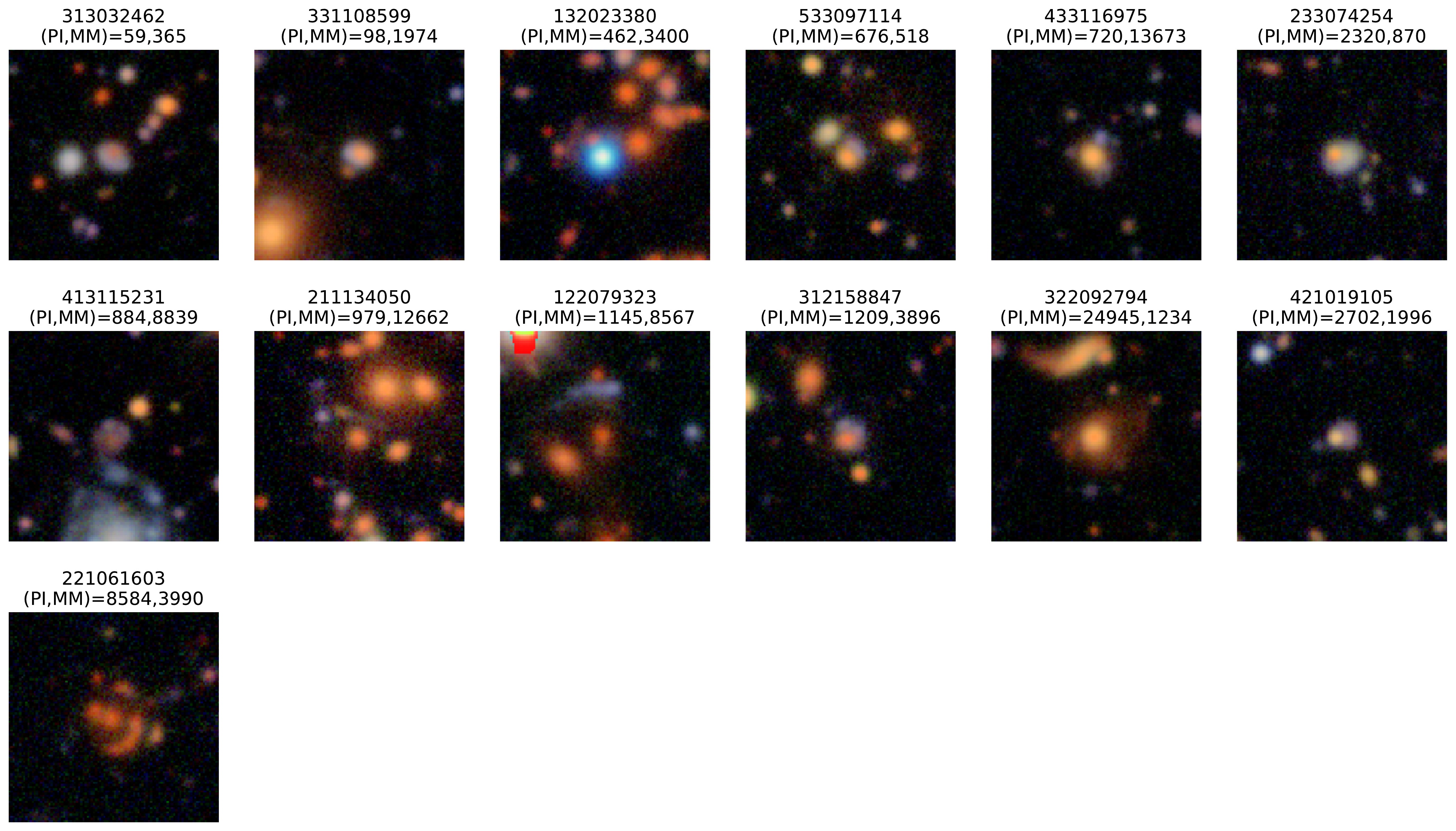}
    \caption{Grade-B lenses found along with the rank (Section~\ref{sec:catalog}) assigned to them by GAN+MixMatch(MM) and GAN+PiModel(PI) models. Targets in this category have either a tentative nebulous arc-like feature surrounding a massive galaxy, or have approximately linear extended morphology near an apparent galaxy group or cluster. It is hard to discern if these features correspond to lensed arcs or are caused by blending of multiple sources, hence the uncertain Grade-B classification. }\label{fig:GradeB-lensesfonud}

\end{figure*}

\subsubsection{Larger non-lens training samples can degrade the classifier's performance}\label{subsec:largerNonLenses}

To understand why our larger training set (TrainingV1) led to poorer generalization, we also performed a test where we fixed the number of simulated Lenses and varied the number of NonLenses in the dataset (see \citetalias{ShengAISTATS}, Table~5). As we gradually increased the number of NonLenses in the training data from 0 to 256,000, we saw that precision gradually increased and peaked at around 8000-16000 NonLenses, then started to significantly decrease to around $\sim$6\% precision for nearly all recall levels. One possible explanation for this effect is that as we increase the number of NonLenses in training, we also increase the number of NonLens false positives which appear similar to real lenses in the survey data (and perhaps even more similar to real lenses than the simulations we use). 
As a result, the decision boundary for non-lenses overlaps more with the regions occupied by real lenses, leading to higher levels of misclassification. Therefore, care must be taken in constructing training data based on simulations. Arbitrarily increasing the size of the training data can evidently lead to significantly worse performance than using a smaller well-curated training set.
 
To summarize, we find that models trained with a semi-supervised learning approach using TrainingV2 and GAN-generated images along with all of our proposed list of data augmentations have high precision values at all recall values. In particular, among the models tested, the top two performing models are GAN+MixMatch and GAN+$\Pi$-model. In the following subsection, we turn to apply these models to the full set of DLS survey images (i.e.,  SurveyCatalog in Section~\ref{sec:making-png})

\subsection{Catalog of Lens candidates found}\label{sec:catalog}
Having established which of our trained models perform best on our test set in terms of PR curves, we now turn to the key question of how many lenses are identified in the DLS and importantly, how much human inspection effort is required to find them. 
\begin{table*}
	\centering
	\begin{tabular}{|c|c||c|c|c||c|c|c|} 
	    \hline 
	    Rank & Number of unique &  Number of  & Number of  & Total lenses &  Number of  &  Number of &  Number of  \\
	    threshold & lenses investigated   &  Grade-A lenses  &   Grade-A lenses  & Grade-A  & Grade-A lenses  & Grade-A lenses  & Grade-A lenses  \\
	    & (G+MM,G+PI) & (G+MM) &  (G+PI) & (both models)  & (SupervisedV2) & (SupervisedV2+DA) & (SupervisedV2+DA+GAN) \\
	    \hline 
	    \hline 
        12 & 9, 9         & 1 & 1 & 1  & 0 & 0 & 0 \\  
        25 & 19, 16       & 1 & 3 & 3  & 0 & 0 & 0 \\ 
        100 & 67, 56      & 2 & 3 & 3 & 0 & 1 & 2\\  
        800 & 513, 430    & 4 & 3 & 4  & 1 & 2 & 3 \\  
        2800 & 1735, 1459  & 6 & 5  & 8 & - & - & - \\ 
        4000 & 2459, 2076  & 7 & 5  & 9 & - & - & - \\ 
        \hline
	\end{tabular}
	\caption{Comparison of the number of Grade-A lenses found by different models tested. The predictions from the models are ranked such that the most likely predicted lens has rank=1. The rank threshold value sets the number of lenses that an investigator has to visually inspect. The left two columns show the chosen rank threshold and the number of unique lenses that it corresponds to (removing duplicates as described in Section~\ref{sec:catalog}).	Our best performing models GAN+MixMatch (G+MM) and GAN+PiModel (G+PI) find 4 and 3 lensed candidates each among the top $\sim$500 unique images (top 800 ranks), and 7 lensed candidates each when the top $\sim2300$ images are investigated. Combining the results from both the models, we find 9 Grade-A candidates (shown in Figure~\ref{fig:GradeA-lensesfonud}). The right three columns show the number of lenses found from the SupervisedV2, SupervisedV2+Data Augmentation(DA) and SupervisedV2+DA+GAN. Although they find fewer ($\lesssim50\%$) lens candidates than our best performing models, we can see that DA and GANs are able to boost the number of lenses found from 1 to 3 at a rank threshold of 800. 
	}\label{tab:model-survey-performance}
	
	\centering
	\begin{tabular}{|c|c||c|c|c||c|} 
	    \hline 
	    Rank & Number of unique &  Number of  & Number of  & Total lenses & Total lenses   \\
	    threshold & lenses investigated   &  Grade-B lenses  &    Grade-B lenses  &  Grade-B lenses & Grade-A+B lenses  \\
	    & (G+MM,G+PI) & (G+MM) &  (G+PI) & (both models) & (both models)  \\
	    \hline 
	    \hline 
        12 & 9, 9         & 0 & 0 & 0 & 1 \\  
        25 & 19, 16       & 0 & 0 & 0 & 3 \\ 
        100 & 67, 56      & 0 & 2 & 2 & 5 \\  
        800 & 513, 430    & 2 & 5 & 5 & 9 \\  
        2800 & 1735, 1459  & 6 & 11 & 12 & 20  \\ 
        4000 & 2459, 2076  & 9 & 11 & 13 & 22 \\ 
        \hline
	\end{tabular}
	\caption{ Similar to Table~\ref{tab:model-survey-performance} but for Grade-B lenses. 
	}\label{tab:model-survey-performance-gradeb}
	
\end{table*}

We obtain a $\sim$97\% and $\sim$86\% precision at 50\% recall (i.e., to find 50\% lenses from our test set) for the GAN + MixMatch and  GAN + $\Pi$-model respectively. On the other hand, if we needed to reach $100\%$ recall (i.e., find all the lenses from our test set), the precision drops to $\sim8\%$  and $\sim22\%$ respectively (Table~\ref{tab:best_recall100}). Based on the results from \citet{Milad-HST} who searched for gravitational lenses in the COSMOS field with excellent image quality from the Hubble Space Telescope, we expect a maximum of $\sim$7 Grade-A lenses per square degree. This gives an upper bound of $\lesssim$140 lens candidates in the 20 degree$^2$ of the DLS, where the number of detectable lenses will be smaller since many of the COSMOS lenses have Einstein radii which are too small to resolve in ground-based DLS data, and because the COSMOS data are more sensitive. We estimate that half of the COSMOS lenses are unresolved in DLS based on the distribution of Einstein radii of the sample, with median $\simeq$1\farcs2 reported by \cite{Milad-HST}, such that we would expect $\sim$70 detectable lenses in the DLS survey area.
At 100\% recall, 8\% precision, and a TP$\approx$70, the number of false positive (FP) images that an investigator has to look at to find 70 lenses is $\sim 850$. If the number of detectable lenses in DLS is much lower, as suggested by samples reported from large ground-based campaigns such as the Dark Energy Survey (DES), then the total number of images and false positives which must be searched is correspondingly smaller.
We also note that these estimates are based on the assumption that the precision values obtained from our test set also apply to the survey data. A decrease in this precision value would increase the number of FPs. Therefore, considering these uncertainties, for this work we visually examine the top 12, 25, 100, 800, 2800, and 4000 predictions from the GAN+Mixmatch and GAN+$\Pi$-model, and investigate the number of lenses found.  Throughout this paper, we focus only on using relative ranks (i.e., top $n$ prediction) to assess model performance since the distribution of absolute prediction threshold values (such as those employed in \citealt{colin-des-2019}) can vary significantly between different models (Appendix~\ref{fig:model-perfomance-allfields}). The absolute prediction values obtained from different models can be calibrated, for example by scaling the obtained model weights to the softmax layer, but this is beyond the scope of our study. We note that the relative ranks which we use in this study will be unaffected under such scaling transformations.

One substantial caveat when looking at the top $n$ predictions is that, due to the density of galaxies in the sky and our image selection method, the top predictions are not necessarily unique. For example, the top 25 predictions from the GAN+$\Pi$-Model contain 17 unique sources and 8 duplicates centered on different nearby objects (shown in Figure~\ref{fig:top25-pimodel} in the Appendix). 
For the top 2800 predictions, the number of unique candidates is $\sim1600$ on average (i.e., $\sim 40\%$ are repeated). Since this is a significant portion of the number of images and would increase human effort during labeling, we remove such repetitions based on their sky coordinates. Given our image size, we remove duplicates within a radius of 26 arcseconds of each object in the top $n$ predictions. 

The remaining images are then replaced with a larger field of view, ensuring that a given region of the sky needs to be visually inspected only once. We note that removing duplicates is strictly a post-processing step. 
Two of us (KVGC and TJ) visually inspected the lens candidates and classified them into confidence categories: Grade-A, Grade-B, Grade-C, and non-lenses.
Grade-A indicates a high likelihood of being a strong lens system, on the basis of a clear arc morphology and/or coincidence with a moderately massive group of galaxies. Grade-B lenses generally have a nebulous arc-like feature surrounding a massive galaxy and/or have approximated linear extended arc morphology near a group or cluster of galaxies. It is uncertain if these features are from the lens or the effect of blending multiple sources. Grade-C lenses (not discussed in this paper) are the lowest-confidence candidates which typically show blended arc-like features likely arising from spiral arms, tidal features, or asymmetric diffuse light from the onset of mergers.

Figures~\ref{fig:GradeA-lensesfonud} and \ref{fig:GradeB-lensesfonud} show the color composite images for the 9 Grade-A and 13 Grade-B lenses found from the survey upon visually inspecting $\sim$ 2500 unique candidates (the top 4000 by rank). Their sky coordinates are listed in Table~\ref{tab:all-lenses-DLS} in the Appendix. Several of the Grade-A lenses appear to be compound lenses or part of a moderately massive group or cluster of galaxies. This is interesting since our training data consists of only galaxy-galaxy lenses. This is likely due to the addition of GAN-generated images to our training data, as the GAN-generated images (Figure~\ref{fig:pipeline}) include irregularly shaped arcs. 

\begin{figure}
\centerline{
DLS212072337 ($z=1.81$)
}
\centerline{
\includegraphics[width=\linewidth]{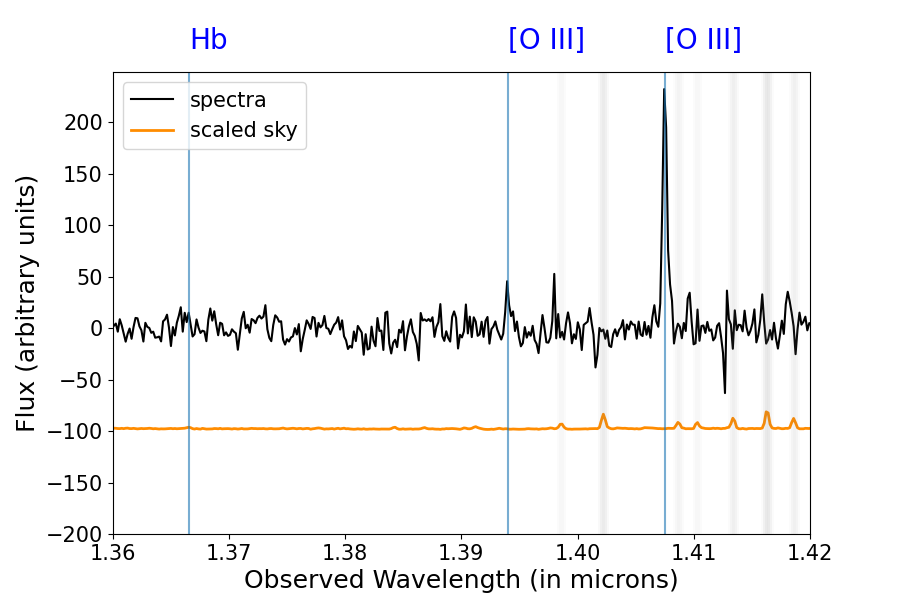}
}
\centerline{
DLS432021848 ($z=1.94$; tentative)
}
\centerline{
\includegraphics[width=\linewidth]{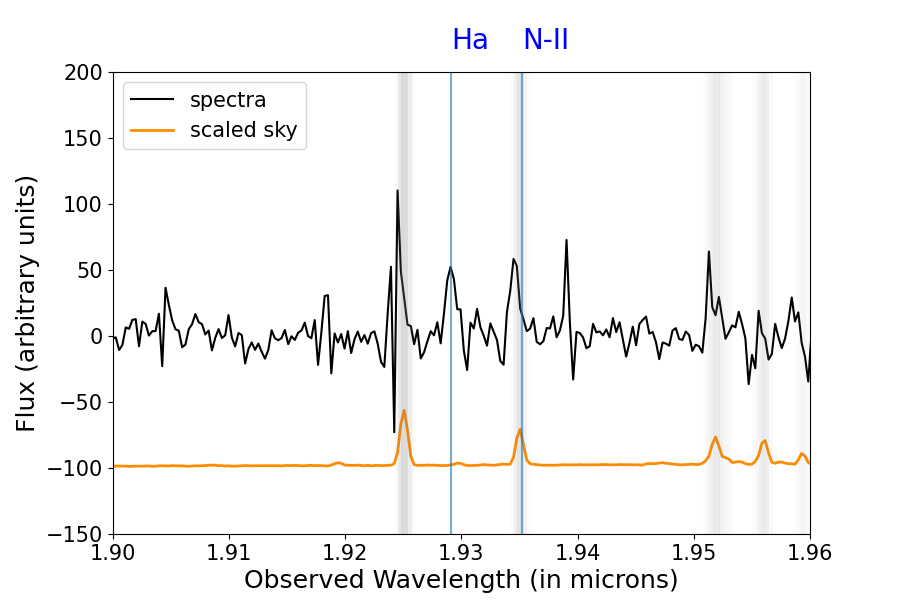}
}
\caption{{\emph (Top)}: NIRES spectra of Grade-A lens DLS212072337 at a redshift of $z = 1.81$ with prominent [O~{\sc iii}] emission lines marked in blue. {\emph (Bottom)}: NIRES spectra of DLS432021848 showing the single emission line detected at $1.93 \mu m$ which we tentatively identify as \Ha at $z = 1.94$. In both panels the scaled sky spectrum is shown in orange (offset by -100), with gray shading denoting regions affected by strong sky lines. } \label{fig:keck-spectra}
\end{figure}

\begin{figure*}
\centerline{
\includegraphics[width=0.9\linewidth]{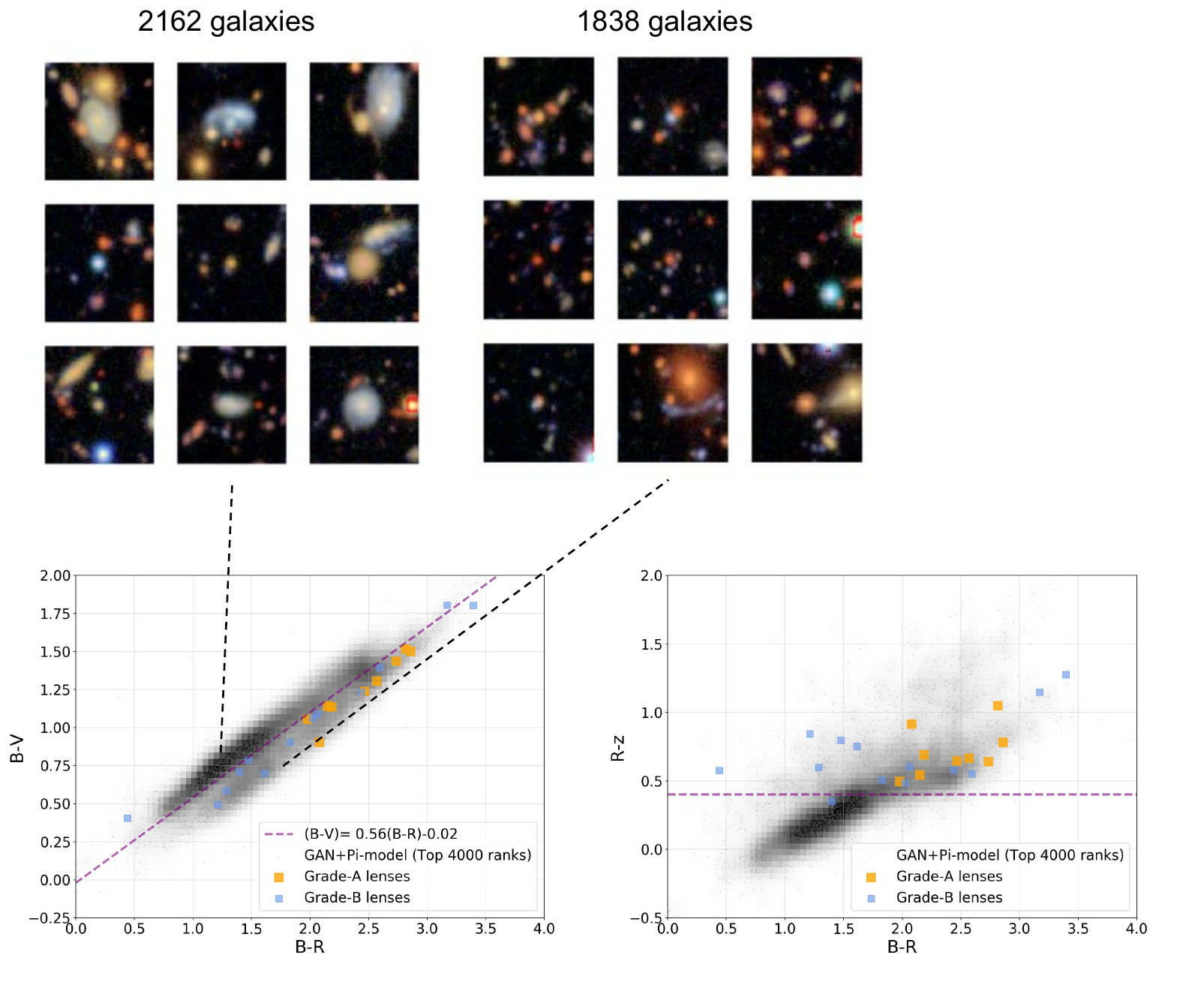}
}
\caption{Distribution of the top 4000 lenses found by GAN+$\Pi$-model in color-color space. The left panel shows $B-R$ vs $B-V$ and the right panel shows $B-R$ vs $R-z$. The images above show examples of galaxies found in the two regions of the left panel separated by the purple line. Low-z galaxy candidates are clustered in the region above the trend line whereas all of the Grade-A lens candidates are below it. 
The right panel additionally shows that lens candidates are typically redder in $R-z$ colors ($\gtrsim 0.5$). A color selection based on the purple lines in each panel would yield higher precision in our lens candidate samples while retaining nearly all of the most probable lenses. 
} \label{fig:prediction-distribution}
\end{figure*}

\subsubsection{Human inspection effort}

We now examine how much human effort is required to find the 22 Grade-A and B lens candidates. 
To quantify the effort we consider the number of lenses found at different ranks, listed in Table~\ref{tab:model-survey-performance}. The rank threshold determines the number of unique images which must be visually inspected. 
Looking at the top 800 predictions from the GAN+MixMatch and GAN+$\Pi$-model (corresponding to 513 and 430 unique lens candidates respectively), we find 4 and 3 Grade-A lenses, and 2 and 5 Grade-B lenses respectively. This is several times ($\gtrsim$3$\times$) higher sky density than has been found from the shallower ground-based DES survey, and smaller than the density found in COSMOS with HST, as expected. 
The number of lens candidates found increases to 9 Grade-A and 13 Grade-B candidates when the top 4000 candidates ($\sim$2500 unique images) are considered. This corresponds to $\sim$1~lens per deg$^2$ searched, which is $\gtrsim$10$\times$ higher sky density of lenses compared to previous shallower ground-based surveys (as we discuss in Section~\ref{sec:FutureSurveys}). 

In comparison, our supervised models (e.g., SupervisedV1, SupervisedV2) find $\lesssim 50\%$ of these top lens candidates. They also have lower precision values (Table~\ref{tab:best_recall100}), with no compelling lenses found within the top $17$ candidates inspected (whereas G+PI finds 3 within this threshold range). This again highlights the value of adding data augmentation and GAN images. The SupervisedV2+DA+GAN model finds 3 times more lenses than SupervisedV2 within the same threshold range.
These results demonstrate the efficiency with which the models explored in this work can find strong lenses.

\subsubsection{Spectroscopic confirmation of two Grade-A lenses}\label{sec:specz}

While image morphology can provide compelling evidence for strong gravitational lensing, spectroscopic redshifts are the standard to unambiguously establish the lensing nature of a system. We have obtained spectroscopy with Keck Observatory to confirm the lensing nature of two Grade-A systems presented herein: DLS212072337 and DLS432021848 (Figure~\ref{fig:keck-spectra}). Observations of the arcs were conducted with NIRES \citep{KeckNIRES} on the Keck~II telescope. Full details of the observations and data reduction are described in \citet[][]{AGELpaper}, along with spectroscopic redshifts for DLS212072337 (reported as AGEL091935+303156). We find a secure redshift of $z_{\text{arc}}=1.81$ for DLS212072337 from detection of \Ha$\lambda$6564 and \OIII$\lambda\lambda$4960,5008 emission lines. The deflector galaxy is at a redshift of  $z_{\text{def}}=0.43$, based on stellar absorption features from optical SDSS/BOSS spectra. 

We observed DLS432021848 with NIRES on 12 January 2022 using the same methodology. We obtained 6 exposures of 300 seconds each. We detect a single emission line at $~\lambda=1.93\mu\text{m}$ which we tentatively identify as either \Ha at $z_{\text{arc}}=1.94$ or \OIII$\lambda$5008 at $z_{\text{arc}}=2.85$. However, we are unable to confirm the redshift with other strong lines, which fall in regions of poor atmospheric transmission at both potential redshifts. We find further support for the lensing nature of DLS432021848 from its morphology in follow-up HST imaging (discussed in Section~\ref{sec:FutureSurveys}), which shows clear kurtosis and evidence of multiple lensed images. Thus we are reasonably confident that this is indeed a strong lensing system on the basis of high-resolution imaging, despite the limited spectroscopic information. 
Together with DLS212072337, these results give additional confidence in the sample of lens candidates presented in this paper and demonstrate that our methods are successful. 

We note that redshifts are known for two additional Grade-A candidate deflectors (DLS212148326, DLS421095124) from archival data. 
DLS212148326 is at $z_{\text{def}}=0.424$ from SDSS/BOSS spectra, while DLS421095124 is part of a massive galaxy cluster spectroscopically confirmed at $z_{\text{def}}=0.680$ \citep[][reported as  DLSCL J1055.2-0503]{Wittman2003Cluster,Wittman2006}. 
These redshifts are promising, as the distances and approximate masses are consistent with the deflection angles implied by the strong lensing interpretation of these images.

\subsubsection{Distribution of lensed candidates in color-color space}\label{subsec:colorcolorspace}

The analysis and model performance described thus far in the paper is based on a source selection using an intentionally simple $R$ band magnitude cut and SExtractor flags (Section~\ref{sec:making-png}). We have demonstrated in the above sections that such cuts are sufficient to search for lensed candidates in the DLS. 
However, more sophisticated selections can increase the efficiency of lens searches. 
Here we briefly consider how color selection can provide higher-purity samples. 

In Figure~\ref{fig:prediction-distribution} we show the distribution of Grade-A and B lenses from Section~\ref{sec:catalog} in various color-color spaces, along with the top 4000 ranked images from the GAN-$\Pi$-model as an example. These colors generally correspond to the central (candidate deflector) galaxy. The top lens candidates are not distributed uniformly, and we demonstrate two color-color selections where the top candidates are clustered: $(B-V) < 0.56(B-R) - 0.02$ (purple line in left panel), and $R-z\gtrsim0.4$ (right panel). Such simple color cuts can retain all Grade-A lenses while removing the majority of false positives, thereby reducing the required human inspection effort. Physically, these colors are indicative of 4000~\AA\ breaks at redshifts $z\gtrsim0.25$ (i.e. in the $V$ or $R$ band) whereas lower-$z$ galaxies are less likely to act as strong lenses.

The distribution of lens candidates in color space suggests that the precision of our models can be further improved by adopting color criteria as a pre- or post-processing step, with minimal loss of the best candidates. Using photometric redshift and mass estimates is a similar and potentially even more promising method \citep[][]{sam-DLS-BVRz-paper} although it is beyond the scope of this paper. 
Alternatively, a state-of-the-art automated means to address this would be by using self-similarity based approaches \citep[e.g.,][]{SelfSimilarity2021}, wherein a CNN further classifies the lens probabilities based on their similarity with each other.

\begin{figure}
        \centering
    \begin{tabular}{ccc}
       & Simulated Lens & \\
       \includegraphics[scale=0.7]{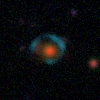}   & \includegraphics[scale=0.97]{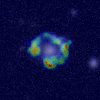}   & \includegraphics[scale=0.97]{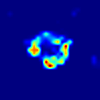}    \\

                 & Lenses found in DLS & \\
       \includegraphics[scale=0.7]{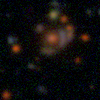}   & \includegraphics[scale=0.97]{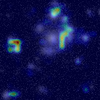}   & \includegraphics[scale=0.97]{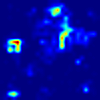}    \\

       \includegraphics[scale=0.7]{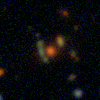}   & \includegraphics[scale=0.97]{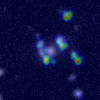}   & \includegraphics[scale=0.97]{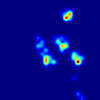}    \\   
     
        & False Positive & \\
       \includegraphics[scale=0.7]{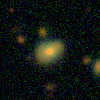}   & \includegraphics[scale=0.97]{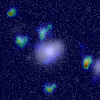}   & \includegraphics[scale=0.97]{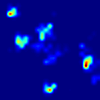}    \\
    
       \includegraphics[scale=0.7]{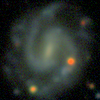}   & \includegraphics[scale=0.97]{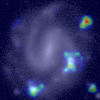}   & \includegraphics[scale=0.97]{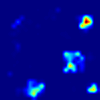}    \\
           
         & NonLens & \\
       \includegraphics[scale=0.7]{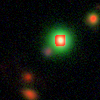}   & \includegraphics[scale=0.97]{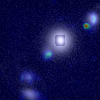}   & \includegraphics[scale=0.97]{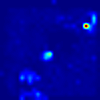}    \\  
  
    \end{tabular}

    \caption{Grad-CAM++ heatmaps for an example simulated lens, two Grade-A lenses, two false positive lenses, and a NonLens. The left column shows the color composite image obtained from HumVI and passed to the model. The right column shows the Gradcam++ heatmaps. The red and green shading indicates regions of high and moderate importance to the model, respectively, whereas blue represents low importance. The middle column shows the heatmaps superimposed on input images for visualization purposes. For the simulated lens, we can clearly see that the entire lensed arc region is taken into consideration. For the Grade-A lens candidates found in DLS, we also find that the lensed arc features are considered important by the model, despite a range of lensing morphologies and colors. This suggests that models have indeed successfully generalized to the survey data. Notably, the massive deflector (i.e., the luminous red galaxy) causing the lensing effect is not highlighted in the simulated or candidate lens systems. 
    Additional objects in the field are also highlighted in heatmaps for the Grade-A lenses, which is also apparent in the False Positive and NonLens examples. In the case of the False Positives, the highlighted object distributions resemble an ``Einstein cross'' lens configuration. 
    Heatmaps for all the Grade-A lenses are provided in Figure~\ref{fig:allgradcamImages} in the appendix. 
    }\label{fig:gradcam-heatmaps}   
\end{figure}

\subsection{Lensing signatures identified by the models}\label{subsec:Gracam}

We now examine which features of the lens candidate images are most relevant for the model predictions. 
Deep neural networks (such as ResNetV2 used in this work) are often considered as ``black boxes'' with all input information collapsed to a simple prediction for the user to interpret. 
Having only a single output, it is impossible to discern which distinguishing features of a gravitational lens are actually being identified and considered by the models. Fortunately, in the past few years, there have been a variety of methods proposed to alleviate this such as occlusion methods, Guided Backprop \citep{guided_backprop}, CAM \citep{cam}, Grad-CAM \citep{grad-cam}, Grad-CAM++ \citep{grad-cam++}, and DeepSHAP \citep{Fernando_2019}.

Gradient-based interpretation methods (e.g., Grad-CAM++) effectively compute gradients on intermediate feature maps of the network to determine the importance of a feature. These gradient maps can then be overlaid on top of the original input image, in order to assess which image regions are contributing most to the predicted output from the classifier. These methods are not without drawbacks \citep[e.g.,][]{sanity_checks_saliency_maps} but can provide valuable insight. Here we use Grad-CAM++ to analyze some of our trained models.

Figure~\ref{fig:gradcam-heatmaps} shows Grad-CAM++ heatmaps obtained for a few illustrative examples. We consider a simulated lens from the training data, real Grade-A lenses from the survey, false positive images (i.e., images which are classified as lenses but show no visual evidence of lensing), and a non-lens. In the case of the simulated lens, it is clear that the model is indeed making its prediction based on the lensed arc features. For the Grade-A lenses, the model does indeed discern the lensed arcs, but there are additional unrelated regions within the images that also influence its decision. Curiously, the central massive deflector galaxy is not highlighted in these cases. In the case of the false positives, the model encouragingly is not misled by the extended central galaxies, but rather the heatmap highlights multiple sources of similar color which surround the central galaxy. For example in the spiral galaxy false-positive image, it is clear that the model picks up on the three nearby red objects. The location and color of these nearby objects is indeed similar to plausible multiple-image lensing configurations. It thus appears that the model has successfully learned to identify the astrophysical signatures of strong lensing.

\subsubsection{Finding red arcs}\label{subsec:redarcs}
As discussed in Section~\ref{sec:simulating_arcs}, our Lens dataset used for training only consists of lensed arcs with blue optical colors. However, it is encouraging that the models have also identified red arcs such as the system DLS212148326 (Figure~\ref{fig:GradeA-lensesfonud}). The network may be learning to identify red arcs through color augmentations (Figure~\ref{fig:augmentation-example}). Although red-lensed arcs are known to exist, presumably a training dataset consisting of only blue arcs is not ideal to robustly search for and quantify them. It could be the case that adding more augmentations or fine-tuning existing ones might suffice to search for arcs of various colors. Alternatively, a broader range of arc colors could be used in the simulated training set, or a separate classifier could be constructed from a training set of red arcs. 
Given our adopted training set, we consider the number of red-lensed arcs found from this work to be a lower limit (relative to the blue arcs). Additionally, there are likely many fainter blue or red arcs which our training set does not represent, although the detection of fainter objects is naturally more challenging.

\subsection{Implications for future large-area sky surveys: sensitivity and angular resolution}\label{sec:FutureSurveys}

The next generation of wide-area sky surveys is expected to uncover $\gtrsim 10^5$ strong lens systems \citep[e.g.,][]{oguri2010,Collett2015}. Here we consider the gain in lens detection with survey depth and angular resolution based on our DLS sample from Section~\ref{sec:catalog}. We compare the sky density of detected lens candidates with two other illustrative examples of CNN-based searches in Table~\ref{tab:different-survey-comparision}. 
In our DLS search, we find $\sim$0.5 Grade-A lenses per square degree (or $\sim$1 Grade-A+B lenses per square degree). This is considerably larger than found in shallower surveys such as SDSS and DES, which have uncovered $\sim$0.1 lenses per square degree (in regions far from the galactic plane). 
While these surveys have a comparable seeing-limited resolution, sharper image quality enables more lenses to be found. 
An example is the search of COSMOS HST imaging by \cite{Milad-HST} using a CNN approach, which found 13 Grade-A candidates and 70 Grade-A+B candidates in the 2 square degree field (i.e., $\sim$35 per square degree). Therefore, we see that the sky density of detectable strong lens systems increases by $\sim$10 times when going from shallower ground-based surveys (e.g., SDSS) to the DLS, and by another factor of $\gtrsim$10 when the angular resolution is improved by an order of magnitude with space-based HST imaging at modest depth. 
These results generally support the predictions of large lens samples which will become detectable with near-future surveys planned with the Rubin \citep{LSST2009}, Roman \citep{Roman2015}, and Euclid \citep{euclid2011} observatories.

\begin{figure}
\centerline{
\includegraphics[width=\linewidth]{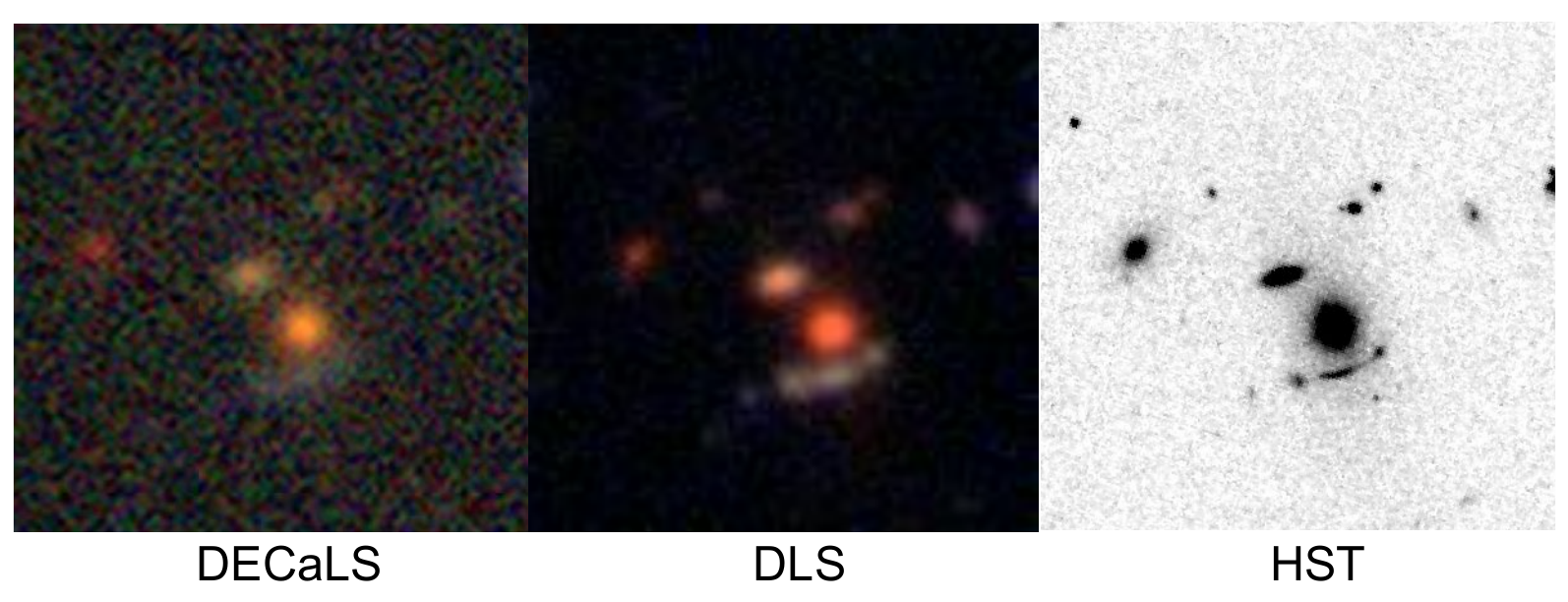}
}
\caption{Comparison of the image quality from different observations of the lens system DLS432021848, which shows a prominent blue arc in DLS imaging (below center of images; all panels show the same field of view). \emph{Left}: The arc is apparent but not well detected in DECaLS imaging, which has modest sensitivity. This image would likely be flagged in a low-confidence category and indeed was not identified in previous lens searches \citep[e.g.,][]{Huang_DESI-decam_2020}. 
\emph{Middle}: DLS image of the target showing a prominent blue arc-like feature below the red deflector galaxy, characteristic of a gravitational lens system. The increased sensitivity of DLS compared to DECaLS imaging (Table~\ref{tab:different-survey-comparision}) enables clear arc detection. 
\emph{Right}: Near-infrared image of the same target observed with HST, with a diffraction-limited angular resolution approximately 6 times sharper than DLS or DECaLS images. The HST image reveals the lensed arc morphology at a high signal-to-noise ratio. This demonstrates the capabilities of a ground-based telescope at good depth (e.g., DLS), and a diffraction-limited space-based telescope with moderate exposure time (e.g., HST).} \label{fig:dls-decals-hst-comparison}
\end{figure}

To visually illustrate the detection of lenses at different depths and angular resolutions, Figure~\ref{fig:dls-decals-hst-comparison} compares DECaLS, DLS, and HST imaging\footnote{The HST image was secured as part of program HST-GO-16773 targeting lens candidates identified primarily in DES and DECaLS imaging \citep{AGELpaper}. In brief, the HST image in Figure~\ref{fig:dls-decals-hst-comparison} was taken with WFC3-IR in the F140W filter with $\sim$30 minutes of exposure time ($<$1 orbit), and reduced using standard procedures. Details of the HST program will be described in a forthcoming paper.}
for the Grade-A lens candidate DLS432021848 found in this work. 
A blue arc is clearly visible in the DLS image and shows typical lensing morphology in the high-resolution HST image. However, the arc is only marginally visible in shallower DECaLS imaging. Indeed, most (if not all) of the Grade-A lens candidates found from this work would be difficult to detect in shallower imaging surveys (e.g., DECaLS; hence for example they are not included in the catalog of \citealt{Huang_DESI-decam_2020}).

\begin{table}
	\centering
	\begin{tabular}{|c|c|c|c|c|} 
	    \hline 
	    Survey & Lenses found & 5$\sigma$ point & FWHM & References \\
	    & per sq.deg & source detection & & \\
	    &            & (r/R/F814W-band  & & \\
	    &            &  magnitude) & & \\

	    \hline 
        DES/DECaLS & $\sim0.1$ & 23.6 (r) & 0\farcs98 & J19\\
        DLS & 1 & 26.7 (R) & 0\farcs9 & This work \\  
        COSMOS & $\sim 35$ &  27.2 (F814W) & 0\farcs07 & P18,K07\\
        \hline
	\end{tabular}
	\caption{ Number of lenses found using machine learning methods per square degree of sky in different surveys, along with the $5\sigma$ point source detection depth and median angular resolution (given as the FWHM: full-width at half maximum). We note that CNN and grading methods employed to find lenses in each survey are different; the density of lenses should thus be treated as an approximate comparison. References are as follows. 
	J19: \citet{colin-des-2019}, P18: \citet{Milad-HST}, K07: \citet{COSMOS2007}. 
	}
	\label{tab:different-survey-comparision}
\end{table}

Given the detectability of many lens systems with upcoming surveys, it is clear that machine learning approaches (such as those we have explored here) will be vitally important for the efficient selection of large samples. We have also demonstrated the feasibility of spectroscopically following up on these moderately faint arc systems (Section~\ref{sec:specz}), which will be vital for confirmation and subsequent analyses.


\section{Conclusions} \label{sec:conclusion}

In this paper, we have evaluated the performance of different CNN learning approaches  and data augmentations on their ability to efficiently find gravitational lens candidates in the Deep Lens Survey. We make use of the deep learning architecture ResNet for our experiments, along with a training dataset consisting of simulated Lenses and survey image NonLenses. We demonstrate that by using these state-of-the-art semi-supervised learning approaches, we can greatly reduce the human effort required to find lensed candidates from a survey. We summarize our key results below.

\begin{enumerate}
    \item Among 17 variants of learning approaches tested in this work, we find that our best performing models (i.e., those which have high precision and minimize false positives during human inspection) are GAN+MixMatch and GAN+$\Pi-$model. They have a precision of $\sim86\%$ and $\sim97\%$ at 50\% recall and, $\sim22\%$ and $\sim8\%$ at 100\% recall respectively. In comparison, our supervised models have a precision of $\sim3\%$ at 100\% recall. This increase in the performance of the best models can be attributed largely to three factors. (1) They leverage data augmentation (Table~\ref{tab:dataAugmentationUsed}) during training, which helps them to generalize better. (2) The datasets used to train these models to contain simulated Lenses as well as GAN-generated images (Section~\ref{sec:methods}), which serves as an additional form of data augmentation. (3) Both of these top models employ a semi-supervised learning approach (MixMatch, $\Pi$-model) which enables our methods to adapt to distributional shift (Section~\ref{sec:learning-methods}). 
    These results indicate that data augmentation, GANs, and semi-supervised learning are highly effective approaches for building an efficient lens classifier. 
    
    \item We investigated the Grad-CAM++ feature maps (Section~\ref{subsec:Gracam}) used by our best performing models to make their predictions, finding that they indeed are influenced mostly by lensed arc regions and are generally not misled by other galaxies/artifacts (e.g., diffraction spikes) in the images. This supplements our results presented above that salient information regarding the arcs needed for classification has been successfully learned by the models through our methods. This is encouraging for future lens searches, since simulated Lenses used in this work are generated without relying on photometric data of the deflector galaxy (Section~\ref{sec:TrainingData}), making it simpler to automate the task of generating a training dataset.

    \item Applying the GAN+MixMatch and GAN+$\Pi$-model to the entire DLS survey, and visually inspecting the top $\sim2500$ lens candidates, we find 9 Grade-A and 13 Grade-B lensed candidates (22 in total). 3 out of the 9 Grade-A candidates are found within the top 17 ranked images. The number of lenses found in the DLS corresponds to $\sim10\times$ higher sky density of lenses per deg$^2$ compared to the shallower DES/DECaLS survey imaging and supports predictions that vast numbers of lens systems ($\gtrsim10^5$) will be detectable in the upcoming generation of sky surveys. We further confirmed the lensed nature of 2 Grade-A candidates with spectroscopy and high-resolution imaging, demonstrating that our methods are successful.
    
\end{enumerate}

We have generally explored methods intended to find as many lenses as possible while minimizing human inspection effort. While there are likely additional detectable lenses beyond those we have identified, it is encouraging that our models have been able to identify lenses that are not represented in the training set. In particular, our training set focused on blue lensed arcs, while our models also find red arc candidates such as DLS212072337 (Section~\ref{subsec:redarcs}), although at a lower rank compared to the bluer lenses. 
Additional augmentation methods and/or training datasets may be able to provide further improvement for diverse lens system properties. Another straightforward improvement to our lens search efficiency is to include simple cuts in color-color space as demonstrated in Section~\ref{subsec:colorcolorspace}. Such cuts can help increase the model precision by excluding sources that are not likely to act as strong lenses based on their color and magnitude (which is physically related to their mass and distance). 
Since our sample is agnostic to color information, our results are well-suited for assessing the color space distribution of the best lens candidates. 

The scope of our models is currently limited to the DLS. However, our methodology can be adapted for other data sets, and we note that the DLS fields overlap with wide-area surveys such as DECaLS and SDSS. Exploring ways to translate these models across surveys would be greatly beneficial. 
Finally, confirming the lensing nature of new candidates either through spectroscopy (Section~\ref{sec:specz}) or via arc morphology (Section~\ref{sec:FutureSurveys}) is essential for a variety of investigations, including probes of galaxy evolution and cosmology. We have demonstrated the feasibility of confirming moderately faint arcs in our sample. 
Accomplishing confirmation for the thousands of lenses that will be discovered in forthcoming surveys (such as with Rubin/LSST, Roman, and Euclid) will aid in our understanding of the formation and evolution of galaxies and the contents of the Universe. 

\section*{Acknowledgements}
We thank Imran Hasan, Sam Schmidt, David Wittman, Tony Tyson and Anupreeta More for their helpful discussions which greatly improved this work. We are immensely grateful to Brian Lemaux and Debora Pelliccia for manually labeling a subset of the survey data. 
We thank the referee for many helpful comments which improved the content and clarity of this manuscript. 
TJ and KVGC gratefully acknowledge financial support from NASA through grant HST-GO-16773, the Gordon and Betty Moore Foundation through Grant GBMF8549, the National Science Foundation through grant AST-2108515, and from a Dean’s Faculty Fellowship.
Some of the data presented herein were obtained at the W. M. Keck Observatory, which is operated as a scientific partnership among the California Institute of Technology, the University of California and the National Aeronautics and Space Administration. The Observatory was made possible by the generous financial support of the W. M. Keck Foundation.
The authors wish to recognize and acknowledge the very significant cultural role and reverence that the summit of Maunakea has always had within the indigenous Hawaiian community.  We are most fortunate to have the opportunity to conduct observations from this mountain. 
Some of the results herein are based on observations with the NASA/ESA Hubble Space Telescope obtained from the Mikulski Archive for Space Telescopes at the Space Telescope Science Institute, which is operated by the Association of Universities for Research in Astronomy, Incorporated, under NASA contract NAS 5-26555. Support for program number HST-GO-16773 was provided through a grant from the STScI under NASA contract NAS5-26555.

\section*{Data availability}
The data underlying this article are available on our GitHub repository (\href{https://github.com/sxsheng/SHLDN}{https://github.com/sxsheng/SHLDN}).
\bibliographystyle{mnras}
\bibliography{references}

\begin{thebibliography}{}
\makeatletter
\relax
\def\mn@urlcharsother{\let\do\@makeother \do\$\do\&\do\#\do\^\do\_\do\%\do\~}
\def\mn@doi{\begingroup\mn@urlcharsother \@ifnextchar [ {\mn@doi@}
  {\mn@doi@[]}}
\def\mn@doi@[#1]#2{\def\@tempa{#1}\ifx\@tempa\@empty \href
  {http://dx.doi.org/#2} {doi:#2}\else \href {http://dx.doi.org/#2} {#1}\fi
  \endgroup}
\def\mn@eprint#1#2{\mn@eprint@#1:#2::\@nil}
\def\mn@eprint@arXiv#1{\href {http://arxiv.org/abs/#1} {{\tt arXiv:#1}}}
\def\mn@eprint@dblp#1{\href {http://dblp.uni-trier.de/rec/bibtex/#1.xml}
  {dblp:#1}}
\def\mn@eprint@#1:#2:#3:#4\@nil{\def\@tempa {#1}\def\@tempb {#2}\def\@tempc
  {#3}\ifx \@tempc \@empty \let \@tempc \@tempb \let \@tempb \@tempa \fi \ifx
  \@tempb \@empty \def\@tempb {arXiv}\fi \@ifundefined
  {mn@eprint@\@tempb}{\@tempb:\@tempc}{\expandafter \expandafter \csname
  mn@eprint@\@tempb\endcsname \expandafter{\@tempc}}}

\bibitem[\protect\citeauthoryear{Adebayo, Gilmer, Muelly, Goodfellow, Hardt  \&
  Kim}{Adebayo et~al.}{2018}]{sanity_checks_saliency_maps}
Adebayo J.,  Gilmer J.,  Muelly M.,  Goodfellow I.,  Hardt M.,   Kim B.,  2018,
  in Proceedings of the 32nd International Conference on Neural Information
  Processing Systems. NIPS'18.
Curran Associates Inc., Red Hook, NY, USA, p. 9525–9536

\bibitem[\protect\citeauthoryear{{Alard}}{{Alard}}{2006}]{threshold-elongationmap}
{Alard} C.,  2006, arXiv e-prints, \href
  {https://ui.adsabs.harvard.edu/abs/2006astro.ph..6757A} {pp
  astro--ph/0606757}

\bibitem[\protect\citeauthoryear{Arjovsky, Chintala  \& Bottou}{Arjovsky
  et~al.}{2017}]{WGAN-2017}
Arjovsky M.,  Chintala S.,   Bottou L.,  2017, in Precup D.,  Teh Y.~W.,  eds,
  Proceedings of Machine Learning Research Vol. 70, Proceedings of the 34th
  International Conference on Machine Learning. PMLR, International Convention
  Centre, Sydney, Australia, pp 214--223, \url
  {http://proceedings.mlr.press/v70/arjovsky17a.html}

\bibitem[\protect\citeauthoryear{{Ascaso}, {Wittman}  \& {Dawson}}{{Ascaso}
  et~al.}{2014}]{WittmanDLSClusters}
{Ascaso} B.,  {Wittman} D.,   {Dawson} W.,  2014, \mn@doi [\mnras]
  {10.1093/mnras/stu074}, \href
  {https://ui.adsabs.harvard.edu/abs/2014MNRAS.439.1980A} {439, 1980}

\bibitem[\protect\citeauthoryear{{Belokurov}, {Evans}, {Hewett}, {Moiseev},
  {McMahon}, {Sanchez}  \& {King}}{{Belokurov} et~al.}{2009}]{Belokurov2009}
{Belokurov} V.,  {Evans} N.~W.,  {Hewett} P.~C.,  {Moiseev} A.,  {McMahon}
  R.~G.,  {Sanchez} S.~F.,   {King} L.~J.,  2009, \mn@doi [\mnras]
  {10.1111/j.1365-2966.2008.14075.x}, \href
  {https://ui.adsabs.harvard.edu/abs/2009MNRAS.392..104B} {392, 104}

\bibitem[\protect\citeauthoryear{Berthelot, Carlini, Goodfellow, Papernot,
  Oliver  \& Raffel}{Berthelot et~al.}{2019}]{MixMatch}
Berthelot D.,  Carlini N.,  Goodfellow I.,  Papernot N.,  Oliver A.,   Raffel
  C.~A.,  2019, in Wallach H.,  Larochelle H.,  Beygelzimer A.,
  d\textquotesingle Alch\'{e}-Buc F.,  Fox E.,   Garnett R.,  eds, , Advances
  in Neural Information Processing Systems 32.
Curran Associates, Inc., pp 5049--5059

\bibitem[\protect\citeauthoryear{Berthelot, Roelofs, Sohn, Carlini  \&
  Kurakin}{Berthelot et~al.}{2021}]{berthelot2021adamatch}
Berthelot D.,  Roelofs R.,  Sohn K.,  Carlini N.,   Kurakin A.,  2021, arXiv
  preprint arXiv:2106.04732

\bibitem[\protect\citeauthoryear{{Bertin} \& {Arnouts}}{{Bertin} \&
  {Arnouts}}{1996}]{SExtractor1996}
{Bertin} E.,  {Arnouts} S.,  1996, \mn@doi [\aaps] {10.1051/aas:1996164}, \href
  {https://ui.adsabs.harvard.edu/abs/1996A&AS..117..393B} {117, 393}

\bibitem[\protect\citeauthoryear{{Bolton}, {Burles}, {Koopmans}, {Treu},
  {Gavazzi}, {Moustakas}, {Wayth}  \& {Schlegel}}{{Bolton}
  et~al.}{2008}]{SLACS}
{Bolton} A.~S.,  {Burles} S.,  {Koopmans} L. V.~E.,  {Treu} T.,  {Gavazzi} R.,
  {Moustakas} L.~A.,  {Wayth} R.,   {Schlegel} D.~J.,  2008, \mn@doi [\apj]
  {10.1086/589327}, \href
  {https://ui.adsabs.harvard.edu/abs/2008ApJ...682..964B} {682, 964}

\bibitem[\protect\citeauthoryear{{Brada{\v{c}}}, {Schneider}, {Steinmetz},
  {Lombardi}, {King}  \& {Porcas}}{{Brada{\v{c}}} et~al.}{2002}]{Bradac2002}
{Brada{\v{c}}} M.,  {Schneider} P.,  {Steinmetz} M.,  {Lombardi} M.,  {King}
  L.~J.,   {Porcas} R.,  2002, \mn@doi [\aap] {10.1051/0004-6361:20020559},
  \href {https://ui.adsabs.harvard.edu/abs/2002A&A...388..373B} {388, 373}

\bibitem[\protect\citeauthoryear{{Ca{\~n}ameras} et~al.,}{{Ca{\~n}ameras}
  et~al.}{2020}]{panstaars-2020}
{Ca{\~n}ameras} R.,  et~al., 2020, \mn@doi [\aap]
  {10.1051/0004-6361/202038219}, \href
  {https://ui.adsabs.harvard.edu/abs/2020A&A...644A.163C} {644, A163}

\bibitem[\protect\citeauthoryear{{Chattopadhay}, {Sarkar}, {Howlader}  \&
  {Balasubramanian}}{{Chattopadhay} et~al.}{2018}]{grad-cam++}
{Chattopadhay} A.,  {Sarkar} A.,  {Howlader} P.,   {Balasubramanian} V.~N.,
  2018, in 2018 IEEE Winter Conference on Applications of Computer Vision
  (WACV). pp 839--847, \mn@doi{10.1109/WACV.2018.00097}

\bibitem[\protect\citeauthoryear{{Chiba}}{{Chiba}}{2002}]{Masashi2002}
{Chiba} M.,  2002, \mn@doi [\apj] {10.1086/324493}, \href
  {https://ui.adsabs.harvard.edu/abs/2002ApJ...565...17C} {565, 17}

\bibitem[\protect\citeauthoryear{{Collett}}{{Collett}}{2015}]{Collett2015}
{Collett} T.~E.,  2015, \mn@doi [\apj] {10.1088/0004-637X/811/1/20}, \href
  {https://ui.adsabs.harvard.edu/abs/2015ApJ...811...20C} {811, 20}

\bibitem[\protect\citeauthoryear{{Diehl} et~al.,}{{Diehl}
  et~al.}{2009}]{Diehl2009}
{Diehl} H.~T.,  et~al., 2009, \mn@doi [\apj] {10.1088/0004-637X/707/1/686},
  \href {https://ui.adsabs.harvard.edu/abs/2009ApJ...707..686D} {707, 686}

\bibitem[\protect\citeauthoryear{Erhan, Courville, Bengio  \& Vincent}{Erhan
  et~al.}{2010}]{pmlr-v9-erhan10a}
Erhan D.,  Courville A.,  Bengio Y.,   Vincent P.,  2010, in Teh Y.~W.,
  Titterington M.,  eds,  Proceedings of Machine Learning Research Vol. 9,
  Proceedings of the Thirteenth International Conference on Artificial
  Intelligence and Statistics. PMLR, Chia Laguna Resort, Sardinia, Italy, pp
  201--208, \url {http://proceedings.mlr.press/v9/erhan10a.html}

\bibitem[\protect\citeauthoryear{{Fassnacht}, {Moustakas}, {Casertano},
  {Ferguson}, {Lucas}  \& {Park}}{{Fassnacht}
  et~al.}{2004}]{manualsearch-GOODS}
{Fassnacht} C.~D.,  {Moustakas} L.~A.,  {Casertano} S.,  {Ferguson} H.~C.,
  {Lucas} R.~A.,   {Park} Y.,  2004, \mn@doi [\apjl] {10.1086/379004}, \href
  {https://ui.adsabs.harvard.edu/abs/2004ApJ...600L.155F} {600, L155}

\bibitem[\protect\citeauthoryear{Fernando, Singh  \& Anand}{Fernando
  et~al.}{2019}]{Fernando_2019}
Fernando Z.~T.,  Singh J.,   Anand A.,  2019, in Proceedings of the 42nd
  International {ACM} {SIGIR} Conference on Research and Development in
  Information Retrieval. {ACM}, \mn@doi{10.1145/3331184.3331312}, \url
  {https://doi.org/10.1145%2F3331184.3331312}

\bibitem[\protect\citeauthoryear{{Garvin}, {Kruk}, {Cornen}, {Bhatawdekar},
  {Ca{\~n}ameras}  \& {Mer{\'\i}n}}{{Garvin}
  et~al.}{2022}]{HubbleCitizenscience2022}
{Garvin} E.~O.,  {Kruk} S.,  {Cornen} C.,  {Bhatawdekar} R.,  {Ca{\~n}ameras}
  R.,   {Mer{\'\i}n} B.,  2022, arXiv e-prints, \href
  {https://ui.adsabs.harvard.edu/abs/2022arXiv220706997G} {p. arXiv:2207.06997}

\bibitem[\protect\citeauthoryear{{Gavazzi}, {Marshall}, {Treu}  \&
  {Sonnenfeld}}{{Gavazzi} et~al.}{2014}]{ringfinder}
{Gavazzi} R.,  {Marshall} P.~J.,  {Treu} T.,   {Sonnenfeld} A.,  2014, \mn@doi
  [\apj] {10.1088/0004-637X/785/2/144}, \href
  {https://ui.adsabs.harvard.edu/abs/2014ApJ...785..144G} {785, 144}

\bibitem[\protect\citeauthoryear{{Gilman}, {Birrer}, {Treu}, {Nierenberg}  \&
  {Benson}}{{Gilman} et~al.}{2019}]{Daniel2019}
{Gilman} D.,  {Birrer} S.,  {Treu} T.,  {Nierenberg} A.,   {Benson} A.,  2019,
  \mn@doi [\mnras] {10.1093/mnras/stz1593}, \href
  {https://ui.adsabs.harvard.edu/abs/2019MNRAS.487.5721G} {487, 5721}

\bibitem[\protect\citeauthoryear{Goodfellow, Pouget-Abadie, Mirza, Xu,
  Warde-Farley, Ozair, Courville  \& Bengio}{Goodfellow et~al.}{2014}]{GAN2014}
Goodfellow I.~J.,  Pouget-Abadie J.,  Mirza M.,  Xu B.,  Warde-Farley D.,
  Ozair S.,  Courville A.,   Bengio Y.,  2014, in Proceedings of the 27th
  International Conference on Neural Information Processing Systems - Volume 2.
  NIPS'14.
MIT Press, Cambridge, MA, USA, p. 2672–2680

\bibitem[\protect\citeauthoryear{Gu et~al.,}{Gu et~al.}{2018}]{GU2018354}
Gu J.,  et~al., 2018, \mn@doi [Pattern Recognition]
  {https://doi.org/10.1016/j.patcog.2017.10.013}, 77, 354

\bibitem[\protect\citeauthoryear{Gulrajani, Ahmed, Arjovsky, Dumoulin  \&
  Courville}{Gulrajani et~al.}{2017}]{wgan_gp}
Gulrajani I.,  Ahmed F.,  Arjovsky M.,  Dumoulin V.,   Courville A.~C.,  2017,
  in Guyon I.,  Luxburg U.~V.,  Bengio S.,  Wallach H.,  Fergus R.,
  Vishwanathan S.,   Garnett R.,  eds, ~ Vol. 30, Advances in Neural
  Information Processing Systems. Curran Associates, Inc., \url
  {https://proceedings.neurips.cc/paper/2017/file/892c3b1c6dccd52936e27cbd0ff683d6-Paper.pdf}

\bibitem[\protect\citeauthoryear{He, Zhang, Ren  \& Sun}{He
  et~al.}{2016a}]{he2016identity}
He K.,  Zhang X.,  Ren S.,   Sun J.,  2016a, in European conference on computer
  vision. pp 630--645

\bibitem[\protect\citeauthoryear{He, Zhang, Ren  \& Sun}{He
  et~al.}{2016b}]{he2016deep}
He K.,  Zhang X.,  Ren S.,   Sun J.,  2016b, in Proceedings of the IEEE
  conference on computer vision and pattern recognition. pp 770--778

\bibitem[\protect\citeauthoryear{Huang et~al.,}{Huang
  et~al.}{2020}]{Huang_DESI-decam_2020}
Huang X.,  et~al., 2020, \mn@doi [The Astrophysical Journal]
  {10.3847/1538-4357/ab7ffb}, 894, 78

\bibitem[\protect\citeauthoryear{Ioffe \& Szegedy}{Ioffe \&
  Szegedy}{2015}]{ioffe2015batch}
Ioffe S.,  Szegedy C.,  2015, in International conference on machine learning.
  pp 448--456

\bibitem[\protect\citeauthoryear{{Ivezi{\'c}} et~al.,}{{Ivezi{\'c}}
  et~al.}{2019}]{Ivezic2019}
{Ivezi{\'c}} {\v{Z}}.,  et~al., 2019, \mn@doi [\apj]
  {10.3847/1538-4357/ab042c}, \href
  {https://ui.adsabs.harvard.edu/abs/2019ApJ...873..111I} {873, 111}

\bibitem[\protect\citeauthoryear{{Jacobs}, {Glazebrook}, {Collett}, {More}  \&
  {McCarthy}}{{Jacobs} et~al.}{2017}]{Colin-CFHTLS}
{Jacobs} C.,  {Glazebrook} K.,  {Collett} T.,  {More} A.,   {McCarthy} C.,
  2017, \mn@doi [\mnras] {10.1093/mnras/stx1492}, \href
  {https://ui.adsabs.harvard.edu/abs/2017MNRAS.471..167J} {471, 167}

\bibitem[\protect\citeauthoryear{{Jacobs} et~al.,}{{Jacobs}
  et~al.}{2019}]{colin-des-2019}
{Jacobs} C.,  et~al., 2019, \mn@doi [\apjs] {10.3847/1538-4365/ab26b6}, \href
  {https://ui.adsabs.harvard.edu/abs/2019ApJS..243...17J} {243, 17}

\bibitem[\protect\citeauthoryear{Kingma \& Welling}{Kingma \&
  Welling}{2014}]{kingma2014autoencoding}
Kingma D.~P.,  Welling M.,  2014, Auto-Encoding Variational Bayes (\mn@eprint
  {arXiv} {1312.6114})

\bibitem[\protect\citeauthoryear{{Koekemoer} et~al.,}{{Koekemoer}
  et~al.}{2007}]{COSMOS2007}
{Koekemoer} A.~M.,  et~al., 2007, \mn@doi [\apjs] {10.1086/520086}, \href
  {https://ui.adsabs.harvard.edu/abs/2007ApJS..172..196K} {172, 196}

\bibitem[\protect\citeauthoryear{{Koopmans}, {Treu}, {Bolton}, {Burles}  \&
  {Moustakas}}{{Koopmans} et~al.}{2006}]{SLACS2006}
{Koopmans} L. V.~E.,  {Treu} T.,  {Bolton} A.~S.,  {Burles} S.,   {Moustakas}
  L.~A.,  2006, \mn@doi [\apj] {10.1086/505696}, \href
  {https://ui.adsabs.harvard.edu/abs/2006ApJ...649..599K} {649, 599}

\bibitem[\protect\citeauthoryear{{Kormann}, {Schneider}  \&
  {Bartelmann}}{{Kormann} et~al.}{1994}]{kormann-SIE-paper}
{Kormann} R.,  {Schneider} P.,   {Bartelmann} M.,  1994, \aap, \href
  {https://ui.adsabs.harvard.edu/abs/1994A&A...284..285K} {284, 285}

\bibitem[\protect\citeauthoryear{Krizhevsky}{Krizhevsky}{2009a}]{Krizhevsky09learningmultiple}
Krizhevsky A.,  2009a, Technical report, Learning multiple layers of features
  from tiny images

\bibitem[\protect\citeauthoryear{Krizhevsky}{Krizhevsky}{2009b}]{CIFAR10-dataset}
Krizhevsky A.,  2009b.

\bibitem[\protect\citeauthoryear{Krizhevsky, Sutskever  \& Hinton}{Krizhevsky
  et~al.}{2012}]{alex2012}
Krizhevsky A.,  Sutskever I.,   Hinton G.~E.,  2012, in Pereira F.,  Burges C.
  J.~C.,  Bottou L.,   Weinberger K.~Q.,  eds, ~ Vol. 25, Advances in Neural
  Information Processing Systems. Curran Associates, Inc., \url
  {https://proceedings.neurips.cc/paper/2012/file/c399862d3b9d6b76c8436e924a68c45b-Paper.pdf}

\bibitem[\protect\citeauthoryear{{Kubo} \& {Dell'Antonio}}{{Kubo} \&
  {Dell'Antonio}}{2008}]{kubo_deeplenssurvey_ML}
{Kubo} J.~M.,  {Dell'Antonio} I.~P.,  2008, \mn@doi [\mnras]
  {10.1111/j.1365-2966.2008.12880.x}, \href
  {https://ui.adsabs.harvard.edu/abs/2008MNRAS.385..918K} {385, 918}

\bibitem[\protect\citeauthoryear{{LSST Science Collaboration} et~al.,}{{LSST
  Science Collaboration} et~al.}{2009}]{LSST2009}
{LSST Science Collaboration} et~al., 2009, arXiv e-prints, \href
  {https://ui.adsabs.harvard.edu/abs/2009arXiv0912.0201L} {p. arXiv:0912.0201}

\bibitem[\protect\citeauthoryear{Laine \& Aila}{Laine \& Aila}{2017}]{pi_model}
Laine S.,  Aila T.,  2017, ArXiv, abs/1610.02242

\bibitem[\protect\citeauthoryear{{Laureijs} et~al.,}{{Laureijs}
  et~al.}{2011}]{euclid2011}
{Laureijs} R.,  et~al., 2011, arXiv e-prints, \href
  {https://ui.adsabs.harvard.edu/abs/2011arXiv1110.3193L} {p. arXiv:1110.3193}

\bibitem[\protect\citeauthoryear{{LeCun}, {Boser}, {Denker}, {Henderson},
  {Howard}, {Hubbard}  \& {Jackel}}{{LeCun} et~al.}{1989}]{lecun1989}
{LeCun} Y.,  {Boser} B.,  {Denker} J.~S.,  {Henderson} D.,  {Howard} R.~E.,
  {Hubbard} W.,   {Jackel} L.~D.,  1989, \mn@doi [Neural Computation]
  {10.1162/neco.1989.1.4.541}, 1, 541

\bibitem[\protect\citeauthoryear{Lee}{Lee}{2013}]{pseudo-label}
Lee D.-H.,  2013, ICML 2013 Workshop : Challenges in Representation Learning
  (WREPL)

\bibitem[\protect\citeauthoryear{{Leethochawalit}, {Jones}, {Ellis}, {Stark},
  {Richard}, {Zitrin}  \& {Auger}}{{Leethochawalit}
  et~al.}{2016}]{Leethochawalit2016}
{Leethochawalit} N.,  {Jones} T.~A.,  {Ellis} R.~S.,  {Stark} D.~P.,  {Richard}
  J.,  {Zitrin} A.,   {Auger} M.,  2016, \mn@doi [\apj]
  {10.3847/0004-637X/820/2/84}, \href
  {https://ui.adsabs.harvard.edu/abs/2016ApJ...820...84L} {820, 84}

\bibitem[\protect\citeauthoryear{{Li} et~al.,}{{Li}
  et~al.}{2020}]{kids-lens-search}
{Li} R.,  et~al., 2020, \mn@doi [\apj] {10.3847/1538-4357/ab9dfa}, \href
  {https://ui.adsabs.harvard.edu/abs/2020ApJ...899...30L} {899, 30}

\bibitem[\protect\citeauthoryear{Litjens et~al.,}{Litjens
  et~al.}{2017}]{LITJENS201760}
Litjens G.,  et~al., 2017, \mn@doi [Medical Image Analysis]
  {https://doi.org/10.1016/j.media.2017.07.005}, 42, 60

\bibitem[\protect\citeauthoryear{{Lupton}, {Blanton}, {Fekete}, {Hogg},
  {O'Mullane}, {Szalay}  \& {Wherry}}{{Lupton} et~al.}{2004}]{Lupton2003}
{Lupton} R.,  {Blanton} M.~R.,  {Fekete} G.,  {Hogg} D.~W.,  {O'Mullane} W.,
  {Szalay} A.,   {Wherry} N.,  2004, \mn@doi [\pasp] {10.1086/382245}, \href
  {https://ui.adsabs.harvard.edu/abs/2004PASP..116..133L} {116, 133}

\bibitem[\protect\citeauthoryear{{Madireddy}, {Li}, {Ramachandra}, {Butler},
  {Balaprakash}, {Habib}  \& {Heitmann}}{{Madireddy}
  et~al.}{2019}]{Madireddy2019}
{Madireddy} S.,  {Li} N.,  {Ramachandra} N.,  {Butler} J.,  {Balaprakash} P.,
  {Habib} S.,   {Heitmann} K.,  2019, arXiv e-prints, \href
  {https://ui.adsabs.harvard.edu/abs/2019arXiv191103867M} {p. arXiv:1911.03867}

\bibitem[\protect\citeauthoryear{{Marshall}, {Sandford}, {More}  \&
  {Buddelmeijerr}}{{Marshall} et~al.}{2015}]{phil-humvi}
{Marshall} P.,  {Sandford} C.,  {More} A.,   {Buddelmeijerr} H.,  2015, {HumVI:
  Human Viewable Image creation} (\mn@eprint {ascl} {1511.014})

\bibitem[\protect\citeauthoryear{{Miranda} \& {Macci{\`o}}}{{Miranda} \&
  {Macci{\`o}}}{2007}]{Marco2007}
{Miranda} M.,  {Macci{\`o}} A.~V.,  2007, \mn@doi [\mnras]
  {10.1111/j.1365-2966.2007.12440.x}, \href
  {https://ui.adsabs.harvard.edu/abs/2007MNRAS.382.1225M} {382, 1225}

\bibitem[\protect\citeauthoryear{{Miyato}, {Maeda}, {Koyama}  \&
  {Ishii}}{{Miyato} et~al.}{2019}]{VAT}
{Miyato} T.,  {Maeda} S.,  {Koyama} M.,   {Ishii} S.,  2019, IEEE Transactions
  on Pattern Analysis and Machine Intelligence, 41, 1979

\bibitem[\protect\citeauthoryear{{More} et~al.,}{{More}
  et~al.}{2016}]{spacewarps-ii}
{More} A.,  et~al., 2016, \mn@doi [\mnras] {10.1093/mnras/stv1966}, \href
  {https://ui.adsabs.harvard.edu/abs/2016MNRAS.455.1191M} {455, 1191}

\bibitem[\protect\citeauthoryear{{Moustakas} et~al.,}{{Moustakas}
  et~al.}{2007}]{MoustakasAEGIS}
{Moustakas} L.~A.,  et~al., 2007, \mn@doi [\apjl] {10.1086/517930}, \href
  {https://ui.adsabs.harvard.edu/abs/2007ApJ...660L..31M} {660, L31}

\bibitem[\protect\citeauthoryear{Netzer, Wang, Coates, Bissacco, Wu  \&
  Ng}{Netzer et~al.}{2011}]{SVHN-dataset}
Netzer Y.,  Wang T.,  Coates A.,  Bissacco A.,  Wu B.,   Ng A.~Y.,  2011, in
  NIPS Workshop on Deep Learning and Unsupervised Feature Learning 2011. \url
  {http://ufldl.stanford.edu/housenumbers/nips2011_housenumbers.pdf}

\bibitem[\protect\citeauthoryear{{Oguri}}{{Oguri}}{2010}]{glafic}
{Oguri} M.,  2010, \mn@doi [\pasj] {10.1093/pasj/62.4.1017}, \href
  {https://ui.adsabs.harvard.edu/abs/2010PASJ...62.1017O} {62, 1017}

\bibitem[\protect\citeauthoryear{{Oguri} \& {Marshall}}{{Oguri} \&
  {Marshall}}{2010}]{oguri2010}
{Oguri} M.,  {Marshall} P.~J.,  2010, \mn@doi [\mnras]
  {10.1111/j.1365-2966.2010.16639.x}, \href
  {https://ui.adsabs.harvard.edu/abs/2010MNRAS.405.2579O} {405, 2579}

\bibitem[\protect\citeauthoryear{{Paraficz} et~al.,}{{Paraficz}
  et~al.}{2016}]{pca-lensfinder}
{Paraficz} D.,  et~al., 2016, \mn@doi [\aap] {10.1051/0004-6361/201527971},
  \href {https://ui.adsabs.harvard.edu/abs/2016A&A...592A..75P} {592, A75}

\bibitem[\protect\citeauthoryear{{Pettini}, {Rix}, {Steidel}, {Adelberger},
  {Hunt}  \& {Shapley}}{{Pettini} et~al.}{2002}]{Pettini2002}
{Pettini} M.,  {Rix} S.~A.,  {Steidel} C.~C.,  {Adelberger} K.~L.,  {Hunt}
  M.~P.,   {Shapley} A.~E.,  2002, \mn@doi [\apj] {10.1086/339355}, \href
  {https://ui.adsabs.harvard.edu/abs/2002ApJ...569..742P} {569, 742}

\bibitem[\protect\citeauthoryear{{Pourrahmani}, {Nayyeri}  \&
  {Cooray}}{{Pourrahmani} et~al.}{2018}]{Milad-HST}
{Pourrahmani} M.,  {Nayyeri} H.,   {Cooray} A.,  2018, \mn@doi [\apj]
  {10.3847/1538-4357/aaae6a}, \href
  {https://ui.adsabs.harvard.edu/abs/2018ApJ...856...68P} {856, 68}

\bibitem[\protect\citeauthoryear{Quinonero-Candela, Sugiyama, Schwaighofer  \&
  Lawrence}{Quinonero-Candela et~al.}{2008}]{quinonero2008dataset}
Quinonero-Candela J.,  Sugiyama M.,  Schwaighofer A.,   Lawrence N.~D.,  2008,
  Dataset shift in machine learning.
Mit Press

\bibitem[\protect\citeauthoryear{{Schmidt} \& {Thorman}}{{Schmidt} \&
  {Thorman}}{2013}]{sam-DLS-BVRz-paper}
{Schmidt} S.~J.,  {Thorman} P.,  2013, \mn@doi [\mnras] {10.1093/mnras/stt373},
  \href {https://ui.adsabs.harvard.edu/abs/2013MNRAS.431.2766S} {431, 2766}

\bibitem[\protect\citeauthoryear{{Seidel} \& {Bartelmann}}{{Seidel} \&
  {Bartelmann}}{2007}]{Arcfinder}
{Seidel} G.,  {Bartelmann} M.,  2007, \mn@doi [\aap]
  {10.1051/0004-6361:20066097}, \href
  {https://ui.adsabs.harvard.edu/abs/2007A&A...472..341S} {472, 341}

\bibitem[\protect\citeauthoryear{Selvaraju, Cogswell, Das, Vedantam, Parikh  \&
  Batra}{Selvaraju et~al.}{2017}]{grad-cam}
Selvaraju R.~R.,  Cogswell M.,  Das A.,  Vedantam R.,  Parikh D.,   Batra D.,
  2017, in Proceedings of the IEEE International Conference on Computer Vision
  (ICCV).

\bibitem[\protect\citeauthoryear{{Shajib} et~al.,}{{Shajib}
  et~al.}{2022}]{Shajib2022}
{Shajib} A.~J.,  et~al., 2022, arXiv e-prints, \href
  {https://ui.adsabs.harvard.edu/abs/2022arXiv221010790S} {p. arXiv:2210.10790}

\bibitem[\protect\citeauthoryear{{Sheng}, {C}, {Choi}, {Sharpnack}  \&
  {Jones}}{{Sheng} et~al.}{2022}]{ShengAISTATS}
{Sheng} S.,  {C} K. V.~G.,  {Choi} C.~P.,  {Sharpnack} J.,   {Jones} T.,  2022,
  \mn@doi [arXiv e-prints] {10.48550/arXiv.2210.11681}, \href
  {https://ui.adsabs.harvard.edu/abs/2022arXiv221011681S} {p. arXiv:2210.11681}

\bibitem[\protect\citeauthoryear{Sohn et~al.,}{Sohn
  et~al.}{2020}]{fixMatch2020}
Sohn K.,  et~al., 2020, in Larochelle H.,  Ranzato M.,  Hadsell R.,  Balcan M.,
    Lin H.,  eds, ~ Vol. 33, Advances in Neural Information Processing Systems.
  Curran Associates, Inc., pp 596--608, \url
  {https://proceedings.neurips.cc/paper/2020/file/06964dce9addb1c5cb5d6e3d9838f733-Paper.pdf}

\bibitem[\protect\citeauthoryear{{Sonnenfeld}, {Treu}, {Gavazzi}, {Suyu},
  {Marshall}, {Auger}  \& {Nipoti}}{{Sonnenfeld} et~al.}{2013}]{SL2S}
{Sonnenfeld} A.,  {Treu} T.,  {Gavazzi} R.,  {Suyu} S.~H.,  {Marshall} P.~J.,
  {Auger} M.~W.,   {Nipoti} C.,  2013, \mn@doi [\apj]
  {10.1088/0004-637X/777/2/98}, \href
  {https://ui.adsabs.harvard.edu/abs/2013ApJ...777...98S} {777, 98}

\bibitem[\protect\citeauthoryear{{Sonnenfeld} et~al.,}{{Sonnenfeld}
  et~al.}{2018}]{hsc-2018}
{Sonnenfeld} A.,  et~al., 2018, \mn@doi [\pasj] {10.1093/pasj/psx062}, \href
  {https://ui.adsabs.harvard.edu/abs/2018PASJ...70S..29S} {70, S29}

\bibitem[\protect\citeauthoryear{{Spergel} et~al.,}{{Spergel}
  et~al.}{2015}]{Roman2015}
{Spergel} D.,  et~al., 2015, arXiv e-prints, \href
  {https://ui.adsabs.harvard.edu/abs/2015arXiv150303757S} {p. arXiv:1503.03757}

\bibitem[\protect\citeauthoryear{Springenberg, Dosovitskiy, Brox  \&
  Riedmiller}{Springenberg et~al.}{2015}]{guided_backprop}
Springenberg J.~T.,  Dosovitskiy A.,  Brox T.,   Riedmiller M.~A.,  2015, CoRR,
  abs/1412.6806

\bibitem[\protect\citeauthoryear{{Stein}, {Harrington}, {Blaum}, {Medan}  \&
  {Lukic}}{{Stein} et~al.}{2021}]{SelfSimilarity2021}
{Stein} G.,  {Harrington} P.,  {Blaum} J.,  {Medan} T.,   {Lukic} Z.,  2021,
  arXiv e-prints, \href {https://ui.adsabs.harvard.edu/abs/2021arXiv211013151S}
  {p. arXiv:2110.13151}

\bibitem[\protect\citeauthoryear{{Swinbank} et~al.,}{{Swinbank}
  et~al.}{2009}]{Swinbank2009}
{Swinbank} A.~M.,  et~al., 2009, \mn@doi [\mnras]
  {10.1111/j.1365-2966.2009.15617.x}, \href
  {https://ui.adsabs.harvard.edu/abs/2009MNRAS.400.1121S} {400, 1121}

\bibitem[\protect\citeauthoryear{Tarvainen \& Valpola}{Tarvainen \&
  Valpola}{2017}]{mean_teacher}
Tarvainen A.,  Valpola H.,  2017, in Guyon I.,  Luxburg U.~V.,  Bengio S.,
  Wallach H.,  Fergus R.,  Vishwanathan S.,   Garnett R.,  eds, , Advances in
  Neural Information Processing Systems 30.
Curran Associates, Inc., pp 1195--1204

\bibitem[\protect\citeauthoryear{{Tran} et~al.,}{{Tran}
  et~al.}{2022}]{AGELpaper}
{Tran} K.-V.~H.,  et~al., 2022, arXiv e-prints, \href
  {https://ui.adsabs.harvard.edu/abs/2022arXiv220505307T} {p. arXiv:2205.05307}

\bibitem[\protect\citeauthoryear{{Treu}}{{Treu}}{2010}]{tomasso-ara-review}
{Treu} T.,  2010, \mn@doi [\araa] {10.1146/annurev-astro-081309-130924}, \href
  {https://ui.adsabs.harvard.edu/abs/2010ARA&A..48...87T} {48, 87}

\bibitem[\protect\citeauthoryear{{Wilson} et~al.,}{{Wilson}
  et~al.}{2004}]{KeckNIRES}
{Wilson} J.~C.,  et~al., 2004, in {Moorwood} A. F.~M.,  {Iye} M.,  eds,
  Society of Photo-Optical Instrumentation Engineers (SPIE) Conference Series
  Vol. 5492, Ground-based Instrumentation for Astronomy. pp 1295--1305,
  \mn@doi{10.1117/12.550925}

\bibitem[\protect\citeauthoryear{{Wittman} et~al.,}{{Wittman}
  et~al.}{2002}]{DLS-Wittman-2002}
{Wittman} D.~M.,  et~al., 2002, in {Tyson} J.~A.,  {Wolff} S.,  eds,  Society
  of Photo-Optical Instrumentation Engineers (SPIE) Conference Series Vol.
  4836, Survey and Other Telescope Technologies and Discoveries. pp 73--82
  (\mn@eprint {arXiv} {astro-ph/0210118}), \mn@doi{10.1117/12.457348}

\bibitem[\protect\citeauthoryear{{Wittman}, {Margoniner}, {Tyson}, {Cohen},
  {Becker}  \& {Dell'Antonio}}{{Wittman} et~al.}{2003}]{Wittman2003Cluster}
{Wittman} D.,  {Margoniner} V.~E.,  {Tyson} J.~A.,  {Cohen} J.~G.,  {Becker}
  A.~C.,   {Dell'Antonio} I.~P.,  2003, \mn@doi [\apj] {10.1086/378344}, \href
  {https://ui.adsabs.harvard.edu/abs/2003ApJ...597..218W} {597, 218}

\bibitem[\protect\citeauthoryear{{Wittman}, {Dell'Antonio}, {Hughes},
  {Margoniner}, {Tyson}, {Cohen}  \& {Norman}}{{Wittman}
  et~al.}{2006}]{Wittman2006}
{Wittman} D.,  {Dell'Antonio} I.~P.,  {Hughes} J.~P.,  {Margoniner} V.~E.,
  {Tyson} J.~A.,  {Cohen} J.~G.,   {Norman} D.,  2006, \mn@doi [\apj]
  {10.1086/502621}, \href
  {https://ui.adsabs.harvard.edu/abs/2006ApJ...643..128W} {643, 128}

\bibitem[\protect\citeauthoryear{{Wuyts}, {Rigby}, {Gladders}  \&
  {Sharon}}{{Wuyts} et~al.}{2014}]{Wuyts2014}
{Wuyts} E.,  {Rigby} J.~R.,  {Gladders} M.~D.,   {Sharon} K.,  2014, \mn@doi
  [\apj] {10.1088/0004-637X/781/2/61}, \href
  {https://ui.adsabs.harvard.edu/abs/2014ApJ...781...61W} {781, 61}

\bibitem[\protect\citeauthoryear{Zhou, Khosla, A., Oliva  \& Torralba}{Zhou
  et~al.}{2016}]{cam}
Zhou B.,  Khosla A.,  A. L.,  Oliva A.,   Torralba A.,  2016, CVPR

\makeatother
\end{thebibliography}

\newpage 

\appendix

\section{Model performance and final lens sample}\label{sec:modelPerformanceByField}

In this appendix, we provide some additional details of the model performance and the top lens candidates identified in this work. 

\begin{figure*}
      \centering
    \begin{tabular}{ccc}
    \includegraphics[scale=0.27]{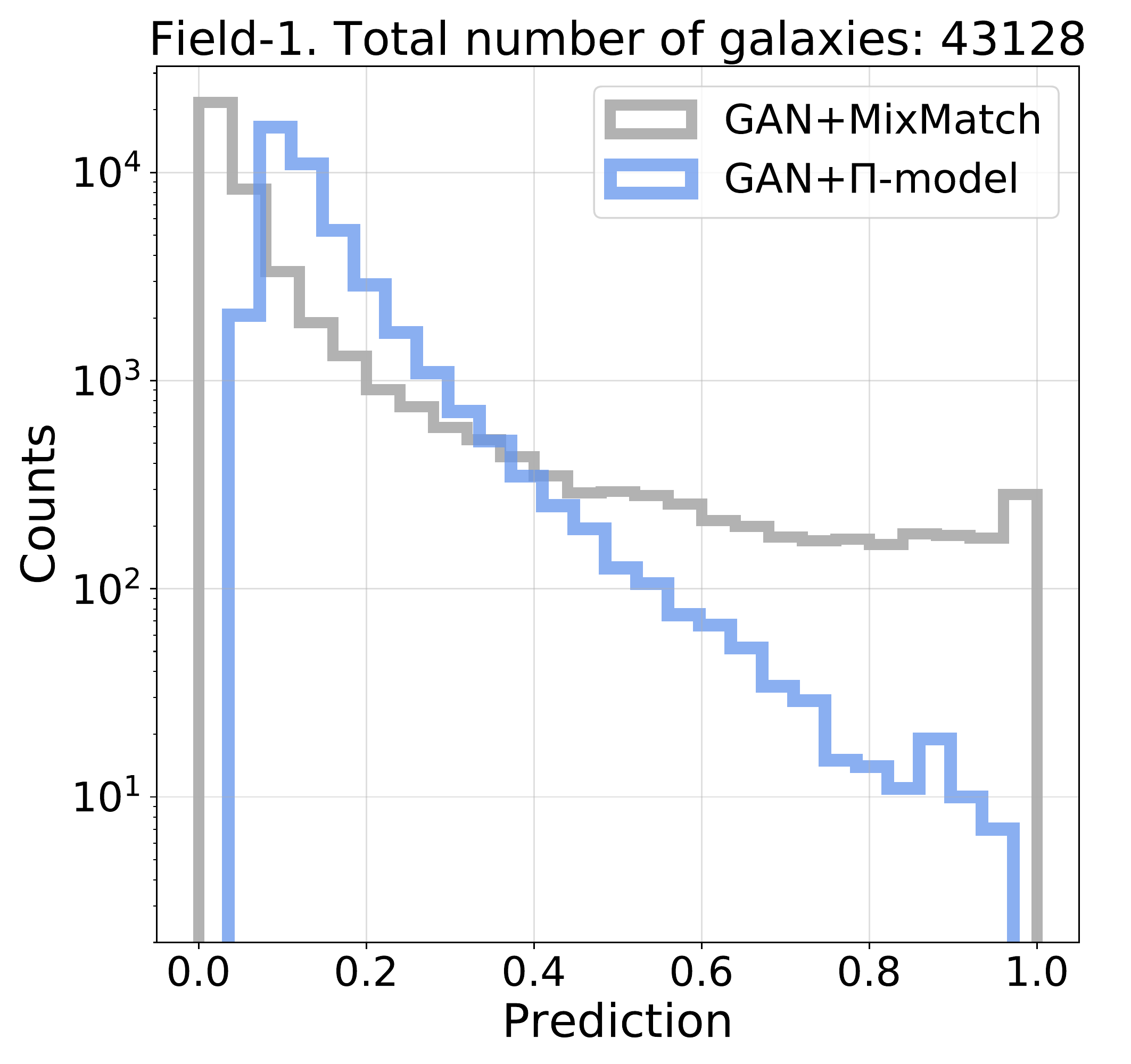} &
    \includegraphics[scale=0.27]{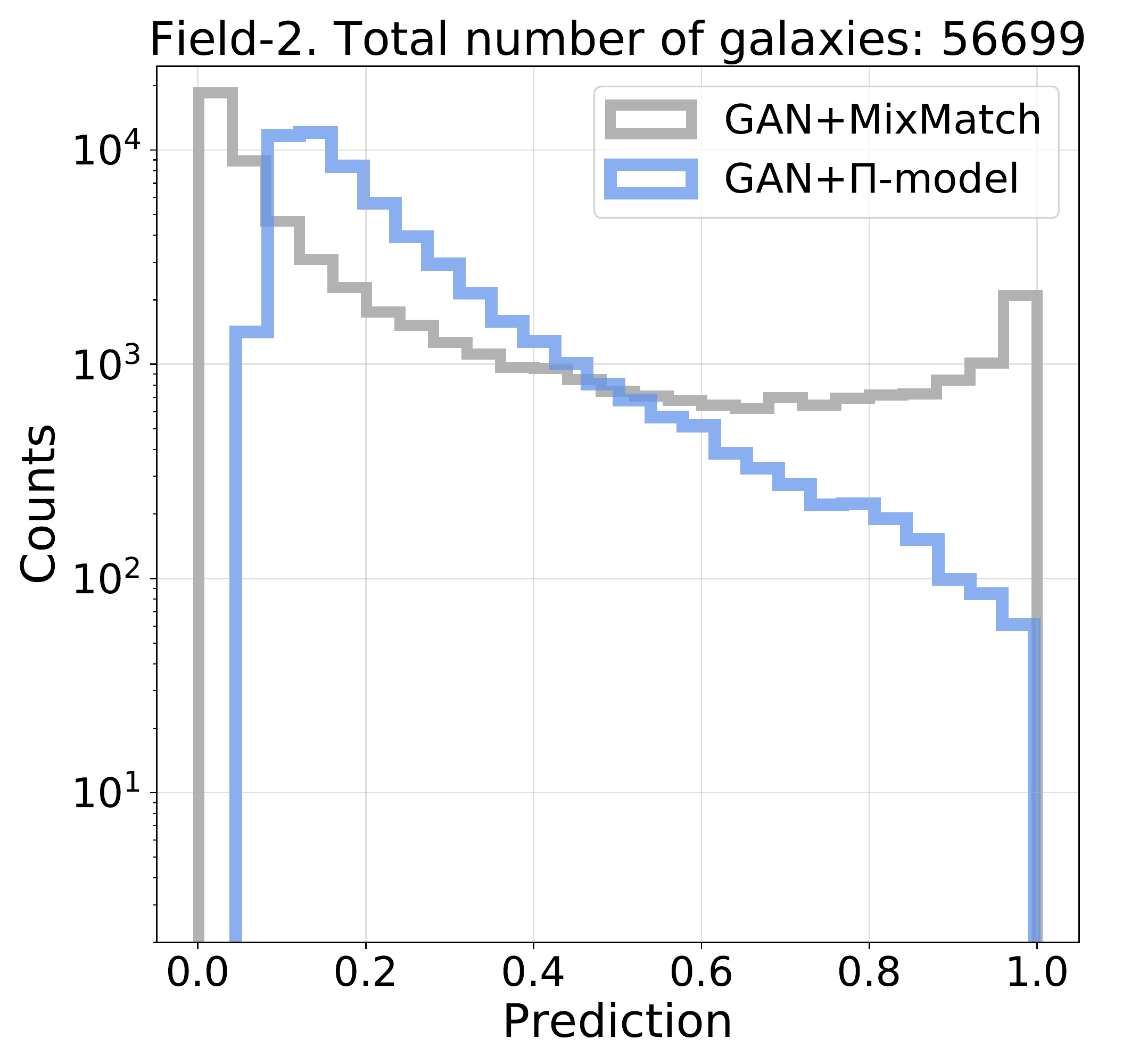} &
    \includegraphics[scale=0.27]{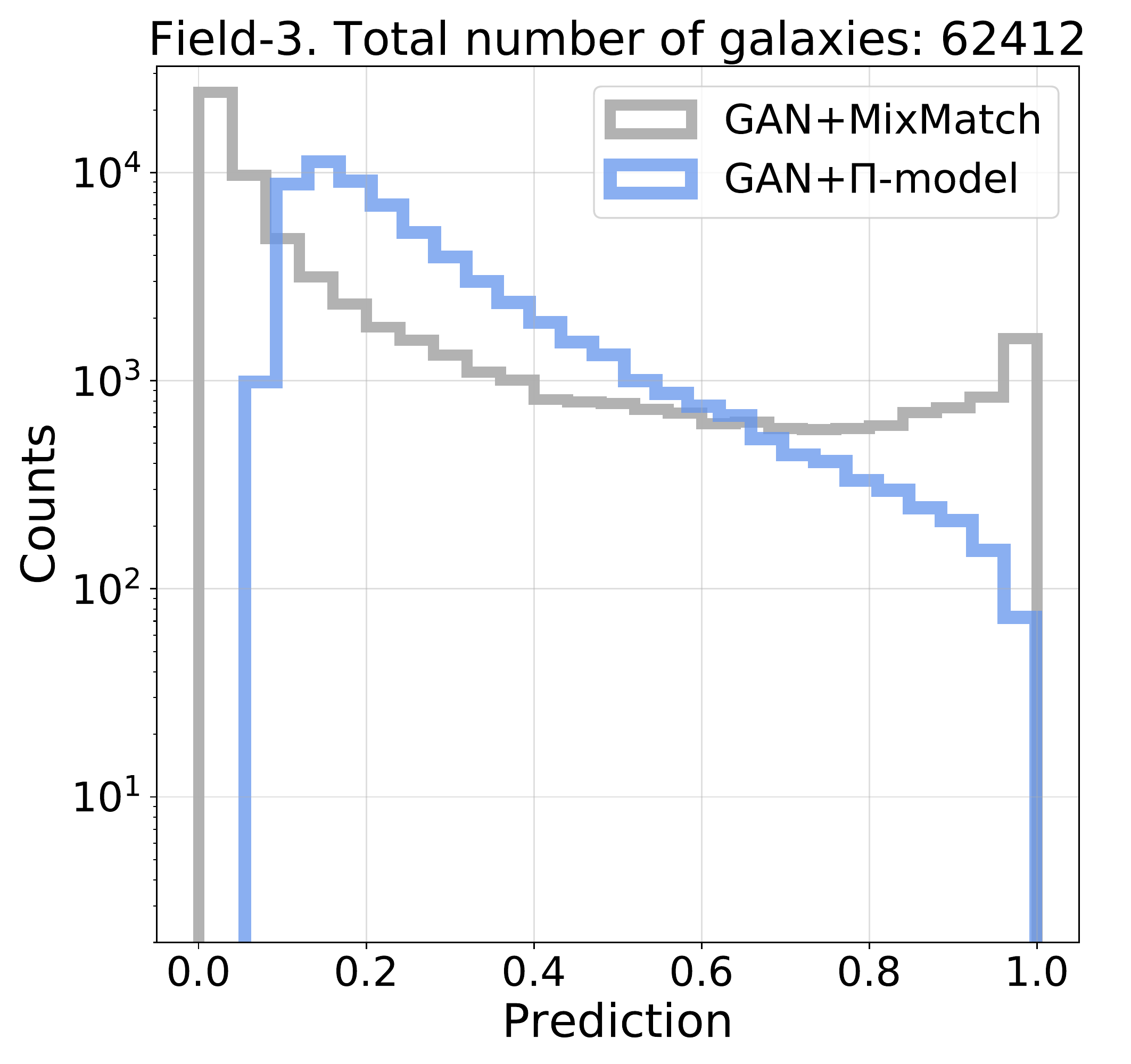} \\
    \includegraphics[scale=0.27]{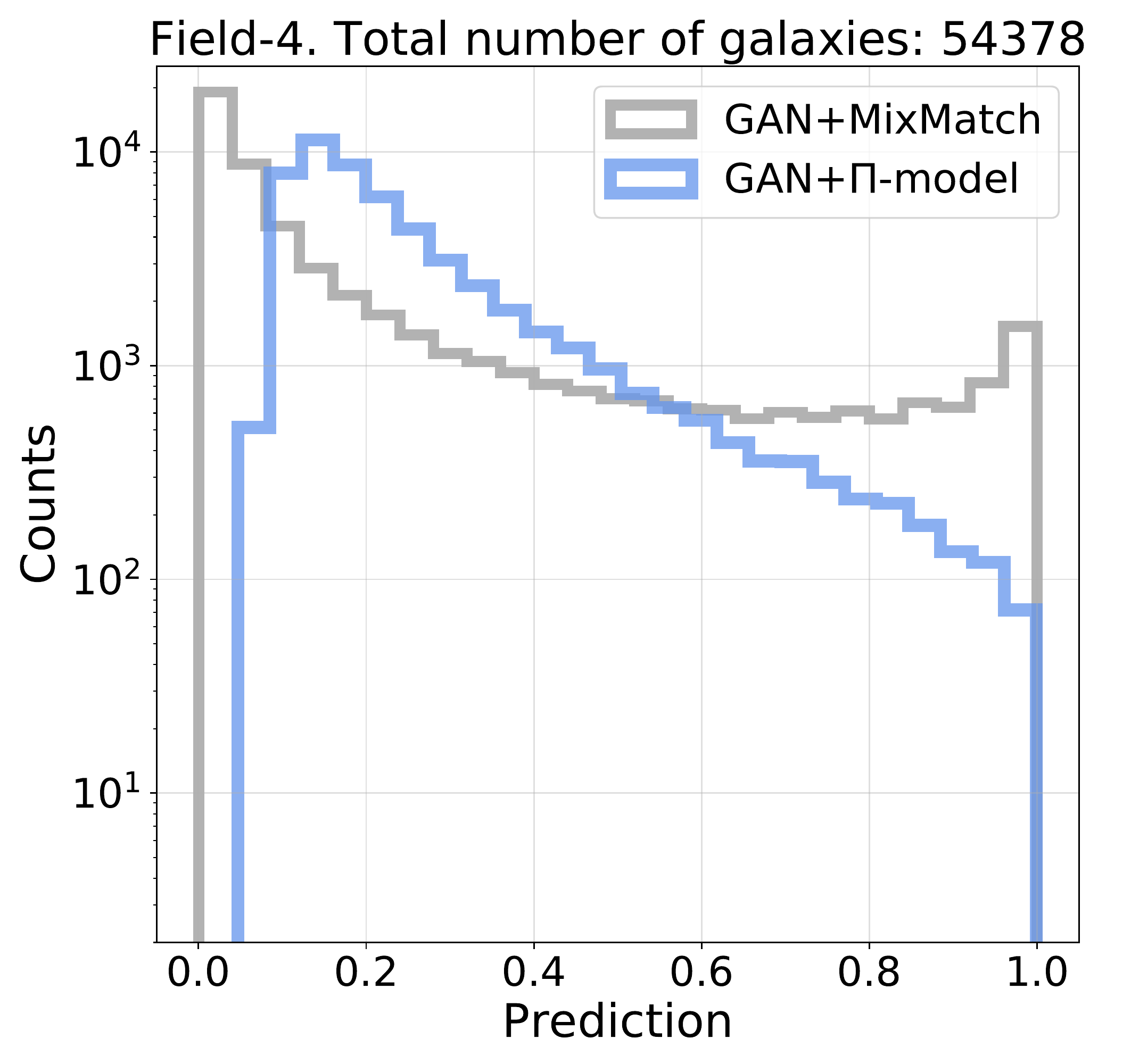} &
    \includegraphics[scale=0.27]{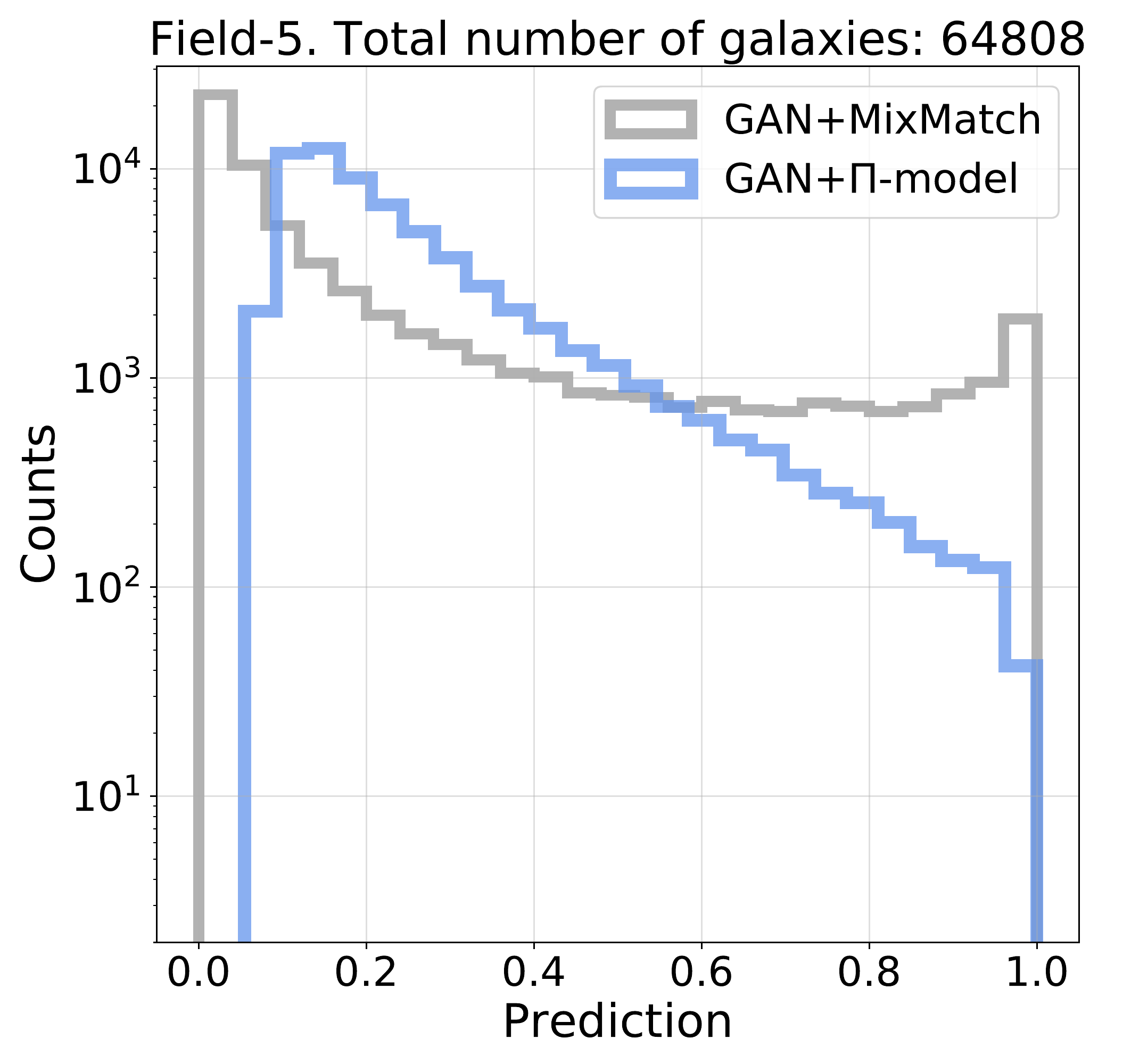} &
    \end{tabular}
\caption{Histogram of the scores obtained by the GAN+MixMatch and GAN+$\Pi$-model in the five independent DLS Fields F1 through F5. As discussed in Section~\ref{sec:TrainingData}, the training set used to train our models (TrainingV2) contains human labeled NonLenses which were randomly sampled only from Field F1. But as we clearly see, the distribution of scores (and performance of the models as a result) is independent of the field chosen.      } \label{fig:model-perfomance-allfields}

\end{figure*}

Figure~\ref{fig:model-perfomance-allfields} shows the distribution of model scores across the different DLS fields (F1 to F5), demonstrating similar performance in each field. This is generally expected given the similar image quality across the DLS survey. Importantly it shows that our use of labeled training data from only F1 does not substantially affect the model performance in the other fields.

Table~\ref{tab:ablation_tests} lists the results from our ablation study discussed in Section~\ref{subsec:ablationstudy}. We show the precision across recall rates from 50-100\%. The performance differences are generally similar across all recall rates. Color augmentation, JPEG quality, and GAN images appear to most prominently improving the model performance (i.e., the models perform significantly worse when these augmentations are removed). 

\begin{figure*}
    \centering
    \includegraphics[width=\textwidth]{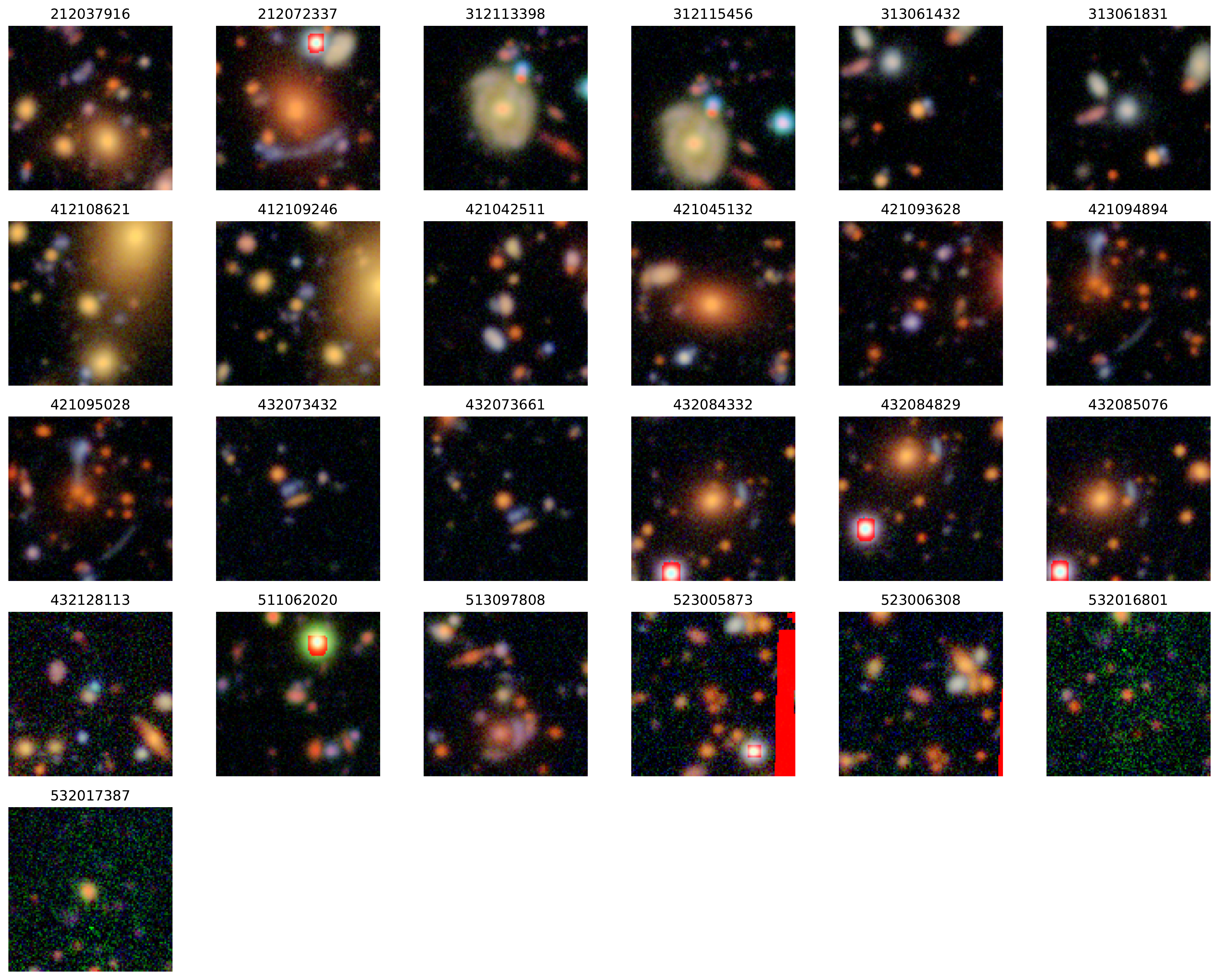}
    \caption{DLS images of the top 25 predictions from GAN+$\Pi$-model. Several show clear evidence of strong lensing, while other images appear to be false positives. We note that many images are duplicates (at overlapping regions of the sky), which we remove before visual inspection. }\label{fig:top25-pimodel}
\end{figure*}

\begin{figure*}
    \centering
    \includegraphics[width=\textwidth]{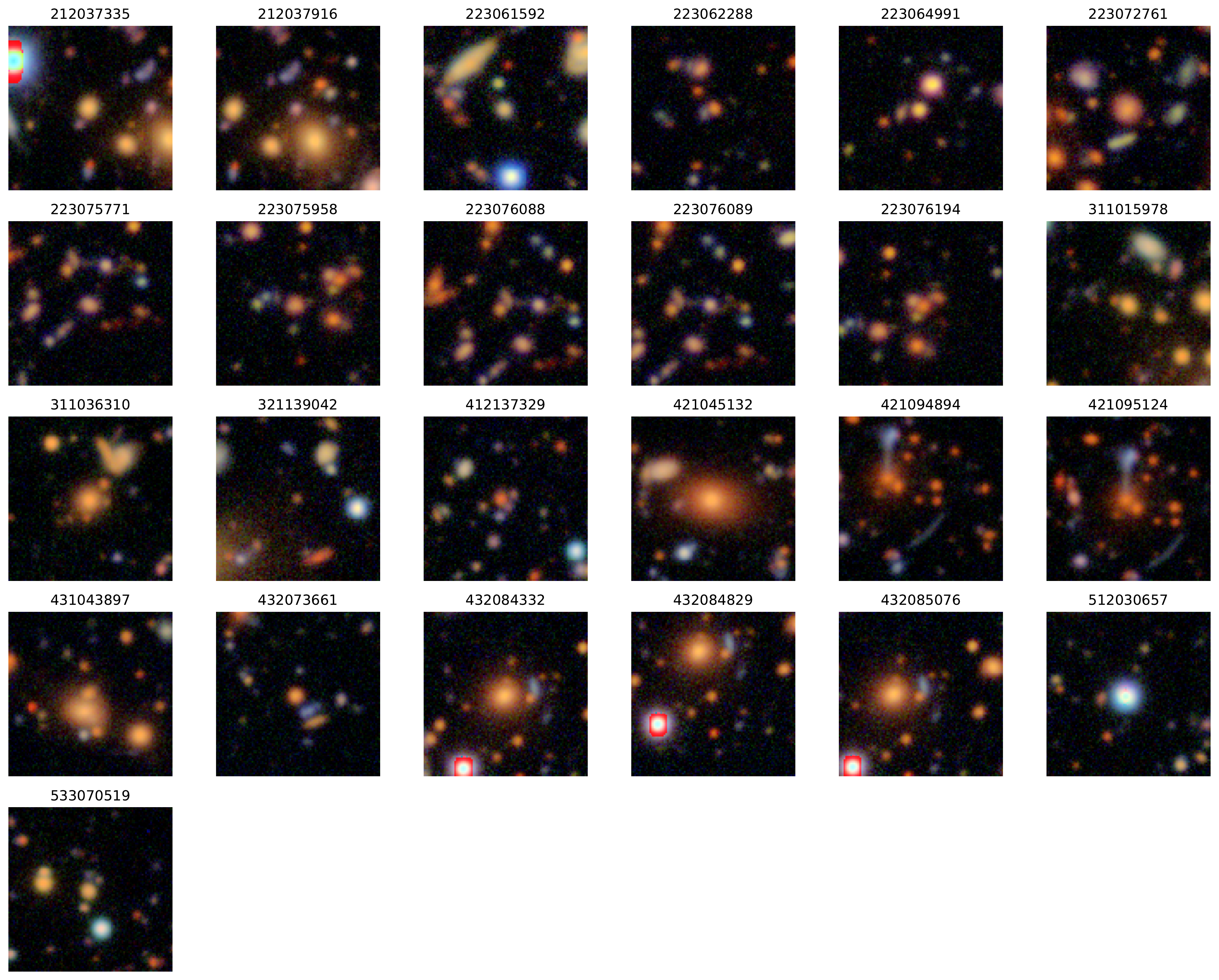}
    \caption{Equivalent to Figure~\ref{fig:top25-pimodel}, showing the top 25 predictions from GAN+Mimatch.}\label{fig:top25-mixmatch}
\end{figure*}

\begin{figure*}
    \centering
    \begin{minipage}{.5\textwidth}
      \centering
      \includegraphics[width=\textwidth]{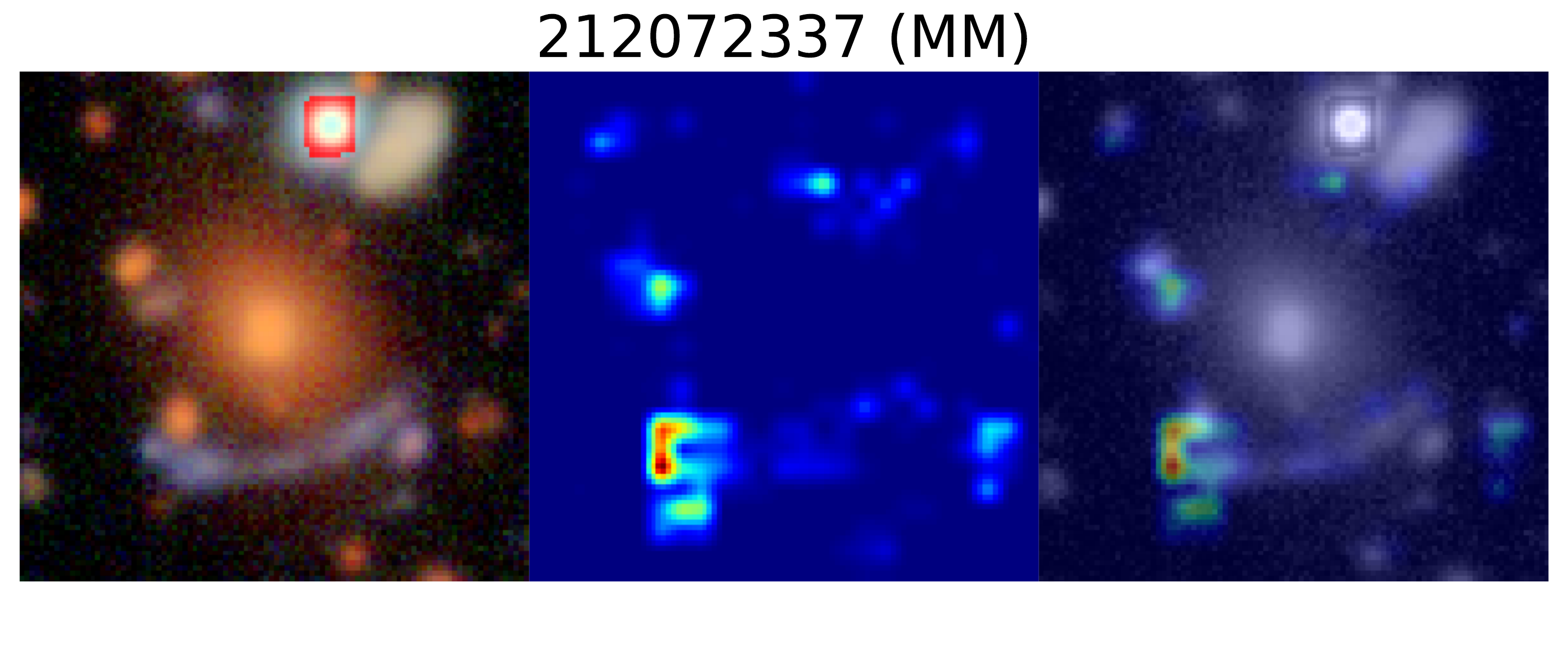}
      \includegraphics[width=\textwidth]{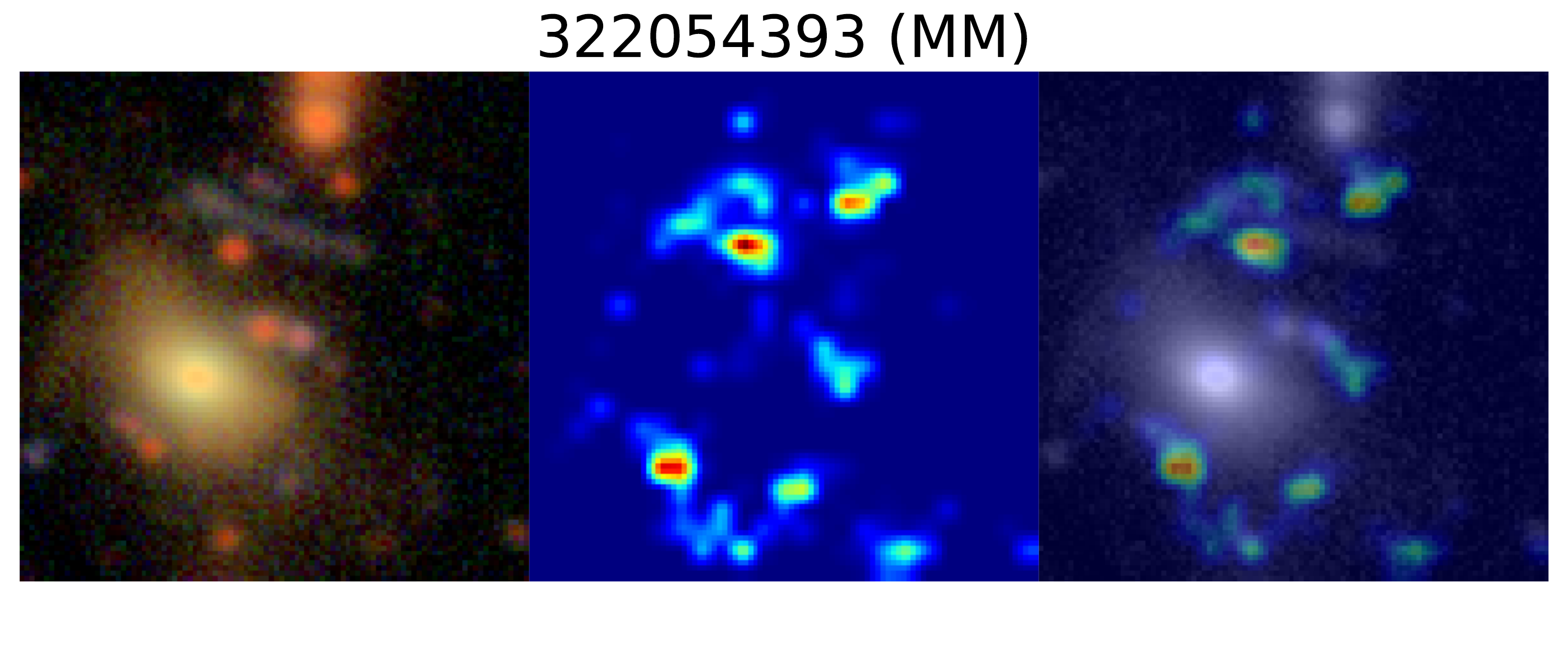}
     \includegraphics[width=\textwidth]{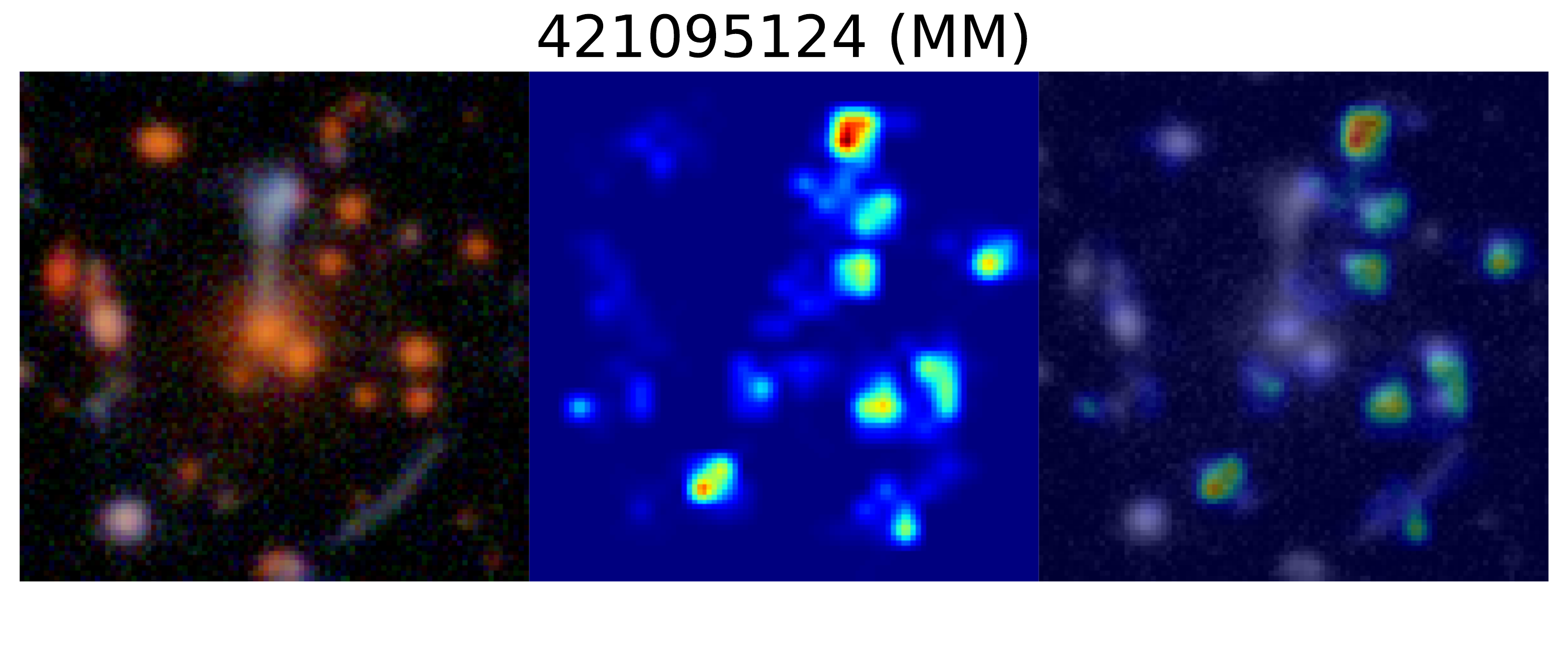}
    \includegraphics[width=\textwidth]{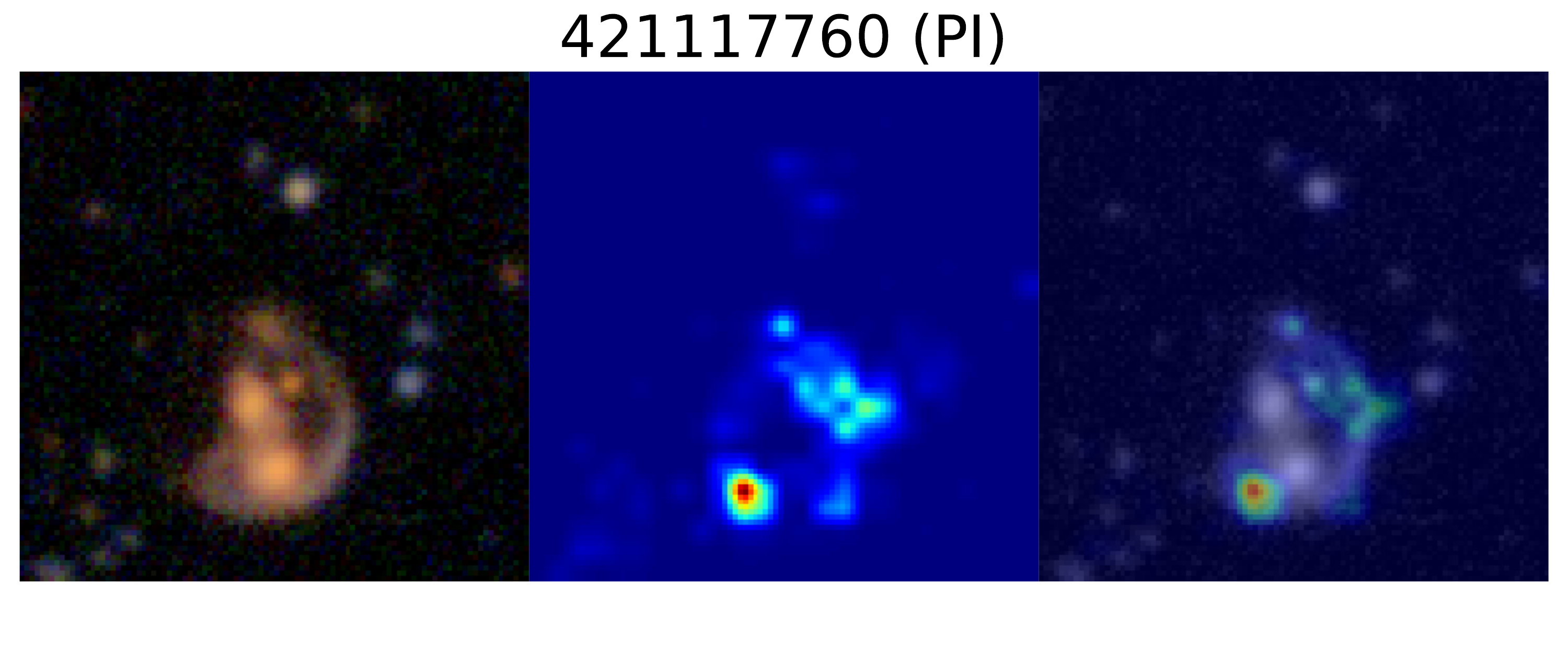}  
    \end{minipage}%
    \begin{minipage}{.5\textwidth}
      \centering
       
     \includegraphics[width=\textwidth]{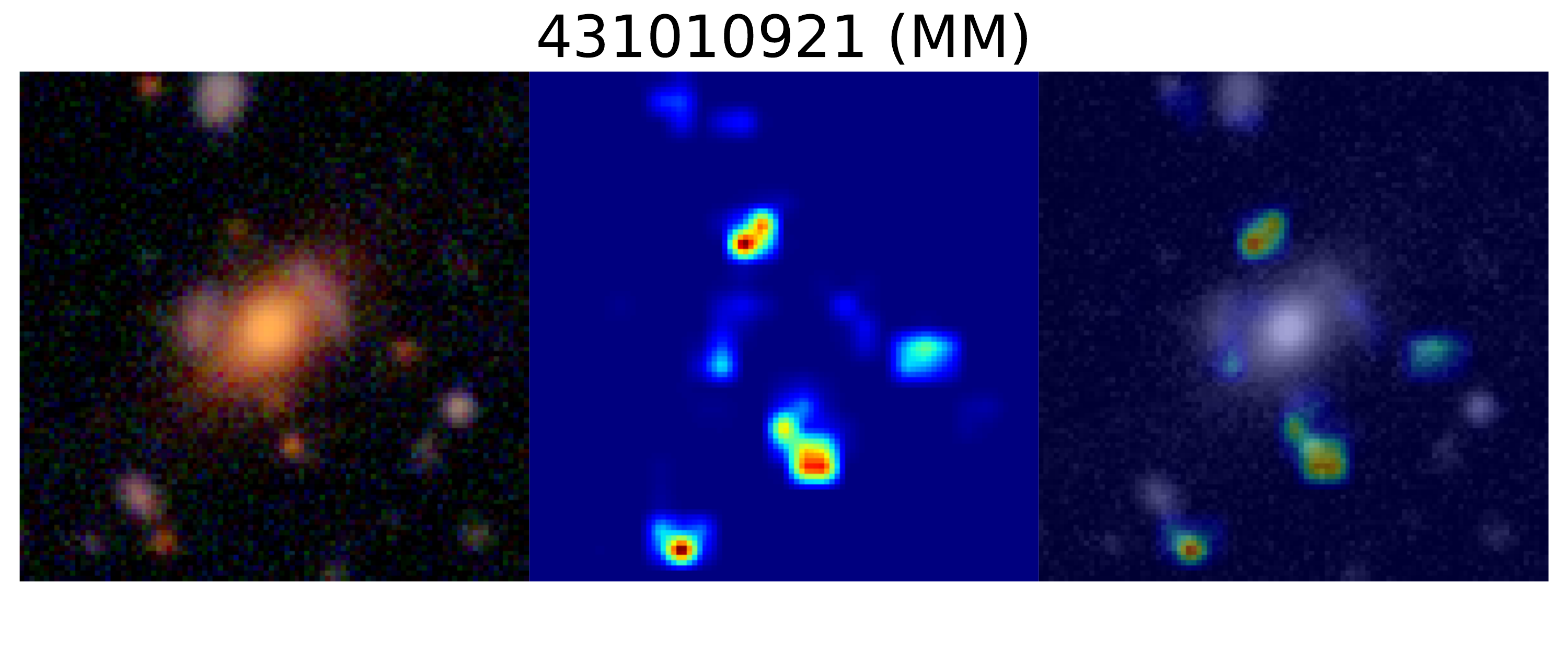}
      \includegraphics[width=\textwidth]{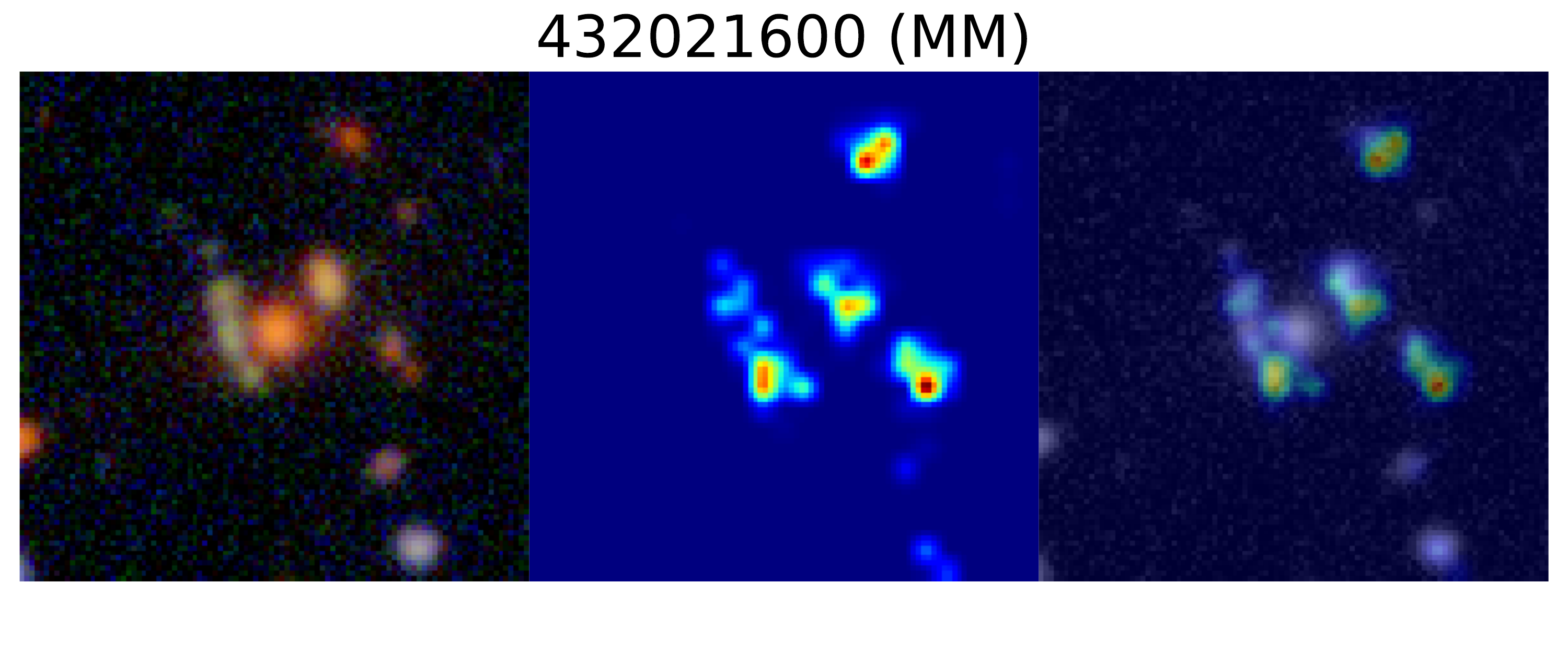}
      \includegraphics[width=\textwidth]{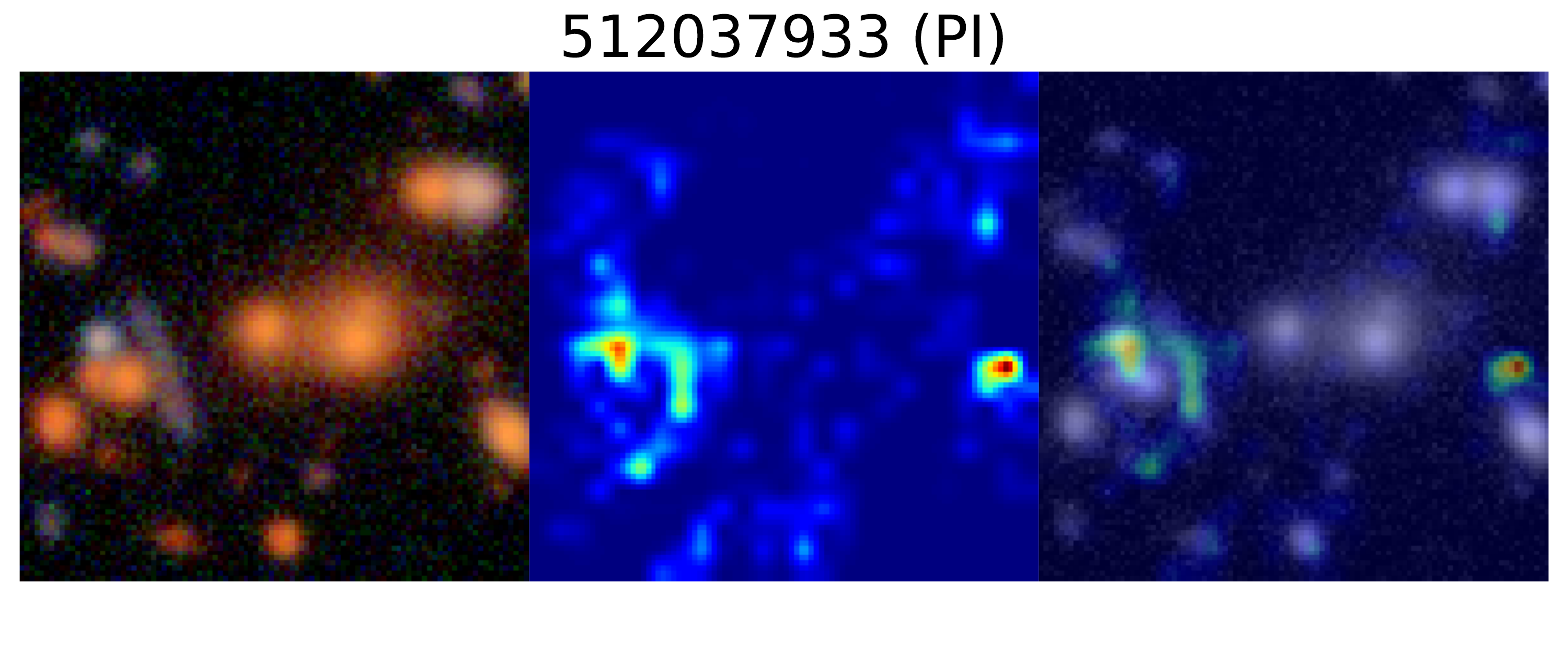}
      \includegraphics[width=\textwidth]{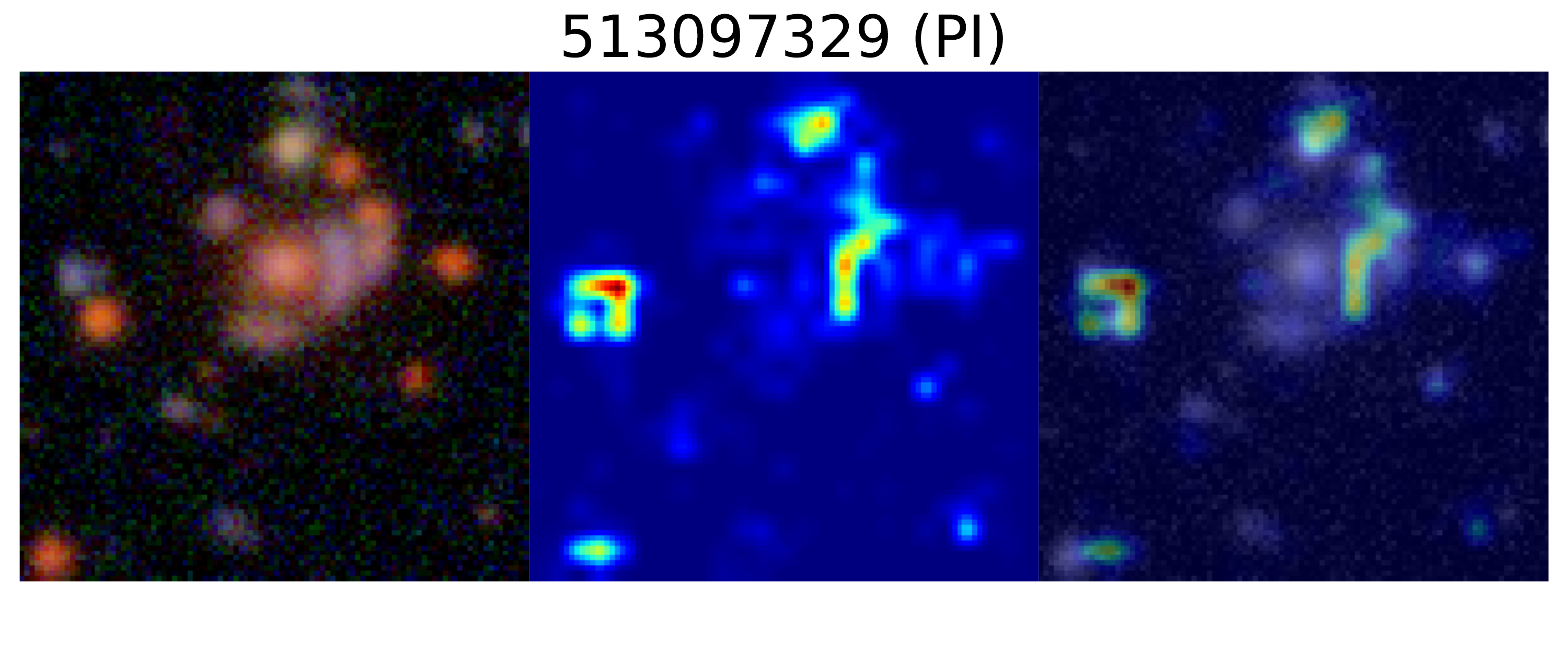}
    \end{minipage}
      \includegraphics[width=0.5\textwidth]{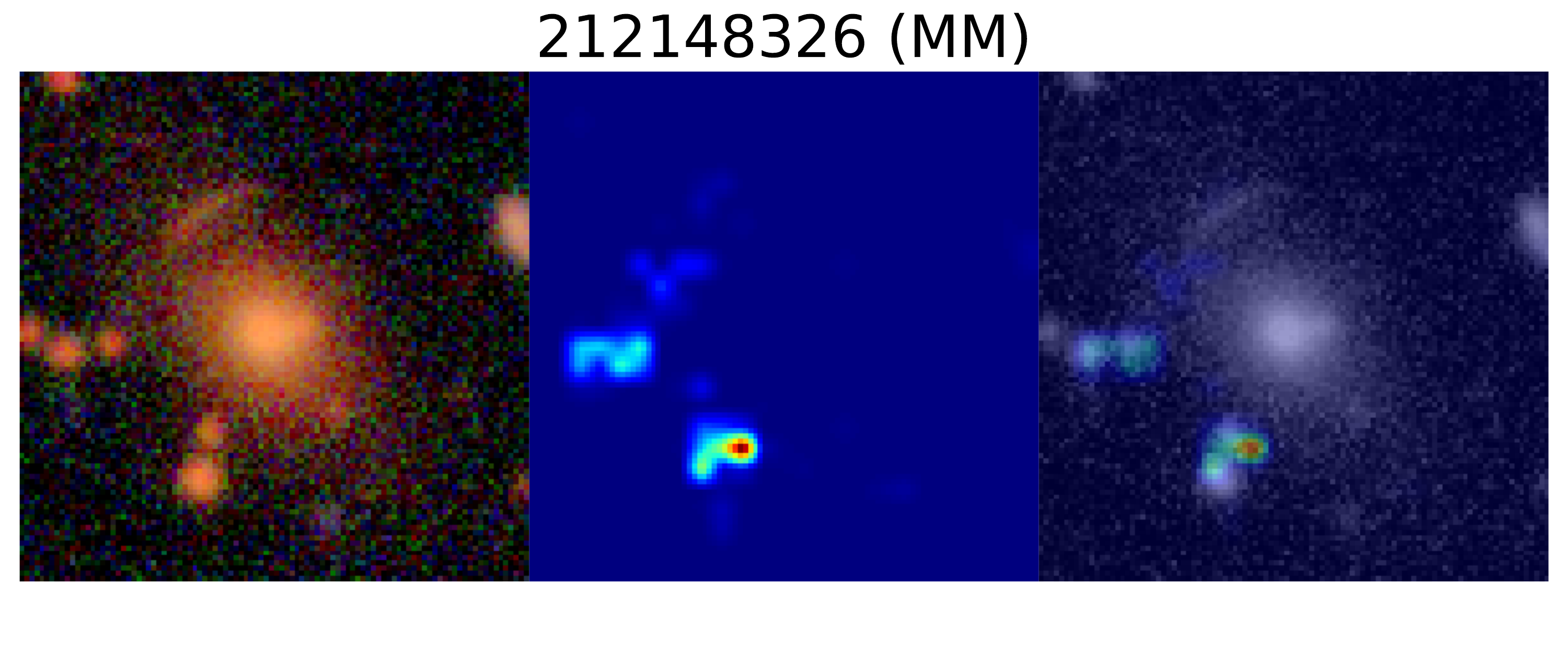}  
 
    \caption{GradCAM++ heatmaps for all Grade-A lenses, equivalent to Figure~\ref{fig:gradcam-heatmaps}.
    Each image is labeled with its object ID, and the model corresponding to the heatmaps (MM = GAN+MixMatch, PI = GAN+$\Pi$-model). }\label{fig:allgradcamImages}
\end{figure*}

We show the top 25 predicted lens candidates from the GAN+$\Pi$-model and GAN+MixMatch models in Figures~\ref{fig:top25-pimodel} and \ref{fig:top25-mixmatch}, respectively. These include several of our top lens candidates based on human inspection (see Figures~\ref{fig:GradeA-lensesfonud}) and \ref{fig:GradeB-lensesfonud}), but many do not show obvious signs of strong lensing. There are several duplicate images at slightly different sky positions as discussed in the main text. 
In Figure~\ref{fig:allgradcamImages} we include the GradCAM++ heatmaps obtained for all the Grade-A candidates (analogous to the example subsets shown in Figure~\ref{fig:gradcam-heatmaps}). These heatmaps were generated using our best performing models: GAN+MixMatch or GAN+$\Pi$-model (discussed in Section~\ref{subsec:Gracam}). 
Finally, we list the sky coordinates of all Grade-A and Grade-B lenses in Table~\ref{tab:all-lenses-DLS}.

\begin{table*}
    \centering
    \small
    \begin{tabular}{|ccccc|}
        \hline

        {Augmentation removed} & {TestV1 Precision(\%)} & {TestV2 Precision(\%)} & {TestV1 baseline difference(\%)} & {TestV2 baseline difference(\%)}\\ 

        \hline
        \hline
\

        & & {\bf Performance at 50\% recall rate} & & \\
         \hline
        None               & $84.95 \pm 8.70$ & $80.19 \pm 17.08$ & - & - \\
        GAN                & \color{red} $65.46 \pm 15.00$ & \color{red} $55.77 \pm 17.43$ & \color{red} -19.49 & \color{red} -24.42 \\
        RGB shuffle        & \color{red} $44.8 \pm 24.32$ & \color{red}$35.45 \pm 21.75$ & \color{red} -40.15 & \color{red} -44.74\\
        JPEG quality       & \color{red} $65.30 \pm 13.84$ & \color{red}$56.84 \pm 14.06$ & \color{red}-19.65 & \color{red} -23.35\\
        Rot90              & \color{blue} $91.81 \pm 4.41$ & \color{blue} $89.96 \pm 6.33$ & \color{blue}+6.86 & \color{blue} +9.77\\
        Translations       & \color{blue}$89.47 \pm 10.34$ & \color{blue}$84.59 \pm 13.04$ & \color{blue}+4.52 & \color{blue} +4.4\\
        Horizontal flips   & \color{red}$84.63 \pm 10.85$ & \color{red}$75.74 \pm 13.96$ & \color{red}-0.32 & \color{red} -4.45\\
        Color augmentation & \color{red}$71.88 \pm 17.35$ & \color{red}$68.23 \pm 15.2$ & \color{red}-13.07 & \color{red} -11.96\\
        
        \hline
        \hline

        & & {\bf Performance at 60\% recall rate} & & \\
         \hline
        None & $79.88 \pm 6.30$ & $78.03 \pm 12.45$ & - & -\\
        GAN & \color{red}$55.54 \pm 13.13$ & \color{red}$44.09 \pm 12.94$ & \color{red} - 24.34 & \color{red} -33.94\\
        RGB shuffle & \color{red}$34.25 \pm 24.49$ & \color{red}$26.14 \pm 15.9$ & \color{red} - 45.63 & \color{red} -51.89\\
        JPEG quality &\color{red} $52.75 \pm 24.68$ & \color{red}$51.69 \pm 17.91$ & \color{red} - 27.13 & \color{red} -26.34\\
        Rot90 & \color{blue} $84.94 \pm 7.04$ & \color{blue} $78.99 \pm 5.23$ & \color{blue} +5.06 & \color{blue} + 0.96\\
        Translations & \color{red}$74.95 \pm 16.11$ & \color{red}$72.26 \pm 24.65$ & \color{red} -4.93 & \color{red} -5.77\\
        Horizontal flips & \color{red}$71.15 \pm 11.46$ & \color{red}$60.64 \pm 12.27$ & \color{red} -8.73 & \color{red} -17.39\\
        Color augmentation & \color{red}$63.24 \pm 15.67$ & \color{red}$50.91 \pm 15.53$ & \color{red} -16.64 & \color{red} -27.12\\

        \hline
        \hline

        & & {\bf Performance at 70\% recall rate} & & \\
         \hline
        None & $69.42 \pm 4.60$ & $68.05 \pm 12.96$ & - & -\\
        GAN & \color{red}$46.45 \pm 14.49$ & \color{red}$37.24 \pm 12.85$ & \color{red} -22.97 & \color{red} -30.81\\
        RGB shuffle & \color{red}$26.68 \pm 17.28$ & \color{red}$18.75 \pm 11.78$ & \color{red} -42.74 & \color{red} -49.3\\
        JPEG quality & \color{red}$40.38 \pm 18.64$ & \color{red}$40.52 \pm 21.74$ & \color{red} - 29.04 & \color{red} -27.53\\
        Rot90 & \color{blue} $79.97 \pm 12.15$ & \color{blue} $73.69 \pm 7.11$ & \color{blue} +10.55 & \color{blue} +5.64\\
        Translations & \color{red}$67.69 \pm 12.67$ & \color{red}$59.31 \pm 22.34$ & \color{red} -1.73 & \color{red}-8.74\\
        Horizontal flips & \color{red}$67.36 \pm 11.95$ & \color{red}$55.94 \pm 14.71$ & \color{red}-2.06 & \color{red}-12.11\\
        Color augmentation & \color{red}$52.79 \pm 14.92$ & \color{red}$45.93 \pm 13.97$ & \color{red}-16.63 & \color{red} -22.12\\
        \hline
        \hline

        & & {\bf Performance at 80\% recall rate} & & \\
         \hline
        None & $54.35 \pm 4.57$ & $40.98 \pm 17.12$ & - & -\\
        GAN & \color{red}$33.02 \pm 10.26$ & \color{red}$24.37 \pm 10.50$ & \color{red} -21.33 & \color{red} -16.61\\
        RGB shuffle & \color{red}$15.34 \pm 7.59$ & \color{red}$11.75 \pm 4.90$ & \color{red} -39.01 & \color{red} -29.23\\
        JPEG quality & \color{red}$21.33 \pm 8.29$ & \color{red}$15.25 \pm 3.41$ & \color{red} -33.02 & \color{red} -25.73\\
        Rot90 & \color{blue} $62.63 \pm 12.43$ & \color{blue} $52.2 \pm 13.65$ & \color{blue} +8.28 & \color{blue} +11.22\\
        Translations & \color{red}$52.17 \pm 12.92$ & \color{red}$35.09 \pm 16.64$ & \color{red} -2.18 & \color{red} -5.89\\
        Horizontal flips & \color{blue} $60.54 \pm 9.78$ & \color{red}$36.46 \pm 9.03$ & \color{blue} +6.19 & \color{red} -4.52\\
        Color augmentation & \color{red}$39.47 \pm 13.24$ & \color{red}$33.84 \pm 10.92$ & \color{red} -14.88 & \color{red} -7.14\\

        \hline
         \hline

        & & {\bf Performance at 90\% recall rate} & & \\
         \hline
        None & $34.12 \pm 7.47$ & $16.33 \pm 8.661$ & - & -\\
        GAN & \color{red}$22.60 \pm 4.78$ & \color{red}$10.99 \pm 4.25$ & \color{red} -11.52 & \color{red} -5.34\\
        RGB shuffle & \color{red}$10.37 \pm 4.68$ & \color{red}$5.88 \pm 2.61$ & \color{red} -23.75 & \color{red} -10.45\\
        JPEG quality & \color{red}$15.15 \pm 6.37$ & \color{red}$6.18 \pm 2.41$ & \color{red} -18.97 & \color{red} -10.15\\
        Rot90 & \color{blue} $40.72 \pm 12.74$ & \color{blue} $21.79 \pm 3.08$ & \color{blue} +6.6 & \color{blue} +5.46\\
        Translations & \color{red}$32.88 \pm 3.35$ & \color{red}$16.14 \pm 6.72$ & \color{red} -1.24 & \color{red} -0.19\\
        Horizontal flips & \color{red}$30.81 \pm 20.44$ & \color{red}$14.72 \pm 6.59$ & \color{red} -3.31 & \color{red} -1.61\\
        Color augmentation & \color{red}$23.86 \pm 8.13$ & \color{red}$14.25 \pm 8.45$ & \color{red} -10.26 & \color{red} -2.08\\
        
        \hline
        \hline

        & & {\bf Performance at 100\% recall rate} & & \\
         \hline
        None & $8.25 \pm 2.85$ & $6.05 \pm 2.69$  & - & -\\
       
        GAN & \color{red} $8.13 \pm 2.49$ & \color{red} $5.79 \pm 3.93$ & \color{red} -0.12 & \color{red} -0.26\\
        RGB shuffle & \color{red} $6.43 \pm 0.97$ & \color{red}$3.84 \pm 1.02$ & \color{red} -1.82 & \color{red} -2.21\\
        JPEG quality & \color{red} $5.88 \pm 0.21$ & \color{red}$4.14 \pm 2.09$ & \color{red} -2.37 & \color{red} -1.91\\
        Rot90 & \color{blue} $14.76 \pm 8.42$ & \color{blue} $13.00 \pm 5.16$ & \color{blue} +6.51 & \color{blue} +6.95\\
        Translations & \color{blue} $10.83 \pm 2.46$ & \color{blue} $7.37 \pm 3.66$ & \color{blue} +2.58 & \color{blue} +1.32\\
        Horizontal flips & \color{blue} $15.74 \pm 3.06$ & \color{blue} \color{blue} $8.86 \pm 1.84$ & \color{blue} +7.49 & \color{blue} +2.81\\
        Color augmentation & \color{blue} $11.42 \pm 7.25$ & \color{blue} $6.84 \pm 4.12$ & \color{blue} +3.17 & \color{blue} +0.79\\
        \hline
    \end{tabular}
    \caption{Ablation performance for 50--100\% recall rates (in steps of 10\%) for the GAN+Supervised model using TrainingV2. The first row of each recall rate shows the baseline precision value obtained from the model on the test sets (TestV1, TestV2) when none of the augmentations are removed. In the subsequent rows, we report the precision obtained when the model was trained without the specified augmentation. For example, the baseline model at 50\% recall has a precision of 80.19\% for TestV2 and decreases to 55.77\% when GAN images are removed during training. The difference in the obtained precision values are quoted in the last two columns. Augmentations which improve model performance (i.e., improve precision when included and decrease decrease precision when removed) are shown in red, while those which decrease model performance are shown in blue. Overall, the models perform worse when color augmentations, JPEG quality and GANs are not included, indicating that these augmentations are important for optimal performance. 
    The errrors quoted here are 1$\sigma$.}
    \label{tab:ablation_tests}
\end{table*}

\begin{table*}
\begin{tabular}{|c|c|c|c|c|c|}
\hline
object ID &       RA &      DEC & Field &  Rank (GAN+MixMatch) &  Rank (GAN+$\Pi$-model) \\
\hline
 & & & {\bf Grade-A candidates} & & \\
 \hline
421095124 &  163.792076 &  -5.070373 &    F4 &              2 &            12 \\
513097468 &  209.340092 & -10.244328 &    F5 &             38 &            13 \\
212072337 &  139.896040 &  30.532355 &    F2 &            181 &            21 \\
322054393 &   79.839914 & -48.949647 &    F3 &            733 &          8326 \\
432021600 &  162.750073 &  -5.941902 &    F4 &           1262 &         12424 \\
431010921 &  163.364259 &  -5.789092 &    F4 &           1279 &         19826 \\
512037933 &  209.677055 & -10.687652 &    F5 &           7461 &          2068 \\
421117552 &  163.897903 &  -5.054885 &    F4 &           4799 &          2768 \\
212148326 &  139.512033 &  30.953524 &    F2 &           3579 &         23223 \\
\hline
\hline
 & & & {\bf Grade-B candidates} & & \\
\hline
313032462 &   78.742878 & -48.149829 &    F3 &            365 &            59 \\
331108599 &   81.300608 & -49.432676 &    F3 &           1974 &            98 \\
132023380 &   13.551513 &  11.794606 &    F1 &           3400 &           462 \\
533097114 &  209.328083 & -11.993324 &    F5 &            518 &           676 \\
433116975 &  162.551767 &  -5.697394 &    F4 &          13673 &           720 \\
233074254 &  139.046712 &  29.298535 &    F2 &            870 &          2320 \\
413115231 &  162.585545 &  -4.498548 &    F4 &           8839 &           884 \\
211134050 &  140.304878 &  30.471131 &    F2 &          12662 &           979 \\
122079323 &   13.182525 &  12.323637 &    F1 &           8567 &          1145 \\
312158847 &   80.455801 & -48.489660 &    F3 &           3896 &          1209 \\
322092794 &   80.115321 & -49.246309 &    F3 &           1234 &         24945 \\
421019105 &  163.411890 &  -4.870280 &    F4 &           1996 &          2702 \\
221061603 &  140.662872 &  29.846367 &    F2 &           3990 &          8584 \\
\hline
\end{tabular}
\caption{Grade-A and Grade-B Lens candidates found from this work with their object ID, RA and DEC coordinates, DLS field (F1 through F5), and their corresponding ranks from GAN+MixMatch and GAN+$\Pi$-models.
The rank is obtained by passing all the survey images (281,425 objects in total; Section~\ref{sec:DLS-details}) through the models and sorting them based on their prediction scores. High-confidence Lens candidates have lower ranks and high prediction scores. For example, the Grade-A lens candidate DLS212072337 whose lensing nature has been spectroscopically confirmed (Section~\ref{sec:specz}) has a rank of 21 from the GAN+$\Pi$-model and a prediction score of $\simeq1$. The ranks quoted here represent an upper bound on the number of images an investigator has to look at to find the lens candidate, as they do not account for duplicated sky regions which we remove before visual inspection (as discussed in Section~\ref{sec:catalog}),  reducing the number of unique lens candidates investigated. 
}\label{tab:all-lenses-DLS}
\end{table*}

\bsp	
\label{lastpage}

\end{document}